\documentclass[10pt]{article}
\usepackage[utf8]{inputenc}
\usepackage [english]{babel}
\usepackage{calc}
\usepackage{microtype}
\usepackage{pifont}
\usepackage{array}
\usepackage{cancel}
\usepackage{tikz}
\newcommand{\cir}[2]{\begin{tikzpicture}
\node[draw,scale=0.7,circle,{#1},{#2}] at (0,0) {}; 
\end{tikzpicture}}
\usepackage{tikz-cd}
\usepackage{bm}
\usepackage{varwidth}
\usepackage{utfsym}
\usepackage{pifont}
\usepackage{graphicx}
\let\oldding\ding
\renewcommand{\ding}[2][1]{\scalebox{#1}{\oldding{#2}}}
\usepackage{geometry}
\usepackage{float}  
\usepackage [autostyle, english = american]{csquotes}  
\MakeOuterQuote{"}
\usepackage[backend=biber, style = numeric-comp, maxbibnames=5, minbibnames=5, safeinputenc,isbn=false,doi=false, giveninits=true, sorting = none,date=year]{biblatex}
\renewbibmacro{in:}{}
\AtEveryBibitem{\clearlist{language}}
\AtEveryBibitem{\clearfield{pages}} 
\usepackage{graphicx}
\usepackage{amsmath}  
\usepackage{amssymb} 
\usepackage{mathtools}
\usepackage{caption}  
\usepackage{enumitem}  
\usepackage{authblk}
\usepackage{multicol}
\usepackage{cellspace}
\usepackage{circuitikz}
\usepackage{multirow}
\title{\textbf{Design of a minimal, allosteric, and ATPase-like machine using mechanical linkages}}

\usepackage[colorlinks,citecolor=blue,linkcolor=black,hypertexnames=true,urlcolor=blue]{hyperref}
\usepackage{xcolor}  

\usepackage{booktabs}

\newcommand{\WidestEntry}{$1.$}
\newcommand{\SetToWidest}[1]{\makebox[\widthof{\WidestEntry}]{$#1$}}

\usepackage{abstract}
\addto{\captionsenglish}{} 

\author{Tosan Omabegho}

\affil{{tosanomabegho@gmail.com}}

\date{\vspace{-5ex}}  

\begin{document}

\maketitle

\begin{abstract}

ATPases cyclically convert chemical energy in the form of ATP gradients into directed motion inside cells. To function, ATPases rely on allosteric communication between at least two binding sites—an internal signaling mechanism that is not well understood. Here, we model an ATPase-like machine by using a system of mechanical linkages to recreate negative allosteric coupling between two binding sites and generate cycles in which the sites alternate occupancy. The ATPase analog has two mechanical degrees of freedom and two discretized binding sites: one for the ATP, Pi and ADP analogs, and one for an allosteric effector analog. The geometry of the ATPase analog allows stepwise binding reactions at each site to capture the two degrees of freedom in a mutually exclusive way. Consequently, the enzyme interconverts between multiple rigid and partially rigid forms, such that neither site can be fully bound when both sites are occupied. Two mechanisms work together to generate an enzymatic cycle: one, in which the tighter-binding ATP analog can bind and displace the effector from the enzyme; and a second, in which flexibility introduced by splitting the ATP analog into two pieces (catalysis) allows the effector to rebind and displace the products (ADP analog). We show that cleavage (forward catalysis) and ligation (reverse catalysis) alter the rigidity of the enzyme complex equivalently to binding and dissociation, respectively, but must do so more slowly for effective cycling to take place. Simple designs for synthetic systems that mimic ATPase monomers can be derived from this work.

\end{abstract}

\begin{refsection}

\section*{Introduction}

ATPases are enzymes that split ATP into Pi and ADP in order to carry out various transport functions. ATPases include cytoskeletal motors \cite{Vale2003-hj}, most membrane pumps \cite{Palmgren2023-um}, translocases and packaging motors \cite{Khan2022-ww} and other families of motors. A long-term goal is to create synthetic molecular machines that mimic the abilities of ATPases. However, ATPases, like many other proteins, rely on allosteric communication to operate \cite{Thirumalai2019-xa, Vologodskii2006-if}, and developing mechanisms and concepts to mimic allostery is a difficult problem \cite{Henzler-Wildman2007-hf}.

Allostery is intramolecular communication between at least two non-overlapping reactive sites such that binding or catalysis of a ligand at one site affects ligand affinity at the other site \cite{Monod1965-pi, Cui2008-nv}. The sites and the ligands that bind to these sites are said to be allosterically "coupled". ATPases mediate coupling between ATP (and ADP and Pi) and a chemically different ligand, or allosteric effector \cite{Vologodskii2006-if, Thirumalai2019-xa}, which is not consumed in the enzymatic process, as is ATP. For example, for myosin, the nonconsumed, coupled ligand is actin; for kinesin and dynein, it is microtubule; and for helicases, it is DNA. The variety of ATP and ligand pairings that exist for ATPases and the variety of structures that have evolved to mediate allostery between them suggest a common language of allosteric communication is at work, a language that can possibly be mimicked if not yet fully understood.

Recently, a number of works have used a mechanical linkage (framework) abstraction to design synthetic allosteric structures \cite{Shen2004-cn, Zhou2012-dt, Ke2016-zr, Rocks2017-rw, Flechsig2017-yf, Yan2017-by, Song2017-iw, Zhang2022-fd, Pillai2024-wp, Madhvacharyula2025-ft, Broerman2025-va}. In the framework abstraction, molecules are viewed as networks of rigid bars or modules connected by flexible hinges, which change shape in response to a signal \cite{Schulze2013-cd, Hermans2017-nw}. These works combine the abstraction with different materials, making use of elastic networks \cite{Rocks2017-rw, Flechsig2017-yf, Yan2017-by}, synthetic DNA \cite{Shen2004-cn, Zhou2012-dt, Ke2016-zr, Zhang2022-fd, Madhvacharyula2025-ft}, or synthetic proteins \cite{Pillai2024-wp, Broerman2025-va}. Most of the experimental works demonstrate allosteric structures that can convert a binding reaction at one site into a chemical signal at a non-overlapping site. For example, in several works they demonstrate how ligand binding can trigger dissociation of an internal bond or of another ligand \cite{Shen2004-cn, Zhou2012-dt, Ke2016-zr, Zhang2022-fd, Pillai2024-wp, Broerman2025-va}. And in one of these works \cite{Pillai2024-wp}, they show how ligand binding in one subunit can increase ligand affinity in repeated subunits. By contrast, the elastic network studies focus on computationally evolving networks to transmit conformational change between non-overlapping sites. In place of reactions with abstracted ligands, strain is used to represent chemical reactivity.

The behavior in all these works can be characterized as switchlike, where in response to some external stimuli, the frameworks change from one conformational state to another to reach some new equilibrium or energy minimum. The initial concept and models of allostery were developed to explain switchlike behavior in signaling proteins \cite{Changeux1961-mk, Changeux2005-oe}, most famously hemoglobin \cite{Monod1965-pi, Koshland1966-ed}. The linkage works explicitly or implicitly take inspiration from these early models, in particular, from the idea that two states, at a minimum \cite{Changeux2005-oe}, can describe allosteric switching in a protein subunit.

However, to reproduce the autonomous and cyclic activity that characterizes ATPases, one-way allosteric switching is not sufficient. To operate cyclically, at a fixed thermodynamic potential of ATP, ATPases go through two phases of allosteric switching in which intermediate states, most importantly the product-bound state, play a critical role. The phases result in the exchange of both nucleotide and effector, where each binds and releases over the course of the cycle, and each acts as an exchange factor for the other in a mutually dependent (reciprocal) manner \cite{Branscomb2017-af, Carter2019-kp, Tafoya2018-iw}. Synthetic systems that emulate ATPases and their ability to autonomously cycle by switching allosterically and reciprocally have not been described.

\begin{figure}[H]
\begin{center}
\includegraphics[width=1\columnwidth]{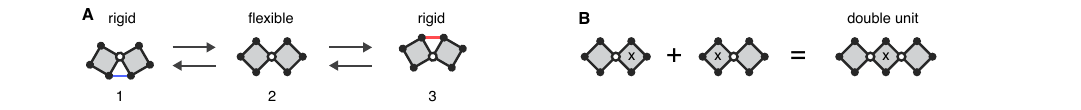}
\caption{\textbf{Three-state switch.} \textbf{A,} A single unit. The single unit consists of two rigid squares connected at a flexible node (white). Three conformations are possible: a flexible one (2), in which no additional bars are connected; a rigid one (1), in which a bar (blue) connects the two bottom nodes to form a rigid triangle; and another rigid one (3), in which a bar (red) connects the two top nodes to form a rigid triangle. \textbf{B,} A double unit. A double unit is formed by merging one square (marked with `x') of two single units. The double unit has four completely rigid conformational states, and four partially rigid conformational states, as described later in more detail.}
\label{fig:zero}
\end{center}
\end{figure}

Here, we use linkages to model an ATPase-like machine. We explore a number of additional concepts previously explored by others: discretized binding pockets \cite{Huxley1971-gv, Koshland1958-fq, Koshland1963-gq}; enzyme flexibility in relation to allostery \cite{Koshland1963-gq} and hydrolysis \cite{Becker1992-ax, Chen2004-ah, Eide2006-lz, Rivoire2024-ur}; timescale separation \cite{Chatzittofi2025-hb, Sing2014-du, Brown2022-ik}; spontaneous dissociation from weak bonds \cite{Chen2004-ah, Bath2005-ih, Sing2014-du, Gibb2014-zh, Brown2022-ik}; facilitated dissociation \cite{Sing2014-du, Gibb2014-zh, Brown2022-ik}; and floppy mechanical modes \cite{Meeussen2020-wo, Liu2025-uw}. The building block of the model is an allosteric switch first described by Koshland to model competitive inhibition \cite{Koshland1963-gq}, which we convert into a linkage (Fig. \ref{fig:zero}). The switch has a single mechanical degree of freedom and three conformational states: two rigid states and one flexible intermediate. Nodes on the top and bottom form two discretized binding sites. The ATPase analog is two connected switch units (Fig. \ref{fig:zero}B). The geometry gives it two mechanical degrees of freedom. Four additional linkage reactants, which are simple chains of bars, complete the ATPase system. These linkages are analogs for ATP, Pi, ADP, and an allosteric effector. The analogs for ATP, Pi, and ADP bind to one site on the ATPase analog, and the effector binds to the other.

We make two basic assumptions that govern how the linkages interact with one another in the model. First, we assume that the ATPase analog (enzyme) and the complexes it forms are mechanically equilibrated and free of elastic bias. This assumption has been made in some models of biomolecular machines \cite{Fox1998-pz, Dean_Astumian2018-gf}, and in the design of mechanical networks with floppy modes \cite{Meeussen2020-wo, Liu2025-uw}. Consequently, a conformational change in the enzyme linkage—to a more flexible or a more rigid state—results only from a change in the connectivity of the network, meaning the removal or addition of bars, rather than from an application of strain. Second, we use the stepwise addition and removal of bars \cite{Bolker1979-gd, Obukhov1995-gq, Ellenbroek2011-es}, which, respectively, create rigidity and flexibility, as analogs for thermally driven bond making and bond breaking. Motivated by an earlier version of this paper, reference \cite{breik2021molecular} used this bonding abstraction to develop a formal model of interacting linkages. They also showed that complex behavior can be achieved in an idealized topological model, which considers solely the graph connectivity of the linkages.

To mimic negative allosteric coupling, the four reactants are designed to have bar lengths that prevent both sites on the enzyme from being fully bound at the same time. The arrangement creates a dynamic situation in which the enzyme interconverts between multiple rigid and partially rigid states when both sites are occupied. By tuning the off rates at each node, the arrangement allows one reactant to bind and displace another from the enzyme in a stepwise and stochastic manner. We refer to the mechanism as an \textit{allosteric displacement}.

To mimic hydrolysis, the ATP analog is allowed to be cleaved in two at a non-binding moiety (node) after it binds. The design is inspired by the use of DNA enzymes in synthetic systems \cite{Chen2004-ah}. Cleaving introduces flexibility and generates the Pi and ADP analogs. Importantly, cleaving does not affect binding nodes, and thus the Pi and ADP analogs bind to the same nodes as does the ATP analog. Using stochastic simulations, we show how the hydrolysis mimic enables continuous cycles of allosteric displacement to take place between the system of five linkages. During each cycle, the ATP analog displaces the effector, the ATP analog is cleaved, and the effector displaces the ADP analog to reset the cycle. We also characterize the multiple ways in which the system can go through futile cycles.

The main result is showing how an autonomous and allosterically operating enzymatic cycle can be constructed with a consistent set of reaction rates for catalysis and for binding to and dissociating from a set of weak bonds. These rates remain the same over the course of the cycle—they are not altered depending on the stage of the cycle the enzyme is in or the identity of the reactant active at that stage. Additionally, we give a simple interpretation for what reversible catalysis accomplishes. Forward catalysis frees a mechanical degree of freedom but must do so slowly enough for the bound effector to be displaced first and not compete for the degree of freedom. Conversely, reverse catalysis traps the degree of freedom but must do so slowly enough for a new effector to bind and capture the degree of freedom.

\section*{Results}

\subsection*{Basic description of the system}

This section describes, in the following order, the chemical cycle of the system, the linkage reactants, the rules that govern complex formation, and the construction of the reaction network.

The chemical cycle of the system is derived from the cycle of a myosin or dynein monomer (see SI \ref{si_myosin_cycle}). There are five reactants: the ATPase analog (the enzyme); the ATP analog (S, for `substrate'); the Pi analog (P1, for `product 1'); the ADP analog (P2, for `product 2'); and the effector analog (L, for `ligand'). With these five reactants, the cycle can be stated this way (note that because every state is a bound state of the enzyme, the enzyme is not listed):

\begin{equation}
    \overbrace{\strut \{\text{L\}} \xrightarrow[]{1} \text{\{S/L\}} \xrightarrow[]{2} \text{\{S}\}}^{\text{S displaces L}} \xrightarrow[]{\protect\usym{2702}, \, 3} \text{\{P1, P2}\} \xrightarrow[]{4} \overbrace{\strut \text{\{P2\}} \xrightarrow[]{5} \text{\{P2/L\}} \xrightarrow[]{6} \{\text{L\}}}^{\text{L displaces P2}} \dashrightarrow
\end{equation}

\noindent There are six reactions: 1, S binds; 2, L dissociates; 3, S is cleaved; 4, P1 dissociates; 5, L binds, and 6, P2 dissociates. Each reaction connects two states, with six states in total (in curly braces). The backslash (/) in each state separates the reactants that bind at the two binding sites (\{substrate site/ligand site\}). No backslash indicates that only one site is occupied. Thus, in the first state, $\{\text{L\}}$, the ligand site is occupied by a single molecule of ligand, denoted by L, and the substrate site is empty. Each displacement sequence (S displaces L, and L displaces P2) includes a binding reaction, followed by a dissociation. The reaction sequence thus says, substrate displaces ligand (1 \& 2), substrate is cleaved (3), P1 spontaneously dissociates (4), and finally ligand displaces P2 (5 \& 6).

\begin{figure}[H]
\begin{center}
\includegraphics[width=1\columnwidth]{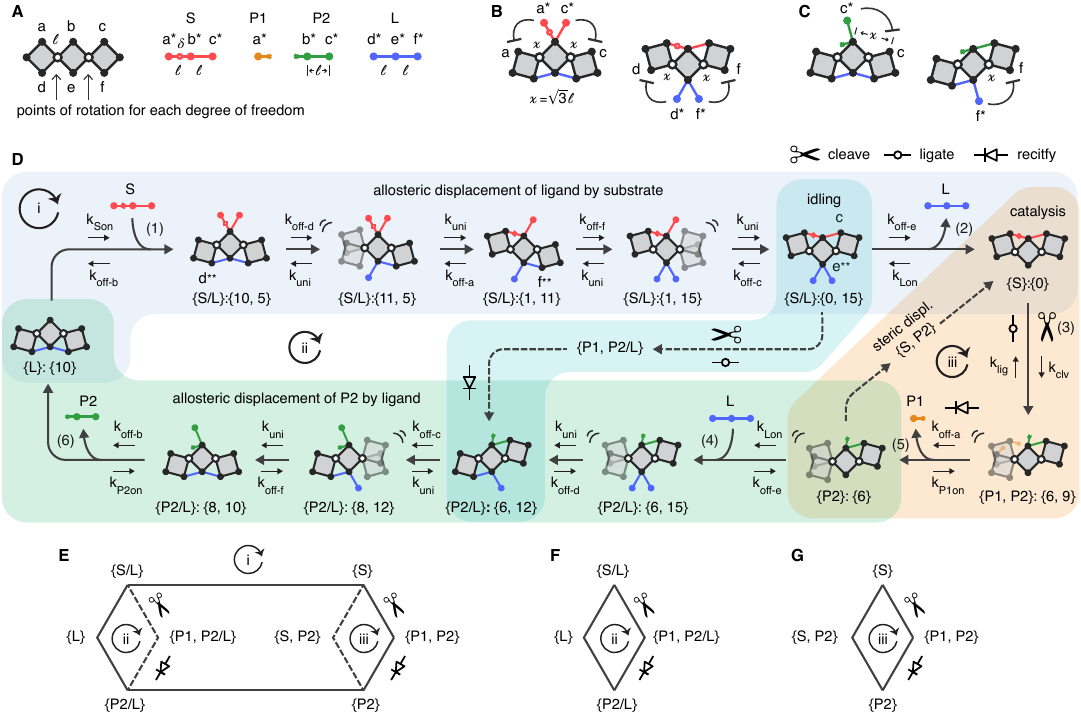}
\caption{\textbf{The linkage system.} \textbf{A,} Five molecules of the linkage system. The enzyme consists of three rigid squares of edge length $\ell$, connected at two points of rotation. Nodes \textsf{a}, \textsf{b} and \textsf{c} form the binding site for the substrate, P1 and P2. Nodes \textsf{d}, \textsf{e} and \textsf{f} from the binding site for the ligand. Complimentary nodes on S, P1, P2 and L, are denoted by the same letter with the superscipt "*". S and L consist of three nodes connected by two bars of length $\ell$, where the center nodes on each are points of flexibility. S is split into P1 and P2 at the special node $\delta$. \textbf{B,} Left, L blocks S. When L is bound at all three nodes, it bends the enzyme down, placing the two outer nodes out of reach for S. The increased distance between S's nodes is $x =\sqrt{3}\ell$. Right, symmetric blocking of L by S. \textbf{C,} Left, L blocks P2. Right, symmetric blocking of L by P2. \textbf{D,} Target cycle and two futile pathways. The target cycle (cycle i; outside black path going clockwise) takes place in thirteen reversible reactions. Each state name (e.g. \{S/L\}:\{10,5\}) indicates its composition (\{S/L\}) and unique linkage state (\{10,5\})(see SI \ref{si_crns}). Forward and reverse rates governing each reaction are labeled on each edge (e.g. between \{S/L\}:\{10,5\} and \{S/L\}:\{11,5\}, the forward rate is $\text{k}_{\text{off-d}}$, and the reverse rate is $\text{k}_{\text{uni}}$). Reactions numbered 1-6 mark the same changes in composition that take place Eq. 1. Starting at \{L\}, the allosteric displacement of L by S (light blue bubble) takes place in four intramolecular steps between 1 and 2, and completes with the enzyme bound to only S (\{S\}). Subsequently (orange bubble), S is cleaved into P1 and P2 (3, \protect\usym{2702}). Directly following cleavage, or rounds of ligation (\textbf{-}\kern-0.1em\textbf{-}\kern-0.1em$\protect\cir{scale=0.55}{thick}{}$\kern-0.1em\textbf{-}\kern-0.1em\textbf{-}) and cleavage, P1 spontaneously dissociates (4), which rectifies catalysis and leaves the enzyme bound to only P2 (\{P2\}). After L binds (5), the allosteric displacement of P2 by L (green bubble) takes place in three intramolecular steps between 5 and 6, and returns the enzyme to the start of the cycle and bound to only L \{L\}. The two futile pathways are \textit{idling} (blue bubble and dashed line) and \textit{steric displacement} (orange bubble and dashed line). Both pathways are abbreviated, and shown without figures. The change in the bound state along each path is stated: \{P1, P2/L\} in idling; and \{S, P2\} in steric displacement. Combined with states along the target path, idling is cycle ii, and steric displacement is cycle iii. \textbf{E,} Simplified network version of the system. Each node here represents a set of states with different intramolecular conformations. \textbf{F,} Idling cycle. \textbf{G,} Steric displacement cycle.}
\label{fig:one}
\end{center}
\end{figure}

The linkage system generates the hidden microstates and reactions required to explain the displacement reactions and explain how the displacements are linked together by catalysis (Fig. \ref{fig:one}). These microstates and reactions lie within the \{S/L\} and \{P2/L\} states in Eq. 1. They reflect \textit{intra}molecular changes that take place when both multivalent binding sites are occupied. In the complete cycle, \{S/L\} and \{P2/L\} are both expanded into a series of microstates. Each microstate in each series has the same reactants but a different multivalent configuration. Microstates are constructed from a set of basis states, where each basis state reflects a unique connection between reactant and enzyme (see SI \ref{si_crns}.1).

As described in the introduction, the enzyme consists of two connected switch units. The arrangement creates a chain of three equivalent rigid squares of edge length $\ell$ that are flexibly connected at two hinge nodes (Fig. \ref{fig:one}A). The resulting linkage has two mechanical degrees of freedom. Chemical specificity is added by defining six nodes on the enzyme to be unique monovalent binding sites for six complementary nodes on the four reactants. The three nodes on top (a, b, and c; red) form the multivalent binding site for substrate, P1, and P2, and the three nodes on the bottom (d, e, and f; blue) form the binding site for the ligand. Chemical specificity means that node binding is specific. Consequently, substrate, P1, and P2 cannot bind to the ligand site, and ligand cannot bind to the substrate site. In addition to its three binding nodes, the substrate contains a catalytic moiety, named $\delta$, where it is split into P1 and P2 when bound to the enzyme. The split takes place on the left side of the enzyme, between the gray-colored squares.

Any single node interaction between the enzyme and a reactant forms a flexible connection. However, two adjacent node interactions create rigidity, as each divalent connection forms a triangle when bracing a hinge (Figs. \ref{fig:one}B \& \ref{fig:one}C). Three adjacent node interactions brace both hinges and thus make the whole enzyme rigid. The different rigid complexes bend one or both sides of the enzyme linkage up, down, or up and down, relative to the center square. These opposing bends create the basis for negative allosteric coupling in the system. When the ligand is fully bound to all three nodes and bends the enzyme down, the nodes of the substrate's binding site are spread out, allowing the substrate to bind only a single node at a time (Fig. \ref{fig:one}B, left). Likewise, trivalent substrate binding bends the enzyme upwards and geometrically inhibits ligand binding (Fig. \ref{fig:one}B, right). The same negative relationship holds for ligand and P2 binding, but only on the right side, where P2 can bind two nodes (Fig. \ref{fig:one}C). The full set of restrictions is described in SI \ref{si_crns}.2.

The geometric restrictions were used to enumerate all states in the system and connect them together to form a reaction network. Any two states connected in the network differ only by one node connection and thus can interconvert to one another (see SI \ref{si_crns}). There are six possible interconversions, or linkage reactions (named before arrow), each of which maps to a chemical reaction (named after arrow):

\begin{enumerate}[label=\Roman*.]
    \item connect linkages $\rightarrow$ intermolecular binding
    \item increase the connection $\rightarrow$ intramolecular binding
    \item decrease the connection $\rightarrow$ intramolecular dissociation
    \item disconnect linkages $\rightarrow$ intermolecular dissociation
    \item split substrate $\rightarrow$ cleavage
    \item make substrate $\rightarrow$ ligation
\end{enumerate}

We assigned reaction rates (Table 1) to the six reaction types in the following way. Intramolecular binding (\RN{2}: $\text{k}_{\text{uni}}$) was assigned to be the fastest rate by several orders of magnitude and serves as the speed limit for all other parameters. This assignment reflects the idea that once a reactant makes an initial connection to an enzyme, additional interactions should take place on a relatively rapid timescale because of the small distances involved, which in turn favors multivalent binding (ref!). Intermolecular binding (\RN{1}: $\text{k}_{\text{Son}}, \text{k}_{\text{Lon}}, \text{k}_{\text{P1on}}, \text{k}_{\text{P2on}}$) ranges from fast to slow, depending on the "concentration" of the reactant binding (denoted by the subscript), where molar concentrations were converted to numbers for the stochastic simulations (see SI \ref{si_stochrates}). Intramolecular dissociation (\RN{3}) and intermolecular dissociation (\RN{4}) are both defined as "off" rates for the nodes ($\text{k}_{\text{off-a}}, \text{k}_{\text{off-b}}, \text{k}_{\text{off-c}}, \text{k}_{\text{off-d}}, \text{k}_{\text{off-e}}, \text{k}_{\text{off-f}}$). These rates represent intramolecular dissociation if the reactant is bound by more than one node and represent intermolecular dissociation if it is bound by a single node. The off rates were assigned to be much slower than $\text{k}_{\text{uni}}$, and all except $\text{k}_{\text{off-b}}$ are faster than cleavage and ligation. Finally, cleavage (\RN{5}: $\text{k}_{\text{clv}}$) and ligation (\RN{6}: $\text{k}_{\text{lig}}$) were assigned to be slow, which helps the system oscillate between substrate and ligand binding and is discussed in detail later.

\begin{table}[H]
\centering
\begin{tabular}{lll}  
rate & value ($\text{s}^{-1}$) & reaction type \\
\midrule
$\text{k}_{\text{reactant-on}}$ & various & \RN{1} \\
$\text{k}_{\text{off-a}}$ & $250$ & \RN{3}, or \RN{4} \\
$\text{k}_{\text{off-b}}$ & $3$ & \RN{3}, or \RN{4}\\
$\text{k}_{\text{off-c}}$ & $680$ & \RN{3}, or \RN{4}\\
$\text{k}_{\text{off-d}}$ & $680$ & \RN{3}, or \RN{4}\\
$\text{k}_{\text{off-e}}$ & $200$ & \RN{3}, or \RN{4}\\
$\text{k}_{\text{off-f}}$ & $680$ & \RN{3}, or \RN{4}\\
$\text{k}_{\text{uni}}$ & $1\times10^6$ & \RN{2}\\
$\text{k}_{\text{clv}}$ & $10$ & \RN{5}\\
$\text{k}_{\text{lig}}$ & $10$ & \RN{6}\\
\end{tabular}
\caption{Rates used for the simulations. These rates are represented in Fig. 2. $\text{k}_{\text{reactant-on}}$ is either $\text{k}_{\text{Son}}, \text{k}_{\text{Lon}}, \text{k}_{\text{P1on}}$, or $\text{k}_{\text{P2on}}$.}
\end{table}

To control which molecule wins in each displacement competition, the off rates were assigned to establish a hierarchy in which substrate binds the tightest, followed by ligand, P2, and finally P1. Because the log of each off rate corresponds to a binding energy, $\varepsilon$, for each node, where $\varepsilon_\text{node} \sim \ln(\text{k}_{\text{off-node}})$ (see SI \ref{si_detailed_balance}), the hierarchy constrains the sums of the binding energies for each binding site or portion of the site, such that they obey the following inequality:

\begin{equation}
    \overbrace{\strut \varepsilon_\text{a} + \varepsilon_\text{b} + \varepsilon_\text{c}}^{\varepsilon_\text{S}} < \overbrace{\strut \varepsilon_\text{d} + \varepsilon_\text{e} + \varepsilon_\text{f}}^{\varepsilon_\text{L}} < \overbrace{\strut \varepsilon_\text{b} + \varepsilon_\text{c}}^{\varepsilon_\text{P2}} < \overbrace{\strut \varepsilon_\text{a}}^{\varepsilon_\text{P1}}
\end{equation}

\noindent where the energies are negative. This hierarchy applies to situations in which two molecules are pitted against each other and compete for binding. The two competing molecules can be substrate and ligand, substrate and P2, or ligand and P2. The hierarchy means that substrate will win both competitions, and ligand will win against P2. When substrate displaces P2, we call this a \textit{steric displacement}. Its regulation is discussed in more detail below, in the context of the cycle and Fig. \ref{fig:one}D.

\subsection*{Operation of the machine}

Fig. \ref{fig:one}D shows the stepwise operation of the machine along the target cycle (cycle i, outside) and along two futile cycles (cycles ii and iii, inside)(see SI \ref{si_find_paths} for a description of all possible pathways). The target cycle (Movies 1 \& 2, SI \ref{si_movies}) follows the same six intermolecular steps defined in Eq. 1 but includes the hidden microstates that take place between them and mechanistically explains the displacement reactions. In this completed cycle, state \{S/L\} is extended to five microstates and \{P2/L\} to four, which extends the sequence in Eq. 1 from six to thirteen states. Each microstate has a macrostate designation (e.g. \{S/L\}), which is the set of bound reactants, and a microstate designation (e.g. \{5, 10\}), which is the set of basis states from which it is composed. The rate for each forward (clockwise) and reverse (counterclockwise) reaction is given along each edge that connects the states together.

Displacement sequences can be viewed as a competition for the two hinges or the available degrees of freedom. During the displacement of ligand by substrate (Fig. \ref{fig:one}D; light blue bubble and clockwise direction), substrate captures the two hinges from ligand. Conversely, in the reverse direction (counterclockwise), ligand captures them from substrate. In both directions of the sequence, the enzyme starts and ends in a completely rigid state. The forward direction is biased by the rate assignments, which allow substrate to hang on to each of its nodes longer than ligand and maintain possession of the hinges. Or, equivalently said, ligand releases its nodes sooner than substrate, freeing up the hinges for capture by substrate. The same competitive process allows ligand to displace P2 (green bubble), but in this case, ligand immediately takes control of the hinge on the left, which is unoccupied when ligand binds (reaction 4). Graphically, it is important to understand that along the target cycle in Fig. 2 (cycle i), the ligand that displaces P2 is a different ligand than the ligand displaced by substrate.

\ctikzset{bipoles/length=.4cm}
\newcommand\esymbol[1]{\begin{circuitikz}
\draw (0,0) to [#1] (1,0); \end{circuitikz}}

The cleavage reaction (Fig. \ref{fig:one}D, orange bubble; reaction 3, \protect\usym{2702}) lies between the two displacement reactions and connects them together. Like intramolecular dissociation from a node, cleaving also releases a hinge and creates flexibility. The flexibility it introduces allows ligand to displace P2, recapture the enzyme's degrees of freedom, and reset the cycle. After cleavage takes place, forward progress is biased by assigning that ligation take place more slowly than P1 dissociation ($\text{k}_{\text{lig}} < \text{k}_{\text{off-a}}$). Whereas ligation reintroduces rigidity in the reverse direction, P1 dissociation rectifies the flexibility created by cleavage in the forward direction (reaction 5, \esymbol{diode}). With flexibility rectified on the left side of the enzyme, ligand can easily bind and displace P2, resetting the cycle.

The two futile pathways depicted—\textit{idling} and \textit{steric displacement}—diverge from the target path just before catalysis begins and right after catalysis ends, respectively. Idling (Fig. \ref{fig:one}D, blue bubble) occurs when ligand stays bound through catalysis, instead of dissociating, and is a natural consequence of the design. Idling forms a cycle (cycle ii) that bridges the very end of the ligand displacement sequence (\{S/L\}:\{0,15\}) to the middle of the P2 displacement sequence (\{P2/L\}:\{6,12\}). Graphically, it is important to understand that the same ligand remains bound along the idling cycle. It does not change as it does along the target cycle. Idling is defined as futile because no exchange of ligand takes place, even though one substrate is consumed in the cycle. Evidence suggests idling takes place in myosin \cite{Stein1979-uk, Stein1981-tq}. To inhibit idling, the cleavage rate was assigned to be slower than the slowest rate at which ligand dissociates from its nodes ($\text{k}_{\text{clv}} < \text{k}_{\text{off-e}}$). Ligand is depicted dissociating from node \textsf{e} in Fig. \ref{fig:one}D, but any of the three is possible. While the ligand meant to be displaced by substrate cannot idle, the second ligand that binds, which displaces P2, can idle if it binds before cleavage. If this second ligand idles, we define this as a special case of a productive cycle (see SI \ref{si_find_paths}, pci). It is productive because ligand exchange still takes place.

Steric displacement (Fig. \ref{fig:one}D, dashed orange path in orange bubble) is a more extreme case of futile behavior in which turnover occurs without any involvement of ligand. Steric displacement occurs if, after P1 dissociates, substrate binds and displaces P2 before ligand displaces P2. Steric displacement forms a futile cycle (cycle iii) with catalysis and rectification. Steric displacement is mechanistically related to facilitated dissociation \cite{Sing2014-du, Brown2022-ik} and DNA strand displacement \cite{Yurke2000-rk, Srinivas2013-qy}. Its definition as futile follows from how ATPases operate. In many ATPases, ATP turnover (ATP binding and Pi and ADP exhaust) drives another binding process by also depending upon that process, often to perform nucleotide exchange (product exhaust) \cite{De_La_Cruz1999-wt, Holzbaur1989-ba, Tafoya2018-iw, Branscomb2017-af, Carter2019-kp}. Along the steric displacement cycle, substrate rather than ligand accomplishes product exhaust, so substrate turnover no longer depends on ligand binding.

To inhibit steric displacement, we assigned node \textsf{b}—the central node at the substrate binding site and one of the two nodes to which P2 binds—to have the slowest off rate ($\text{k}_{\text{off-b}}$). This assignment allows P2 to inhibit invading substrate from binding two adjacent nodes on the enzyme. After P1 dissociates, substrate can begin a steric displacement of P2 by binding at node \textsf{a}. To make a second node connection, substrate must wait for P2 to dissociate from \textsf{b}. However, because substrate's association with node \textsf{a} is weaker than P2's with \textsf{b}, it is more probable for any single substrate that binds to \textsf{a} to spontaneously dissociate before P2 dissociates from node \textsf{b}. This arrangement favors the allosteric displacement of P2 by ligand over steric displacement by substrate. In contrast to substrate, ligand can immediately make two node associations with the enzyme (at \textsf{d} and \textsf{e}) after the dissociation of P1 rectifies catalysis. And once bound, ligand can complete the displacement of P2 either by P2 dissociating from node \textsf{b} or node \textsf{c}.

The two different ways in which the futile cycles are futile—idling, which makes substrate turnover dependent upon a single ligand, and steric displacement, which makes substrate turnover completely independent of ligand—highlight how productive cycling is designed to be a controlled oscillation between ligand turnover and substrate turnover. The futile cycles destroy the oscillation.

Finally, Figs. \ref{fig:one}E, \ref{fig:one}F, and \ref{fig:one}G are graph representations of the target and futile cycles. They chart only intermolecular changes and catalysis and are used in place of the full depiction in the remaining figures of the main text. The target cycle (cycle i), as defined in Eq. 1, can be followed clockwise around the outside path of Fig. \ref{fig:one}E (blue path).

\subsection*{Basic simulations}

Stochastic simulations were done on the system's reaction network, which includes all possible states the enzyme can make with its four reactants. Simulations were performed using the next reaction method \cite{Gibson2000-px} in StochPy \cite{Maarleveld2013-ik}. The simulations represent the behavior of a single molecule of enzyme in solution with multiple copies of the four reactants, where the copy numbers of the reactants are interpreted as concentrations (SI \ref{si_stochrates}). In the current analysis, we focused on characterizing and counting occurrences of the different turnover pathways (target and futile), to confirm particular sequences of operation. We did not analyze dwell times (time spent bound to the enzyme by each reactant) or cycle times.

\begin{figure}[H]
\begin{center}
\includegraphics[width=1\columnwidth]{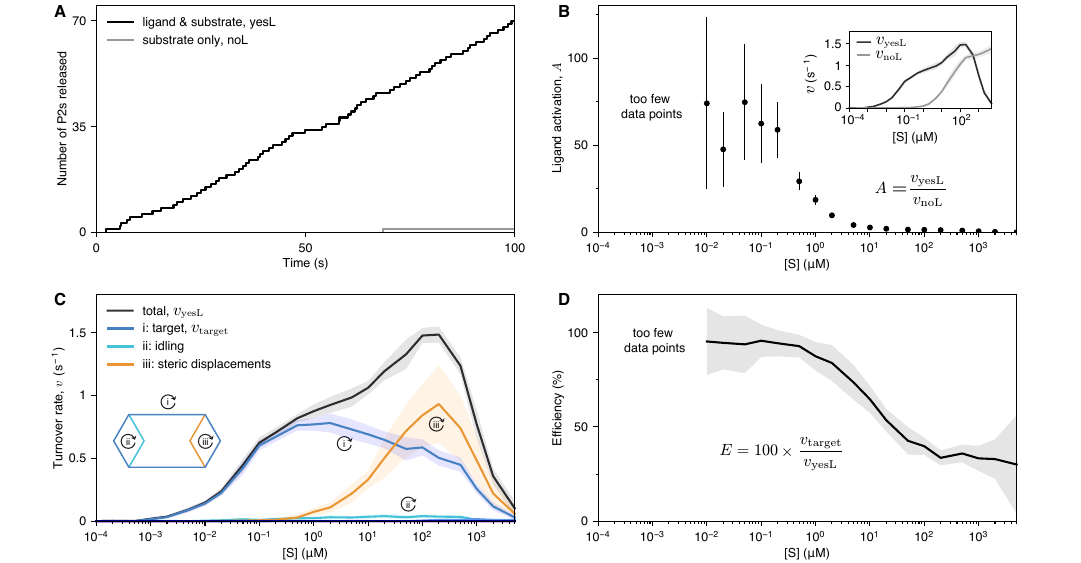}
\caption{\textbf{Plots showing behavior of the system.} \textbf{A,} Time traces of two stochastic simulations (done with StochPy) comparing the number of P2's released when ligand and substrate are present (blue) versus substrate only (red), where [S] = [L] = 100 $\mu$M. \textbf{B,} Ligand activation ($A$) plotted over a range of substrate concentrations. Because of no, or low activity for simulations done without ligand ($v_{\text{noL}}$) at low [S] ($< 0.1$ $\mu$M), the error in $A$ is too large, and these data are left out. \textbf{C,} Turnover rates ($v$) for the most relevant cycles plotted over a range of substrate concentrations. The total rate (black; $v_{\text{total}}$) is the sum of the separated pathways (in color). The target cycle (i, blue; $v_{\text{target}}$) peaks around () and then is overtaken by steric displacements (ii, orange) until the system saturates at high [S]. \textbf{D,} Efficiency ($E$) plotted over a range of substrate concentrations. Efficiency is highest where incidence of the target cycle is highest (compare with \textbf{C}, above).}
\label{fig:two}
\end{center}
\end{figure}

Substrate turnover was measured by tracking the release of P2 from the enzyme. Although we compare the turnover rate across different pathways in the main analysis, the initial aim was to confirm that the presence of ligand stimulated (increased) the rate of turnover, consistent with the operation of a myosin monomer. Simulations performed with and without ligand showed that ligand stimulates the turnover of substrate (Fig. \ref{fig:two}A; and Movie 3, SI \ref{si_movies}). The stimulatory behavior, which is named \textit{ligand activation (A)}, can be quantified:

\begin{equation}
    A = \frac{v_\text{yesL}}{v_\text{noL}}
\end{equation}

\noindent where $v_\text{yesL}$ is the turnover rate with ligand and $v_\text{noL}$ is the turnover rate without ligand. Measured over a range of substrate concentrations, ligand activation peaked at lower substrate concentrations and tended to zero at high substrate concentrations (Fig. \ref{fig:two}B). The shape of the curve follows from the individual contributions of $v_\text{yesL}$ and $v_\text{noL}$ (inset of Fig. \ref{fig:two}B). The turnover rate with ligand ($v_\text{yesL}$) peaks and then tends toward zero at the highest concentration of substrate, and the turnover rate without ligand ($v_\text{noL}$) increases and begins to saturate at high [S].

To verify that ligand activation resulted from incidence of the target cycle, we wrote routines to track each P2 released and bin them by path (Fig. \ref{fig:two}C). The binned data shows that incidence of the target cycle (Movies 1 \& 2) dominated in the same concentration range that ligand activation was highest. At higher concentrations of substrate, steric displacements dominated. Steric displacements dominate at high substrate concentrations because substrate binds more rapidly at high concentrations, allowing substrate to saturate node a after P1 dissociates. When node a saturates with substrate, additional substrate is more likely to bind node b when P2 dissociates from node b and is thus more likely to displace P2 (Movie 4, see SI \ref{si_movies}). In contrast to steric displacements, incidence of idling (Movie 5, see SI \ref{si_movies}) remained low over the entire range of substrate concentrations tested. Other futile pathways occurred at an even lower incidence and thus are not visible in Fig. \ref{fig:two}C, but they do minimally contribute to the overall turnover count (see SI \ref{si_data}).

At the highest concentrations of substrate, all enzymatic activity tended toward zero. The mechanism by which this occurs is inhibitory saturation, where a complex containing multiple substrates and one ligand is repeatedly formed (Movie 6, see SI \ref{si_movies}). Inhibitory saturation is explained in detail in SI, but briefly, it takes place when substrate binds to the enzyme faster than bound ligand dissociates from its nodes. In simulations performed without ligand, substrate saturation alone did not cause inhibition (Fig. \ref{fig:two}B inset, $v_\text{noL}$). Although we did not test it, we believe that inhibitory saturation would be eliminated if the reaction network was restricted to states in which only one substrate is bound at a time.

Finally, the efficiency ($E$) of the target cycle was quantified:

\begin{equation}
    E = 100 \times \frac{v_{\text{target}}}{v_{\text{yesL}}}
\end{equation}

\noindent where $v_\text{target}$ is the turnover rate of the target cycle. Higher efficiency overlaps with occurrences of the target cycle and higher ligand activation (Fig. \ref{fig:two}D).

\subsection*{Siloing simulations}

\begin{figure}[H]
\begin{center}
\includegraphics[width=1\columnwidth]{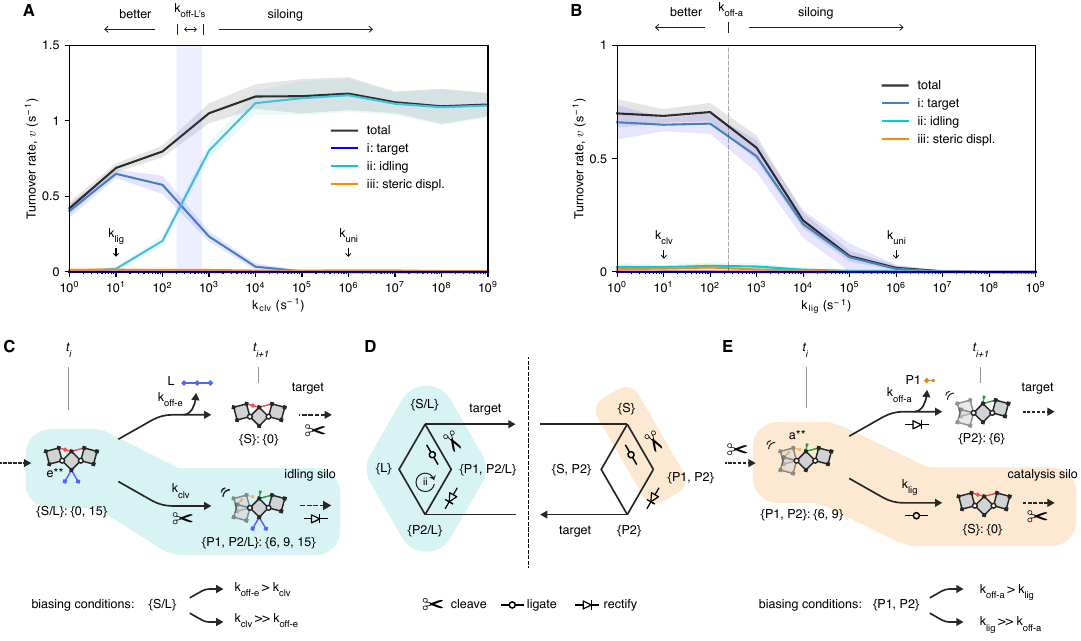}
\caption{\textbf{Idling and catalysis siloing.} \textbf{A,} Plot of idling silo. While keeping the ligation rate, $\text{k}_{\text{lig}}$, constant at $10\,\text{s}^{-1}$, idling becomes the dominant pathway taken by the system as the cleavage rate, $\text{k}_{\text{clv}}$, is increased. The crossover point lies where $\text{k}_{\text{clv}}$ surpasses the ligand's off rates (between $10^{2}$ and $10^{3}\,\text{s}^{-1}$; vertical blue band). Before this point, in particular where $\text{k}_{\text{clv}}$ equals $\text{k}_{\text{lig}}$ (at $10\,\text{s}^{-1}$), the behavior is optimal, and the target cycle dominates. \textbf{B,} Plot of catalysis silo. While keeping the cleavage rate constant at $10\,\text{s}^{-1}$, the turnover rate decreases and approaches zero as the ligation rate, $\text{k}_{\text{lig}}$, approaches the intramolecular binding rate, $\text{k}_{\text{uni}}$. \textbf{C,} Idling silo pathway. Two time points along the target trajectory (top) versus idling silo trajectory (bottom) are shown. Starting at \{S/L\}:\{0, 15\} at $t_i$ (in blue circle), siloing dominates when $\text{k}_{\text{clv}}$ is much greater than the ligand off rates (here, $\text{k}_{\text{off-e}}$), which biases cleavage to take place before L dissociates, and a transition to \{P1, P2/L\}: \{6, 9, 15\} at $t_{i+1}$ (see SI Fig. \ref{fig:seqs3}N for full sequence of idling). By contrast, when $\text{k}_{\text{off-e}}$ is greater than $\text{k}_{\text{clv}}$, L is biased to dissociate before cleavage (top path), and the system transition to \{S\}:\{0\} instead, and stays on the target path. The rate conditions governing the top vs bottom paths are shown below. \textbf{D,} Left, Idling silo on the mini-network. Right, Catalysis silo on the mini-network. The starting state for each silo is circled: \{S/L\} for idling; and \{P1, P2\} for catalysis. Unlike the idling silo, the catalysis silo is not a cycle, but rather interconversion between \{S\} and \{P1, P2\} without turnover. \textbf{E,} Catalysis silo pathway. Two time points along the target trajectory (top) versus catalysis silo (bottom) are shown. Starting at \{P1, P2\}: \{6, 9\}, siloing dominates when $\text{k}_{\text{lig}}$ is much greater $\text{k}_{\text{off-a}}$, allowing ligation (bottom path) to take place before P1 dissociates (top path). The rate conditions governing the top vs bottom paths are shown below.}
\label{fig:three}
\end{center}
\end{figure}

In this last set of results, we discuss simulations that justify the values selected for cleavage and ligation and that help to clarify the role that catalysis plays in the system. These simulations involved varying the rates for cleavage or ligation while keeping the other parameters the same (Fig. \ref{fig:three}).

Varying the cleavage rate showed how the cleavage reaction controls entry into the idling pathway (Fig. \ref{fig:three}A). Faster cleavage increases the overall turnover rate but has the negative effect of funneling turnover events into the idling cycle (Fig. \ref{fig:three}A). This funneling, or siloing, increases as the cleavage rate exceeds the off rates for the ligand's nodes. In this regime, the cleavage of substrate is biased to take place before the displacement of ligand by substrate completes, which biases idling over the target cycle (Figs. \ref{fig:three}A, right; and \ref{fig:three}B, bottom path in blue bubble, between $t_i$ and $t_{i+1}$). Slower cleavage, conversely, decreases the overall turnover rate but has the positive effect of increasing the efficiency of the target cycle by biasing ligand displacement to complete before cleavage takes place (Figs. \ref{fig:three}A, left; and \ref{fig:three}B, top path between $t_i$ and $t_{i+1}$). The ideal range lies where the cleavage rate is slower than that of the ligand $\text{k}_{\text{off}}$'s, but not too slow.

Varying the ligation rate while keeping the cleavage rate at an ideal constant showed that ligation controls when the substrate-bound enzyme can escape from reversible catalysis. No escape means that rectification cannot occur, and thus turnover cannot occur (Fig. \ref{fig:three}C). Rectification, or dissociation of P1, after cleavage, prevents substrate from reforming by ligation and thus liberates a degree of freedom for ligand binding. Slow ligation biases rectification by giving P1 more time to dissociate after cleavage and maximizes entry into the target cycle (Fig. \ref{fig:three}C, left; and \ref{fig:three}B, top path between $t_{i+2}$ and $t_{i+3}$). Faster ligation, by contrast, inhibits enzymatic turnover and eventually stops it altogether (Fig. \ref{fig:three}C, far right).

The inhibition of turnover begins as the ligation rate approaches and then exceeds $\text{k}_{\text{off-a}}$, the rate at which P1 dissociates (Fig. \ref{fig:three}C, middle region). At this point, substrate begins to reform faster, through ligation, than P1 can dissociate. Beyond this threshold, the system becomes increasingly trapped, or siloed, in a stable but catalytically active complex of product and substrate that cannot be dismantled by ligand binding and that permanently sequesters the enzyme's degrees of freedom (Fig. \ref{fig:three}B, bottom path in orange bubble, between $t_{i+2}$ and $t_{i+3}$). In Fig. \ref{fig:three}D, the graph representation is used to highlight states populated by siloing from fast ligation (orange) and those populated by siloing from fast cleavage (blue).

Because cleavage transiently liberates one degree of freedom for ligand binding and dissociation of P1 rectifies this liberation by preventing ligation, the rate of P1 dissociation, $\text{k}_{\text{off-a}}$, is a good upper bound (speed limit) for the rate of ligation, meaning that ligation should be slower than $\text{k}_{\text{off-a}}$ so that rectification can take place before ligation. However, in a system in which P1 does not dissociate spontaneously and must be displaced as is P2, we believe that the intramolecular binding rate, $\text{k}_{\text{uni}}$, is a more fundamental speed limit that the ligation rate must be well below. $\text{k}_{\text{uni}}$ is the speed limit for ligation in this hypothetical system because ligand binding will have to stop ligation, either by displacing a product molecule or by first directly competing for the mechanical degree of freedom that cleavage releases. For one of these two things to take place, ligation should not outcompete the ligand's ability to bind internally, or $\text{k}_{\text{lig}}$ should be less than $\text{k}_{\text{uni}}$.

\section*{Discussion}

\subsection*{The purpose of catalysis}

Due to the design of the linkage system, splitting substrate into two pieces has the same geometric effect on the substrate-enzyme complex as substrate dissociating from a node, which is to introduce flexibility. Based on this mechanical equivalence, we defined the hydrolysis mimic. The hydrolysis mimic allows substrate to be split (cleaved) or joined (ligated) back together at a special "catalytic" node, unique to substrate. The uniqueness allows cleavage to introduce flexibility into the substrate-enzyme complex via a different path and timescale (or "off" rate) than that of normal bond dissociation. In reverse, ligation can remove the introduced flexibility with a different rate than the intramolecular binding rate ($\text{k}_{\text{uni}}$), and thus determine the duration of the flexibility. In this way, substrate binding can be made weaker than ligand binding through catalysis without destroying substrate's ability to displace ligand.

By varying the rates of catalysis, we showed that when cleavage was too fast (flexibility was introduced too soon), a single ligand remained bound through many turnover events. Conversely, when ligation was too fast (flexibility was introduced for too short a time), a single substrate remained bound and inhibited turnover. Only when the flexibility introduced by catalysis existed for the right amount of time did the system oscillate between substrate and ligand. Hence, catalysis (bond splitting), as interpreted here, provides a pathway by which a tight-binding fuel can cede control of a single mechanical degree of freedom in the enzyme complex to its weaker-binding ligand or allosteric partner. This pathway must operate in parallel to normal stochastic binding and dissociation but on a slower timescale.

\subsection*{Relevance to engineering molecular motors}

The chain-like linkage geometry used here is simple compared to the complex three-dimensional geometries of proteins and ribonucleoproteins. Hence, only a small space of allosteric behavior is explored, which is negative allosteric cooperativity between two binding sites. Even in longer versions of these chains, events occurring at any single link in the chain can only affect adjacent links, similar to how DNA strand displacement progressively takes place \cite{Yurke2000-rk, Srinivas2013-qy}. This attribute places limitations on the complexity of allosteric signaling that can be achieved beyond that between two sites, which is a large space \cite{Biddle2021-un,10.2307/j.ctvx5w8pf, Wodak2019-lc}.

For example, what kind of linkage could incorporate a lever-arm-like structure that moves in response to the sequential displacement of two product molecules? In myosin, actin binding triggers the release of both products, and both releases are coupled to the motion of the lever arm. How could a third binding site be incorporated into a linkage that positively or negatively affected allosteric coupling between two other sites? How can self-binding be modeled? In this system, all interactions are between the enzyme and external reactants, whereas an ATPase like kinesin also self-binds with its neck-linker domain to modulate external specificity to ATP and microtubule.

Designing more complex allosteric behaviors probably requires extending the model to two- or three-dimensional networks \cite{Flechsig2017-yf, Rocks2017-rw, Yan2017-by, Bravi2020-zt, Falk2023-vw} or origami-inspired structures \cite{Dudek2025-mu, Overvelde2016-hx}. These systems can have more independent degrees of freedom and more ways to arrange mechanical coupling between them. However, rational design of complex mechanical networks is a challenge. Most previous studies begin with networks that are algorithmically pruned to send signals between target sites. While powerful, extracting design principles from algorithmic solutions is not easy. Whether future work on mechanical network models of molecules involves rational design, algorithmic tuning, or some combination of strategies, it seems important to incorporate mechanisms that allow transduction. In this context, transduction means mechanisms that allow internal signals to be converted into external signals, and vice versa. Here, we showed that simple mechanical networks can transduce signals in ways that mimic chemistry. To do so, we assumed that conformational changes can be made without strain and can be stochastically controlled by the stepwise addition and removal of bars in the network.

A major challenge in studying biomolecular machines is determining which structures are essential for the functions they demonstrate and which may be nonessential \cite{Manhart2015-am, Rivoire2019-ro, Hochberg2020-lx}. Developing allosteric models that can produce the internal steps required to achieve various intermolecular outcomes should make it easier to emulate various biomolecular machines through interpolation, without the requirement of knowing every detail of how the machines may operate. Such allosteric models will also make it easier to use known experimental methods, like DNA or RNA nanotechnology \cite{Seeman2017-xx, Wang2018-ly, Choi2018-ad, Dey2021-th, Yesselman2019-an, Grabow2014-ht}, rotaxane/catenane chemistry \cite{Zhang2018-zh}, or protein engineering \cite{Huang2016-zl, Pillai2024-wp, Broerman2025-va}, to construct more capable synthetic molecular machines and motors. The linkage model is intended to be a contribution towards creating such methodologies.

\section*{Author contributions}

T.O. developed the model, ran the simulations, analyzed the data, and wrote the manuscript.

\section*{Acknowledgements}

I thank Zev Bryant and David Soloveichik for many helpful discussions about the work. In addition, I thank Dean Astumian, Keenan Breik, Miranda Holmes-Cerfon, Rizal Hariadi, Robert Kohn, Muneaki Nakamura, Enrique Rojas, Howard Stone, and Erik Winfree for feedback on the ideas within. This work was supported by a National Institutes of Health (NIH) Fellowship F32GM09442 to Tosan Omabegho, while a postdoctoral scholar at Stanford University. All simulation animations were done by Christian Swinehart (drafting@samizdat.co).

\section*{Data availability}

The movies and trajectory data are available at \url{https://doi.org/10.5281/zenodo.17969328}.

\section*{Declaration of interests}

The authors declare no competing interests.

\section*{References}
 
\printbibliography[heading = none]

\end{refsection}

\newpage

\section*{Supplemental}

\setcounter{section}{1}

\begin{refsection}

\setcounter{figure}{0}

\subsection{Defining a generic ATPase cycle} \label{si_myosin_cycle}

In this study, we define a chemical cycle that is based on the cycle of a myosin and dynein monomer. Below, we first give two reasons for why myosin and dynein were chosen as starting points and then describe how the simplified cyle is derived from the myosin and dynein cycles.

\subsubsection*{Why choose to emulate myosin and dynein monomer cycles?}

The first reason is that myosin and dynein monomers can be viewed as having the same chemical cycle if their allosteric effectors—actin for myosin, and microtubule for dynein—are equated to one another by function and order of reactivity. Actin and microtubule both function as nucleotide release factors, and bind in the same order in their respective cycles. If actin and microtubule are given the same generic name, the myosin cycle (SI Eq. 5) and dynein cycle (SI Eq. 6) become equivalent statements (SI Eq. 7). This generic chemical cycle can thus define a fundamental form of ATPase behavior—behavior that can be satisfied by different structures that bind different effectors.

The second reason is that the chemical behavior defined by myosin and dynein monomers is simple compared to that of other ATPases, even kinesin. In kinesin—the third type of cytoskeletal motor—specificity to nucleotide and microtubule is also conferred by the state of the neck linker. The neck linker connects two monomers together and essentially acts as a third category of reactant. By contrast, myosin and dynein have two categories of reactants, which are nucleotide (ATP, Pi and ADP), and nucleotide release factor (effector).

\subsubsection*{Defining the simplified cycle.}

The chemical cycle of myosin monomer is a series of six states \cite{Lymn1971-ye, Eisenberg1980-by}. When thinking about the chemical cycle abstractly, it is important to keep in mind that all structural and mechanical features are forgotten, which is difficult to do given all that is currently known about myosin. In the chemical cycle, a myosin monomer is essentially an allosteric enzyme that has two binding sites, one which binds nucleotide, and one which binds actin:

\footnotesize

\begin{equation}
    \overbrace{\strut \{\text{actin\}} \xrightarrow[]{1} \text{\{ATP/actin\}} \xrightarrow[]{2} \text{\{ATP}\}}^{\text{ATP displaces actin}} \xrightarrow[]{3} \text{\{Pi, ADP}\} \xrightarrow[]{4} \overbrace{\strut \text{\{Pi, ADP/actin\}} \xrightarrow[]{5} \text{\{ADP/actin\}} \xrightarrow[]{6} \{\text{actin\}}}^{\text{actin displaces Pi, then ADP}} \dashrightarrow
\end{equation}

\normalsize

\noindent Starting with the actin bound state, this cycle says that ATP binds (1) and displaces actin (2), atp is cleaved (3), actin rebinds (4) and displaces Pi (5), and then ADP (6). A dynein monomer follows the same sequence, except that actin is replaced with microtubule (`MT'):

\footnotesize

\begin{equation}
    \overbrace{\strut \{\text{MT\}} \xrightarrow[]{1} \text{\{ATP/MT\}} \xrightarrow[]{2} \text{\{ATP}\}}^{\text{ATP displaces MT}} \xrightarrow[]{3} \text{\{Pi, ADP}\} \xrightarrow[]{4} \overbrace{\strut \text{\{Pi, ADP/MT\}} \xrightarrow[]{5} \text{\{ADP/MT\}} \xrightarrow[]{6} \{\text{MT\}}}^{\text{MT displaces Pi, then ADP}} \dashrightarrow
\end{equation}

\normalsize

\noindent We formally equate and abstract these two cycles by renaming the reactants:

\setlist[description]{font=\normalfont\itshape\space}

\begin{center}
\begin{varwidth}{\textwidth}
\begin{description}[itemsep = 0pt]
\item[i.]
    \text{actin}\,\,=\,\,\text{microtubule}\,\,=\,\,\text{ligand (L)}
\item[ii.]
    \text{ATP}\,\,=\,\,\text{substrate (S)}
\item[iii.]
    \text{Pi}\,\,=\,\,\text{product 1 (P1)}
\item[iv.]
    \text{ADP}\,\,=\,\,\text{product 2 (P2)}
\end{description}
\end{varwidth}
\end{center}

\noindent With these changes, both cycles can be represented by the following cycle:

\normalsize

\begin{equation}
    \overbrace{\strut \{\text{L\}} \xrightarrow[]{1} \text{\{S/L\}} \xrightarrow[]{2} \text{\{S}\}}^{\text{S displaces L}} \xrightarrow[]{3} \text{\{P1, P2}\} \xrightarrow[]{4} \overbrace{\strut \text{\{P1, P2/L\}} \xrightarrow[]{5} \text{\{P2/L\}} \xrightarrow[]{6} \{\text{L\}}}^{\text{L displaces P1, then P2}} \dashrightarrow
\end{equation}

\normalsize

In the final simplification of the cycle, P1 is allowed to dissociate spontaneously, and thus ligand binding is responsible only for the displacement of P2, where it is assumed that P1 spontaneously dissociates before P2 is displaced:

\normalsize

\newcommand{\tallvdots}{%
  \vcenter{%
    \baselineskip=4pt \lineskiplimit=0pt
    \hbox{.}\hbox{.}\hbox{.}
    \hbox{.}\hbox{.}\hbox{.}
    \hbox{.}\hbox{.}\hbox{.}
  }%
}
\makeatother

\begin{equation}
    \overbrace{\strut \{\text{L\}} \xrightarrow[]{1} \text{\{S/L\}} \xrightarrow[]{2} \text{\{S}\}}^{\text{S displaces L}} \xrightarrow[]{3}\rlap{$\underbrace{\strut\phantom{\text{\{P1, P2}\} \xrightarrow[]{4} \text{\{P2\}}}}_{\text{P1 dissoc. spontaneously}}$}\text{\{P1, P2}\} \xrightarrow[]{4} \overbrace{\strut \text{\{P2\}} \xrightarrow[]{5} \text{\{P2/L\}} \xrightarrow[]{6} \text{\{L\}}}^{\text{L displaces P2}} \dashrightarrow
\end{equation}

\begin{figure}[H]
\begin{center}
\includegraphics[width=1\columnwidth]{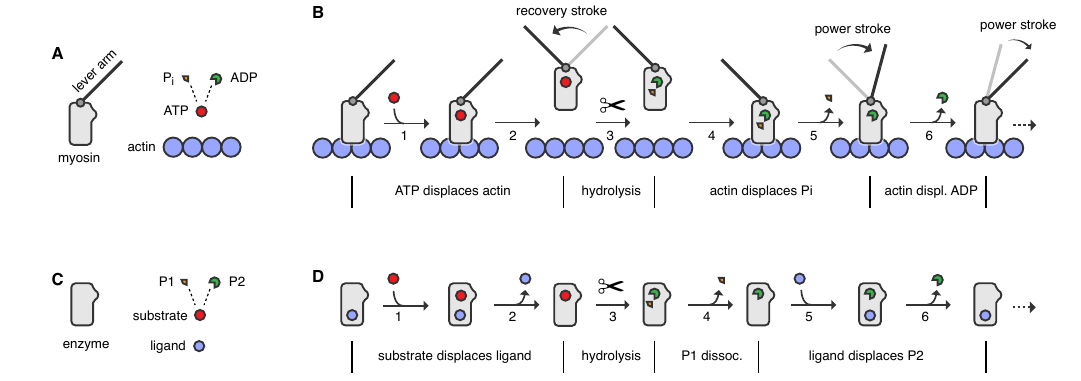}
\caption{\textbf{Myosin chemomechanical cyle vs simplified chemical cycle.} \textbf{A.} Five reactants of a myosin monomer: myosin with a lever arm; polymeric actin; ATP; Pi; and ADP. \textbf{B.} Chemomechanical cycle of a myosin monomer. Myosin goes through six chemical steps and two major mechanical changes of the lever arm, which are the recovery stroke, and the power stroke, where the power stroke is often depicted taking place in two stages. \textbf{C.} Five reactants of the generic ATPase-like machine: the enzyme; the ligand; substrate; P1; and P2. \textbf{D.} Simplified chemical cycle of the generic ATPase-like machine. In this cycle, P1 (the Pi analog) dissociates before ligand (the actin analog) binds.}
\label{fig:myosupp}
\end{center}
\end{figure}

\subsection{Constructing the reaction networks} \label{si_crns}

The reaction networks were constructed in these four steps:

\begin{enumerate}
\itemsep0em
    \item Generate basis states.
    \item Map exclusion rules between basis states.
    \item Generate complete set of states.
    \item Connect states to one another.
\end{enumerate}

\noindent Each of these four steps are described in detail below:

\subsubsection*{1. Generate basis states.} 
\begin{figure}[H]
\begin{center}
\includegraphics[width=1\columnwidth]{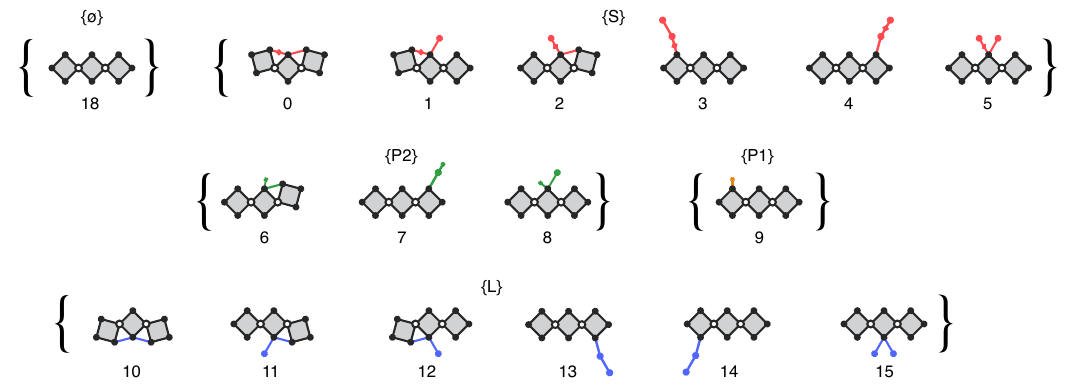}
\caption{\textbf{Basis states.} Subset of seventeen states that are used to generate and name the complete set of 449 states. In each of these states only one molecule of S, P1, P2 or L is bound, and the subset of states enumerates all the different ways these four molecules can bind to the enzyme, barring the two states eliminated by the adjacency rule.}
\label{fig:twobasis}
\end{center}
\end{figure}

\begin{figure}[H]
\begin{center}
\includegraphics[width=1\columnwidth]{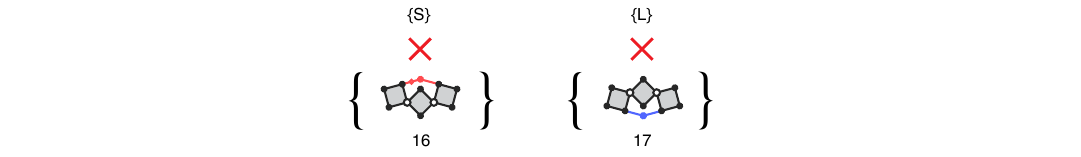}
\caption{\textbf{Adjacency rule.} Only adjacent bonds are allowed to form or dissociate between two interacting linkages. Left, disallowed substrate state in which the two outer nodes are bound, and the central node is dissociated. Right, disallowed ligand state in which the two outer nodes are bound, and the central node is dissociated.}
\label{fig:adjacency}
\end{center}
\end{figure}

The basis states were created by enumerating all the ways the single reactants can bind to the enzyme. There are seventeen basis states including the empty state (SI Fig. \ref{fig:twobasis}). The TRV basis states excludes two geometrically possible states by invoking an ad hoc principle, named the \textit{adjacency rule}, that only adjacent multivalent bonds can form (SI Fig. \ref{fig:adjacency}).

\subsubsection*{2. Map exclusion rules between basis states.} The exclusion rules for each system are mapped out as a binary matrix that matches the three sets of substrate and product basis states against the set of ligand binding basis states. For each matchup, the two states can either coexist together on the enzyme, and thus are an allowed two-molecule-bound state, or they cannot coexist together on the enzyme, and thus are an excluded two-molecule-bound state (symbolized by ). What should be noticed is that excluded states only result from matching one multivalent state to another multivalent state. The system has nine excluded states: seven for S and L (\{1, 12\}, \{0, 12\}, \{1, 10\}, \{0, 10\}, \{2, 10\}, \{0, 11\}, and \{2, 11\}); and two for P2 and L (\{6, 10\} and \{6, 11\}).

\begin{figure}[H]
\begin{center}
\includegraphics[width=1\columnwidth]{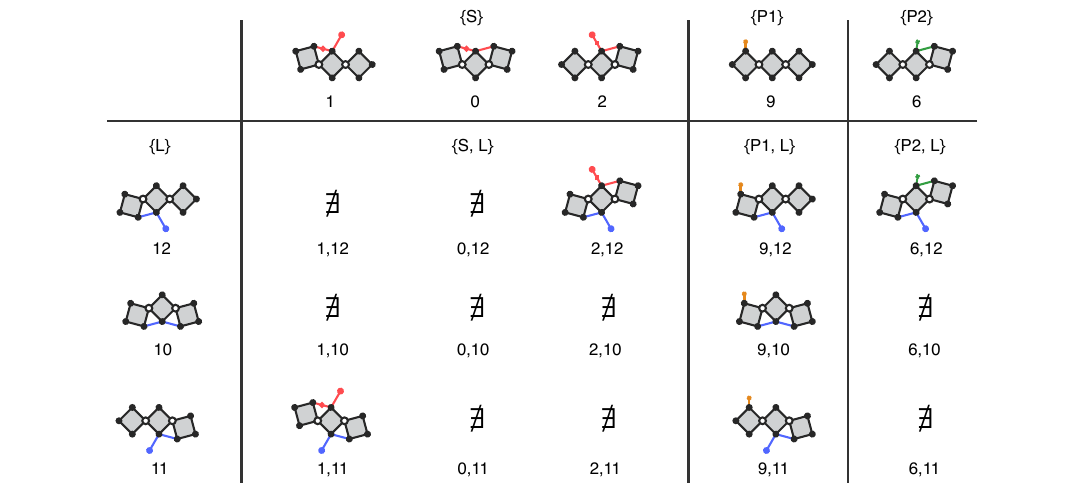}
\caption{\textbf{Rules matrix.} This binary matrix graphically displays the geometric restrictions that exist between the basis states, and which each represent an instance of negative allosteric coupling. Basis states that cannot coexist, and generate a new state, are represented with $\nexists$, for "does not exist". There are nine conflicts denoted with $\nexists$. Note that P1 and the ligand-bound-states (\{L\}) do have any conflicts, because P1 is bound monovalently to the enzyme.} 
\label{fig:tworules}
\end{center}
\end{figure}
    
\subsubsection*{3. Generate complete set of states.} [Notations used: $B^n$, where $n$ = 0 to 6, denotes sets of states with $n$ number of molecules bound to the enzyme. For example, $B^0$ is the empty state, and $B^1$ is the set of one-molecule-bound basis states (see SI Fig. \ref{fig:twobasis}). $\mathbf{B}^n$, for $n > 1$, is the matrix form of each set of bound states, and $\mathbf{B}^n_m$, refers to the row of the matrix, where the subscript $m$ denotes a state written as a set of basis states (e.g. if $m$ = 0, 13, this means state \{0, 13\}). In the matrix, `0' means the states cannot coexist, and `1' means that can coexist. Hence, `0' is equivalent to the $\nexists$ symbol used in Fig. \ref{fig:tworules}.] 
    
    Sets of higher order bound states were successively generated by multiplying (as a binary matrix) the set of states with one molecule less, with the set of basis states $B^1$ (e.g. $B^4 = B^3 \times B^1$). The process starts with the generation of $B^2$, where $B^2 = B^1 \times B^1$:
    
\small
\[
\mathbf{B^2} = \
\begin{array}{*{19}{c}}
\SetToWidest{} \ \ \vline & \SetToWidest{0} & \SetToWidest{1} & \SetToWidest{2} & \SetToWidest{3} & \SetToWidest{4} & \SetToWidest{5} & \SetToWidest{6} & \SetToWidest{7} & \SetToWidest{8} & \SetToWidest{9} & \SetToWidest{10} & \SetToWidest{11} & \SetToWidest{12} & \SetToWidest{13} & \SetToWidest{14} & \SetToWidest{15} \\
\hline
\SetToWidest{0} \ \ \vline & 0 & 0 & 0 & 0 & 0 & 0 & 0 & 0 & 0 & 0 & 0 & 0 & 0 & 1 & 1 & 1 \\
\SetToWidest{1} \ \ \vline & 0 & 0 & 0 & 0 & 1 & 0 & 0 & 1 & 0 & 0 & 0 & 1 & 0 & 1 & 1 & 1 \\
\SetToWidest{2} \ \ \vline & 0 & 0 & 0 & 1 & 0 & 0 & 0 & 0 & 0 & 1 & 0 & 0 & 1 & 1 & 1 & 1 \\
\SetToWidest{3} \ \ \vline & 0 & 0 & 1 & 0 & 1 & 1 & 1 & 1 & 1 & 0 & 1 & 1 & 1 & 1 & 1 & 1 \\
\SetToWidest{4} \ \ \vline & 0 & 1 & 0 & 1 & 0 & 1 & 0 & 0 & 1 & 1 & 1 & 1 & 1 & 1 & 1 & 1 \\
\SetToWidest{5} \ \ \vline & 0 & 0 & 0 & 1 & 1 & 0 & 0 & 1 & 0 & 1 & 1 & 1 & 1 & 1 & 1 & 1 \\
\SetToWidest{6} \ \ \vline & 0 & 0 & 0 & 1 & 0 & 0 & 0 & 0 & 0 & 1 & 0 & 0 & 1 & 1 & 1 & 1 \\
\SetToWidest{7} \ \ \vline & 0 & 1 & 0 & 1 & 0 & 1 & 0 & 0 & 1 & 1 & 1 & 1 & 1 & 1 & 1 & 1 \\
\SetToWidest{8} \ \ \vline & 0 & 0 & 0 & 1 & 1 & 0 & 0 & 1 & 0 & 1 & 1 & 1 & 1 & 1 & 1 & 1 \\
\SetToWidest{9} \ \ \vline & 0 & 0 & 1 & 0 & 1 & 1 & 1 & 1 & 1 & 0 & 1 & 1 & 1 & 1 & 1 & 1 \\
\SetToWidest{10} \ \ \vline & 0 & 0 & 0 & 1 & 1 & 1 & 0 & 1 & 1 & 1 & 0 & 0 & 0 & 0 & 0 & 0 \\
\SetToWidest{11} \ \ \vline & 0 & 1 & 0 & 1 & 1 & 1 & 0 & 1 & 1 & 1 & 0 & 0 & 0 & 0 & 1 & 0 \\
\SetToWidest{12} \ \ \vline & 0 & 0 & 1 & 1 & 1 & 1 & 1 & 1 & 1 & 1 & 0 & 0 & 0 & 1 & 0 & 0 \\
\SetToWidest{13} \ \ \vline & 1 & 1 & 1 & 1 & 1 & 1 & 1 & 1 & 1 & 1 & 0 & 0 & 1 & 0 & 1 & 1 \\
\SetToWidest{14} \ \ \vline & 1 & 1 & 1 & 1 & 1 & 1 & 1 & 1 & 1 & 1 & 0 & 1 & 0 & 1 & 0 & 1 \\
\SetToWidest{15} \ \ \vline & 1 & 1 & 1 & 1 & 1 & 1 & 1 & 1 & 1 & 1 & 0 & 0 & 0 & 1 & 1 & 0 
\end{array}
\]
    
    The resulting $\mathbf{B}^2$ matrix has two uses: 1) it generates the two-molecule-bound states, which are indicated wherever a `1' appears in the matrix; and 2) it generates a set of row vectors $\mathbf{B}^2_m$ to compliment each basis state $m$ in $B^1$, which are in turn used as multipliers to generate the higher order states. For example, the `1' in row 0, column 13, represents state \{0, 13\} in $B^2$. To generate $B^3$ ($B^3 = B^2 \times B^1$), each state in $B^2$ is converted into a row vector for $\mathbf{B}^3$, using the set of basis vectors $\mathbf{B}^2_m$. This is done by performing an element wise multiplication of the basis vectors that comprise each state in $B^2$. Hence, state \{0, 13\} is converted into row vector $\mathbf{B}^3_{0,13}$ in $\mathbf{B}^3$ by multiplying basis vectors $\mathbf{B}^2_{0}$ and $\mathbf{B}^2_{13}$ together element-wise, which is expressed as the Hadamard product, where `$\circ$' means element-wise multiplication:
    
\begin{align*}
\mathbf{B}^3_{0,13} = \mathbf{B}^2_{0} \circ \mathbf{B}^2_{13} = 
[ \ \ 0 \ \ 0 \ \ 0 \ \ 0 \ \ 0 \ \ 0 \ \ 0 \ \ 0 \ \ &0 \ \ 0 \ \ 0 \ \ 0 \ \ 0 \ \ 1 \ \ 1 \ \ 1 \ \ ] \ \circ \\
[ \ \ 1 \ \ 1 \ \ 1 \ \ 1 \ \ 1 \ \ 1 \ \ 1 \ \ 1 \ \ &1 \ \ 1 \ \ 0 \ \ 0 \ \ 1 \ \ 0 \ \ 1 \ \ 1 \ \ ] = \\
[ \ \ 0 \ \ 0 \ \ 0 \ \ 0 \ \ 0 \ \ 0 \ \ 0 \ \ 0 \ \ &0 \ \ 0 \ \ 0 \ \ 0 \ \ 0 \ \ 0 \ \ 1 \ \ 1 \ \ ]
\end{align*}

    Thus, for every `1' in $\mathbf{B}^2$, the Hadamard product is used to construct the rows of $\mathbf{B}^3$: 
    
\[
\mathbf{B^3} = \
\begin{array}{*{19}{c}}
\SetToWidest{} \ \quad \vline & \SetToWidest{0} & \SetToWidest{1} & \SetToWidest{2} & \SetToWidest{3} & \SetToWidest{4} & \SetToWidest{5} & \SetToWidest{6} & \SetToWidest{7} & \SetToWidest{8} & \SetToWidest{9} & \SetToWidest{10} & \SetToWidest{11} & \SetToWidest{12} & \SetToWidest{13} & \SetToWidest{14} & \SetToWidest{15} \\ 
\hline
\SetToWidest{0, 13} \quad \ \vline & 0 & 0 & 0 & 0 & 0 & 0 & 0 & 0 & 0 & 0 & 0 & 0 & 0 & 0 & 1 & 1 \\
\SetToWidest{0, 14} \quad \ \vline & 0 & 0 & 0 & 0 & 0 & 0 & 0 & 0 & 0 & 0 & 0 & 0 & 0 & 1 & 0 & 1 \\
\SetToWidest{0, 15} \quad \ \vline & 0 & 0 & 0 & 0 & 1 & 0 & 0 & 0 & 0 & 0 & 0 & 0 & 0 & 1 & 1 & 0 \\
\SetToWidest{1, 4} \quad \ \vline & 0 & 0 & 0 & 0 & 0 & 0 & 0 & 0 & 0 & 0 & 0 & 1 & 0 & 1 & 1 & 1 \\
\SetToWidest{1, 7} \quad \ \vline & 0 & 0 & 0 & 0 & 0 & 0 & 0 & 0 & 0 & 0 & 0 & 1 & 0 & 1 & 1 & 1 \\
\SetToWidest{1, 11} \quad \ \vline & 0 & 0 & 0 & 0 & 1 & 0 & 0 & 1 & 0 & 0 & 0 & 0 & 0 & 0 & 1 & 0 \\
\SetToWidest{\vdots} \quad \ \vline & \vdots & \vdots & \vdots & \vdots & \vdots & \vdots & \vdots & \vdots & \vdots & \vdots & \vdots & \vdots & \vdots & \vdots & \vdots & \vdots\\
\SetToWidest{14, 15} \quad \ \vline & 1 & 1 & 1 & 1 & 1 & 1 & 1 & 1 & 1 & 1 & 0 & 0 & 0 & 1 & 0 & 0 \\
\end{array}
\]
    
    To generate $\mathbf{B}^4$ the process is repeated. For example, the row vector $\mathbf{B}^4_{0, 13, 14}$ is generated by the product of $\mathbf{B}^3_{0, 13}$ and basis vector $\mathbf{B}^2_{14}$:  

\begin{align*}
\mathbf{B}^4_{0,13,14} = \mathbf{B}^3_{0, 13} \circ \mathbf{B}^1_{14} = 
[ \ \ 0 \ \ 0 \ \ 0 \ \ 0 \ \ 0 \ \ 0 \ \ 0 \ \ 0 \ \ &0 \ \ 0 \ \ 0 \ \ 0 \ \ 0 \ \ 0 \ \ 1 \ \ 1 \ \ ] \ \circ \\
[ \ \ 1 \ \ 1 \ \ 1 \ \ 1 \ \ 1 \ \ 1 \ \ 1 \ \ 1 \ \ &1 \ \ 1 \ \ 0 \ \ 1 \ \ 0 \ \ 1 \ \ 0 \ \ 1 \ \ ] = \\
[ \ \ 0 \ \ 0 \ \ 0 \ \ 0 \ \ 0 \ \ 0 \ \ 0 \ \ 0 \ \ &0 \ \ 0 \ \ 0 \ \ 0 \ \ 0 \ \ 0 \ \ 0 \ \ 1 \ \ ]
\end{align*}

    If $\mathcal{B}$ is used to symbolize the full set of states in the system, then $\mathcal{B} = \{B^0, B^1, B^2, B^3, B^4, B^5, B^6\}$, where $B^0$ is the empty state, in which none of the nodes are occupied, and $B^6$ is the subset in which all six nodes are occupied by a different reactant. The number of states in each set is given below in Table 2. The total list of states is given in....
    
\begin{table}[H]
\centering
\begin{tabular}{cc}  
Set & No. of states \\
\midrule
\(B^0\) & 1 \\
\(B^1\) & 16 \\
\(B^2\) & 74 \\
\(B^3\) & 150 \\
\(B^4\) & 142 \\
\(B^5\) & 58 \\
\(B^6\) & 8 \\
\(\mathcal{B} \, \text{(total)}\) & 449 \\
\end{tabular}
\caption{\textbf{Number of states in each set.}}
\end{table}

\subsubsection*{4. Connect states to one another} The states were connected together in two stages: (i) a unique set was constructed in which each member was a pair of states that could reversibly transition to one another; and (ii) each direction of the pair was again classified by reaction type and assigned with the appropriate rate. The process has some redundancy because pairs were classified by reaction type at both stages. In principle, this could have been done once. First we describe stage i in more detail, and then stage ii:

\textbf{i, Constructing a unique set of transition pairs:} This process was done in three parts: (a) finding pairs that define bimolecular binding/dissociation; (b) finding pairs that define intramolecular binding/dissociation; and (c) finding pairs that define catalysis/ligation. These three categories are described individually below: 
    
\textit{a, Finding bimolecular binding/dissociation pairs:} These pairs were found using a matrix called the binding matrix $\mathbf{b}$, which describes whether two basis states in the $B^1$ set can combine to form a new state by a binding reaction, with the addition of a row for the empty state (\{18\} or $B^0$):  
    
\[
\mathbf{b} = \
\begin{array}{*{19}{c}}
\SetToWidest{} \ \ \vline & \SetToWidest{0} & \SetToWidest{1} & \SetToWidest{2} & \SetToWidest{3} & \SetToWidest{4} & \SetToWidest{5} & \SetToWidest{6} & \SetToWidest{7} & \SetToWidest{8} & \SetToWidest{9} & \SetToWidest{10} & \SetToWidest{11} & \SetToWidest{12} & \SetToWidest{13} & \SetToWidest{14} & \SetToWidest{15} \\
\hline                      
\SetToWidest{0} \ \ \vline  & 0 & 0 & 0 & 0 & 0 & 0 & 0 & 0 & 0 & 0 & 0 & 0 & 0 & 1 & 1 & 1\\
\SetToWidest{1} \ \ \vline  & 0 & 0 & 0 & 0 & 1 & 0 & 0 & 1 & 0 & 0 & 0 & 0 & 0 & 1 & 1 & 1\\
\SetToWidest{2} \ \ \vline  & 0 & 0 & 0 & 1 & 0 & 0 & 0 & 0 & 0 & 1 & 0 & 0 & 0 & 1 & 1 & 1\\
\SetToWidest{3} \ \ \vline  & 0 & 0 & 0 & 0 & 1 & 1 & 0 & 1 & 1 & 0 & 0 & 0 & 0 & 1 & 1 & 1\\
\SetToWidest{4} \ \ \vline  & 0 & 0 & 0 & 1 & 0 & 1 & 0 & 0 & 1 & 1 & 0 & 0 & 0 & 1 & 1 & 1\\
\SetToWidest{5} \ \ \vline  & 0 & 0 & 0 & 1 & 1 & 0 & 0 & 1 & 0 & 1 & 0 & 0 & 0 & 1 & 1 & 1\\
\SetToWidest{6} \ \ \vline  & 0 & 0 & 0 & 1 & 0 & 0 & 0 & 0 & 0 & 1 & 0 & 0 & 0 & 1 & 1 & 1\\
\SetToWidest{7} \ \ \vline  & 0 & 0 & 0 & 1 & 0 & 1 & 0 & 0 & 1 & 1 & 0 & 0 & 0 & 1 & 1 & 1\\
\SetToWidest{8} \ \ \vline  & 0 & 0 & 0 & 1 & 1 & 0 & 0 & 1 & 0 & 1 & 0 & 0 & 0 & 1 & 1 & 1\\
\SetToWidest{9} \ \ \vline  & 0 & 0 & 0 & 0 & 1 & 1 & 0 & 1 & 1 & 0 & 0 & 0 & 0 & 1 & 1 & 1\\
\SetToWidest{10} \ \ \vline & 0 & 0 & 0 & 1 & 1 & 1 & 0 & 1 & 1 & 1 & 0 & 0 & 0 & 0 & 0 & 0\\
\SetToWidest{11} \ \ \vline & 0 & 0 & 0 & 1 & 1 & 1 & 0 & 1 & 1 & 1 & 0 & 0 & 0 & 0 & 1 & 0\\
\SetToWidest{12} \ \ \vline & 0 & 0 & 0 & 1 & 1 & 1 & 0 & 1 & 1 & 1 & 0 & 0 & 0 & 1 & 0 & 0\\
\SetToWidest{13} \ \ \vline & 0 & 0 & 0 & 1 & 1 & 1 & 0 & 1 & 1 & 1 & 0 & 0 & 0 & 0 & 1 & 1\\
\SetToWidest{14} \ \ \vline & 0 & 0 & 0 & 1 & 1 & 1 & 0 & 1 & 1 & 1 & 0 & 0 & 0 & 1 & 0 & 1\\
\SetToWidest{15} \ \ \vline & 0 & 0 & 0 & 1 & 1 & 1 & 0 & 1 & 1 & 1 & 0 & 0 & 0 & 1 & 1 & 0\\
\SetToWidest{18} \ \ \vline & 0 & 0 & 0 & 1 & 1 & 1 & 0 & 1 & 1 & 1 & 0 & 0 & 0 & 1 & 1 & 1\\
\end{array}
\]

    The matrix $\mathbf{b}$ maps out a subset of the pairs in $\mathbf{B^2}$, with entries that reflect only allowed single-node interactions. An entry of a `1' at a given column position means that the basis state at that column position can combine with the basis state for that row to form a new state that contains the row basis state and the column basis state, where the column state is the new molecule that binds at a single node. Hence, each row can be thought of as a basis vector for binding. And thus, for every state in the system, a combined binding vector could be calculated using the Hadamard product in the exact same way it was done to find successive higher order states. For example, to find the binding vector $\mathbf{b_{1, 13}}$ for state \{1, 13\}, the binding vectors for \{1\} and \{13\} are multiplied together:
    
\begin{align*}
\mathbf{b}_{1,13} = \mathbf{b}_{1} \circ \mathbf{b}_{13} = 
[ \ \ 0 \ \ 0 \ \ 0 \ \ 0 \ \ 1 \ \ 0 \ \ 0 \ \ 1 \ \ &0 \ \ 0 \ \ 0 \ \ 0 \ \ 0 \ \ 1 \ \ 1 \ \ 1 \ \ ] \ \circ \\
[ \ \ 0 \ \ 0 \ \ 0 \ \ 1 \ \ 1 \ \ 1 \ \ 0 \ \ 1 \ \ &1 \ \ 1 \ \ 0 \ \ 0 \ \ 0 \ \ 0 \ \ 1 \ \ 1 \ \ ] = \\
[ \ \ 0 \ \ 0 \ \ 0 \ \ 0 \ \ 1 \ \ 0 \ \ 0 \ \ 1 \ \ &0 \ \ 0 \ \ 0 \ \ 0 \ \ 0 \ \ 0 \ \ 1 \ \ 1 \ \ ]
\end{align*}

    \noindent As there is a `1'  at positions $\{4\}, \{6\}, \{14\}$, and $\{15\}$, state $\{1, 13\}$ can make four different binding transitions to new states containing these basis states, and in reverse, dissociations take place to transition to state $\{1, 13\}$:

\begin{equation}
\begin{aligned}
\{1, 13\} &\rightleftharpoons \{1, 4, 13\} \\
\{1, 13\} &\rightleftharpoons \{1, 6, 13\} \\
\{1, 13\} &\rightleftharpoons \{1, 13, 14\} \\
\{1, 13\} &\rightleftharpoons \{1, 13, 15\}
\end{aligned}
\end{equation}

    \noindent Hence, for every state in the system, the process is repeated to find all the possible bimolecular binding/dissociation transitions.
    
    \textit{b. Finding intramolecular binding/dissociation pairs:} These pairs were found using a matrix, the isomerization matrix ($\mathbf{i}$), that describes which states can transition to another by an intramolecular binding/dissociation reaction. For example, when the ligand or substrate are bound at two nodes, and make a third association with the enzyme, and vice versa for dissociation:
    
\[
\mathbf{i} = \
\begin{array}{*{20}{c}}
\SetToWidest{} \ \ \vline & \SetToWidest{0} & \SetToWidest{1} & \SetToWidest{2} & \SetToWidest{3} & \SetToWidest{4} & \SetToWidest{5} & \SetToWidest{6} & \SetToWidest{7} & \SetToWidest{8} & \SetToWidest{9} & \SetToWidest{10} & \SetToWidest{11} & \SetToWidest{12} & \SetToWidest{13} & \SetToWidest{14} & \SetToWidest{15} \\ 
\hline                      
\SetToWidest{0} \ \ \vline  & 0 & 1 & 1 & 0 & 0 & 0 & 0 & 0 & 0 & 0 & 0 & 0 & 0 & 0 & 0 & 0\\
\SetToWidest{1} \ \ \vline  & 1 & 0 & 0 & 1 & 0 & 1 & 0 & 0 & 0 & 0 & 0 & 0 & 0 & 0 & 0 & 0\\
\SetToWidest{2} \ \ \vline  & 1 & 0 & 0 & 0 & 1 & 1 & 0 & 0 & 0 & 0 & 0 & 0 & 0 & 0 & 0 & 0\\
\SetToWidest{3} \ \ \vline  & 0 & 1 & 0 & 0 & 0 & 0 & 0 & 0 & 0 & 0 & 0 & 0 & 0 & 0 & 0 & 0\\
\SetToWidest{4} \ \ \vline  & 0 & 0 & 1 & 0 & 0 & 0 & 0 & 0 & 0 & 0 & 0 & 0 & 0 & 0 & 0 & 0\\
\SetToWidest{5} \ \ \vline  & 0 & 1 & 1 & 0 & 0 & 0 & 0 & 0 & 0 & 0 & 0 & 0 & 0 & 0 & 0 & 0\\
\SetToWidest{6} \ \ \vline  & 0 & 0 & 0 & 0 & 0 & 0 & 0 & 1 & 1 & 0 & 0 & 0 & 0 & 0 & 0 & 0\\
\SetToWidest{7} \ \ \vline  & 0 & 0 & 0 & 0 & 0 & 0 & 1 & 0 & 0 & 0 & 0 & 0 & 0 & 0 & 0 & 0\\
\SetToWidest{8} \ \ \vline  & 0 & 0 & 0 & 0 & 0 & 0 & 1 & 0 & 0 & 0 & 0 & 0 & 0 & 0 & 0 & 0\\
\SetToWidest{9} \ \ \vline  & 0 & 0 & 0 & 0 & 0 & 0 & 0 & 0 & 0 & 0 & 0 & 0 & 0 & 0 & 0 & 0\\
\SetToWidest{10} \ \ \vline & 0 & 0 & 0 & 0 & 0 & 0 & 0 & 0 & 0 & 0 & 0 & 1 & 1 & 0 & 0 & 0\\
\SetToWidest{11} \ \ \vline & 0 & 0 & 0 & 0 & 0 & 0 & 0 & 0 & 0 & 0 & 1 & 0 & 0 & 1 & 0 & 1\\
\SetToWidest{12} \ \ \vline & 0 & 0 & 0 & 0 & 0 & 0 & 0 & 0 & 0 & 0 & 1 & 0 & 0 & 0 & 1 & 1\\
\SetToWidest{13} \ \ \vline & 0 & 0 & 0 & 0 & 0 & 0 & 0 & 0 & 0 & 0 & 0 & 1 & 0 & 0 & 0 & 0\\
\SetToWidest{14} \ \ \vline & 0 & 0 & 0 & 0 & 0 & 0 & 0 & 0 & 0 & 0 & 0 & 0 & 1 & 0 & 0 & 0\\
\SetToWidest{15} \ \ \vline & 0 & 0 & 0 & 0 & 0 & 0 & 0 & 0 & 0 & 0 & 0 & 1 & 1 & 0 & 0 & 0\\
\end{array}
\]

    Like $\mathbf{b}$, $\mathbf{i}$ is a subset of the pairs defined in $\mathbf{B}^2$. The possible isomerizations indicated by $\mathbf{i}$ cannot conflict with any other molecule that is bound to the enzyme, where the conflict can be either steric or allosteric. For example, the state \{1, 14\} does not have any conflicts, but \{1, 12\} does have an allosteric conflict, even though \{14\} can transition to \{12\} when it is on the enzyme alone. Hence, to account for conflicts, $\mathbf{i}$ is used with $\mathbf{B}^2$ in the following way. To test for isomerizations, each basis state in a state is tested individually by taking the Hadamard product of its $\mathbf{i}$ vector and the $\mathbf{B}^2$ basis vectors for all the other basis states in the state. For example, to test state \{3, 12, 13\} for isomerizations, \{3\}, \{12\}, and \{13\} are tested separately. Testing \{3\} first is done by multiplying $\mathbf{i}_{3}$ by $\mathbf{B}^2_{12}$ and $\mathbf{B}^2_{13}$, which shows that \{3\} cannot isomerize when present with \{12\}, and \{13\}; or substrate on node \textit{a} cannot bind any other node when one ligand is bound divalently at \textit{d} and \textit{e}, and another is bound at \textit{f}:

\begin{align*}
\mathbf{i}_{3} \circ \mathbf{B}^2_{12} \circ \mathbf{B}^2_{13} =
[\ \ 0 \ \ 1 \ \ 0 \ \ 0 \ \ 0 \ \ 0 \ \ 0 \ \ 0 \ \ &0 \ \ 0 \ \ 0 \ \ 0 \ \ 0 \ \ 0 \ \ 0 \ \ 0 \ \ ] \ \circ \\
[ \ \ 0 \ \ 0 \ \ 1 \ \ 1 \ \ 1 \ \ 1 \ \ 1 \ \ 1 \ \ &1 \ \ 1 \ \ 0 \ \ 0 \ \ 0 \ \ 1 \ \ 0 \ \ 0 \ \ ] \ \circ \\
[ \ \ 1 \ \ 1 \ \ 1 \ \ 1 \ \ 1 \ \ 1 \ \ 1 \ \ 1 \ \ &1 \ \ 1 \ \ 0 \ \ 0 \ \ 1 \ \ 0 \ \ 1 \ \ 1 \ \ ] = \\
[ \ \ 0 \ \ 0 \ \ 0 \ \ 0 \ \ 0 \ \ 0 \ \ 0 \ \ 0 \ \ &0 \ \ 0 \ \ 0 \ \ 0 \ \ 0 \ \ 0 \ \ 0 \ \ 0 \ \ ]
\end{align*}

    Testing \{12\}, reveals one possible isomerization to state \{3, 13, 14\}: 
    
\begin{align*}
\mathbf{i}_{12} \circ \mathbf{B}^2_{3} \circ \mathbf{B}^2_{13} =
[\ \ 0 \ \ 0 \ \ 0 \ \ 0 \ \ 0 \ \ 0 \ \ 0 \ \ 0 \ \ &0 \ \ 0 \ \ 1 \ \ 0 \ \ 0 \ \ 0 \ \ 1 \ \ 1 \ \ ] \ \circ \\
[ \ \ 0 \ \ 0 \ \ 1 \ \ 0 \ \ 1 \ \ 1 \ \ 1 \ \ 1 \ \ &1 \ \ 0 \ \ 1 \ \ 1 \ \ 1 \ \ 1 \ \ 1 \ \ 1 \ \ ] \ \circ \\
[ \ \ 1 \ \ 1 \ \ 1 \ \ 1 \ \ 1 \ \ 1 \ \ 1 \ \ 1 \ \ &1 \ \ 1 \ \ 0 \ \ 0 \ \ 1 \ \ 0 \ \ 1 \ \ 1 \ \ ] = \\
[ \ \ 0 \ \ 0 \ \ 0 \ \ 0 \ \ 0 \ \ 0 \ \ 0 \ \ 0 \ \ &0 \ \ 0 \ \ 0 \ \ 0 \ \ 0 \ \ 0 \ \ 1 \ \ 1 \ \ ]
\end{align*}

\noindent where in this isomerization the divalently bound ligand is dissociating from node \textit{d} or \textit{e}. Testing \{13\} results in no other isomerizations. Hence the final result is that \{3, 12, 13\} isomerizes to \{3, 13, 14\} and \{3, 13, 15\}, where because basis states are ordered numerically, \{14\} and \{15\} are written after \{13\} in the new states, though they are isomerization of \{12\}: 
    
\begin{equation}
\begin{aligned}
\{3, 12, 13\} &\rightleftharpoons \{3, 13, 14\} \\
\{3, 12, 13\} &\rightleftharpoons \{3, 13, 15\} \\
\end{aligned}
\end{equation}

\noindent To describe the transitions in terms of the molecules: in the forward direction, the divalently bound ligand dissociates from node \textit{d} or \textit{e}, and in the reverse direction, the same ligand rebinds binds at node \textit{d} or \textit{e}.
    
\textit{c. Finding cleavage/ligation pairs:} These pairs were found by looking for all states that contain \{0\} the fully bound substrate, and converting \{0\} to \{6\} and \{9\}, where \{6\} is P1 bound at node \textit{a}, and \{9\} is P2 bound divalently. Any state that contains \{0\} automatically accounts for the reaction in both directions and rules out any geometric conflicts, because the substrate can only be fully bound when there are no geometric conflicts. Thus any state generated by converting \{0\} to \{6\} and \{9\} is a ligation ready state without geometric conflicts, which specifically means they do not contain basis states \{11\} or \{12\}. 
    
\textbf{ii. Assigning rates to each direction of a reaction pair:} Rates were assigned using a series of logical statements that checked to see what kind of forward and reverse reactions a pair of states described, and then assigned rates accordingly: dissociations were assigned rates based on the node from which the dissociation took place, bimolecular binding reactions were assigned the bimolecular binding rate constant multiplied by a factor that reflected the concentration of the molecule binding, intramolecular binding reactions were all assigned the same unimolecular rate ($k_{\text{uni}}$); and cleavage and ligation reactions were assigned the same rate ($k_{\text{cat}}$) (see tables 5, 6, and 7, which are repeated from the main text.)

\begin{table}[H]
\centering
\captionsetup{width=.75\textwidth}
\caption{Categories of reaction rates}
\begin{tabular}{ll}  
Parameter & Value \\
\midrule
Bimolecular rate constant, $k_{\text{bi}}$  & $9\times10^6\,\text{M}^{-1}\text{s}^{-1}$ \\
Concentration, $c_m$ ($m =$ S, P1, P2, L) & varies with $m$; units of M\\
Intermolecular binding rates, $k_{m}$ & $k_{\text{bi}} \times c_m$ $\text{s}^{-1}$\\
Intramolecular binding rate, $k_{\text{uni}}$ & $1\times10^6\,\text{s}^{-1}$ \\
Dissociation rates, $k_{\text{off-}node}$ & varies with \textit{node}; units of $\text{s}^{-1}$\\
Catalysis, $k_{\text{cat}}$ & $100\,\text{s}^{-1}$\\
\end{tabular}
\end{table}

\begin{table}[H]
\centering
\captionsetup{width=.75\textwidth}
\caption{Dissociation rates}
\begin{tabular}{lll}  
Rate & Reactants & Value \\
\midrule
$k_{\text{off-a}}$ & S/P1 & $250\,\text{s}^{-1}$ \\
$k_{\text{off-b}}$ & S/P2 & $3\,\text{s}^{-1}$ \\
$k_{\text{off-c}}$ & S/P2 & $680\,\text{s}^{-1}$ \\
$k_{\text{off-d}}$ & L & $680\,\text{s}^{-1}$\\
$k_{\text{off-e}}$ & L & $200\,\text{s}^{-1}$ \\
$k_{\text{off-f}}$ & L & $680\,\text{s}^{-1}$ \\
\end{tabular}
\end{table}

\subsection{Stochastic formulation of the intermolecular on rates} \label{si_stochrates}

The rates expressed in table 5 in the SI (table 1 in main text) were input as stochastic rates in the simulations (table 8 in SI). To express the rates stochastically, a scaling factor was used, which is a reaction volume multiplied by Avogadro's number ($N_A \times v$). By dividing the bimolecular rate constant ($k_{\text{bi}}$) by this scaling factor, $k_{\text{bi}}$, which has units of $\text{M}^{-1}\text{s}^{-1}$, and which retains some relevance to reaction rates for real biomolecules, was converted into a stochastic version, $k^*_{\text{bi}}$, with units of $\text{s}^{-1}$. Likewise, concentrations, $c_m$'s, which have units of M, were multiplied by the scaling factor to convert them into their stochastic versions, $c^*_m$'s, which have units of "number of molecules" (No.).

\begin{table}[H]
\centering
\captionsetup{width=.75\textwidth}
\caption{Stochastic conversions of rates}
\begin{tabular}{ll}  
Parameter & Value \\
\midrule
Volume, $v$ & $1 \times 10^{-12} \, \text{m}^3$ \\
Scaling factor, $N_A \times v$ & $6.022 \times 10^{11} \, \text{M}^{-1}$ \\
Stoch. bimol. rate constant, $k^*_{\text{bi}} = k_{\text{bi}} / (N_A \times v)$  & $9 / 6.022 \times 10^{-5} \,\text{s}^{-1}$ \\
Stoch. "concentration", $c^*_m = c_m \times (N_A \times v)$ & varies with $m$; units of No. \\
Intermolecular binding rates, $k_{m}$ & $k^*_{\text{bi}}\times c^*_m$ (or $k_{\text{bi}} \times c_m$) \\
\end{tabular}
\end{table}

\subsection{Converting node energies into off rates} \label{si_detailed_balance}

Here, an expression is derived for calculating a node dissociation rate ($k_{\text{off-}node}$) in terms of its node energy ($\varepsilon_{node}$), and vice versa. This is done by satisfying local detailed balance for binding to and dissociation from a single node, as shown in the following figure:

\vspace{-1mm}

\begin{figure}[H]
\begin{center}
\includegraphics[width=1\columnwidth]{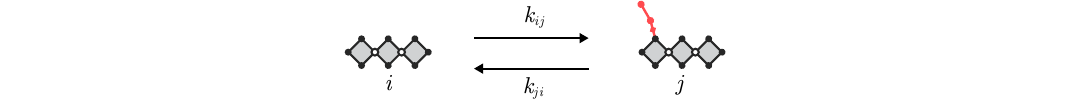}
\label{fig:det_bal}
\end{center}
\end{figure}

\vspace{-10mm}

\noindent State $i$ is the unbound enzyme, and state $j$ is the bound enzyme; $k_{ij}$ is the rate for transitioning from $i$ to $j$, and $k_{ji}$ is the rate for transitioning from $j$ to $i$. Although a specific reaction is shown in the figure, the setup of the derivation is general and can work for any node and molecule that binds that node. 

If the equilibrium probability of being in state $i$ is written as $p_i$, and for state $j$, written as $p_j$, then the equilibrium condition must satisfy:

\begin{equation}
    k_{ij}p_i = k_{ji}p_{j}
\end{equation}

\noindent The equilibrium probabilities are given by the Boltzmann factors for each state, divided by the partition function:

\begin{equation}
    p_i = \frac{e^ {-\beta G_{i}}}{Z}
\end{equation}

\noindent where $G_{i}$ is the free energy of the state. Using the above expression for $p_i$ and $p_j$ in equation 3, allows the transition rates to be expressed in terms of the state energies:

\begin{equation}
    \frac{k_{ij}}{k_{ji}} = e^ {\beta(G_{j} - G_{i})}
\end{equation}

\noindent The goal is to replace the above free energies with expressions relevant to states of the linkage. To do this, we closely follow the modeling done in Marzen et al. \cite{Marzen2013-me} for a "one-site MWC molecule", except we leave out the enzyme's "activated" state and its corresponding energy:

\begin{table}[H]
\centering
\begin{tabular}{lc@{\hspace{1cm}}c}  
\; \; State & Energy & Boltzmann factor \\
\midrule
\begin{minipage}{.15\textwidth}
{\includegraphics{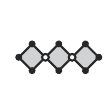}}
\end{minipage}
& $\varepsilon_l$ & $e^ {-\beta\varepsilon_l}$ \\
\begin{minipage}{.15\textwidth}
{\includegraphics{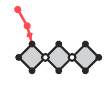}}
\end{minipage}
& $\varepsilon_l + \varepsilon_{node} - \mu$ & $e^ {-\beta(\varepsilon_l + \varepsilon_{node} - \mu)}$ \\
\end{tabular}
\caption{Energies of the unbound and bound enzyme.}
\label{marzen_table}
\end{table}

\noindent In Table \ref{marzen_table}, $\varepsilon_l$ is the conformational energy of the enzyme (or receptor), and $\varepsilon_{node}$ is the binding energy of the node. The full expression for $\mu$, the chemical potential, which is described in Marzen et al. \cite{Marzen2013-me} as the "free energy cost of removing a ligand from dilute solution", is:

\begin{equation}
    \mu = \mu_0 + k_BT \, \text{ln}\frac{c}{c_0}
\end{equation}

\noindent where $\mu_0$ is an "unspecified reference chemical potential", and $c_0$ is an "unspecified reference concentration". 

Equation 5 can be re-expressed in terms of the energies in Table \ref{marzen_table} by letting $G_{i} = \varepsilon_l$ and $G_{j} = \varepsilon_l + \varepsilon_b - \mu$, and by using the full expression for $\mu$ in equation 6:

\begin{equation}
    \frac{k_{ij}}{k_{ji}} = \frac{c}{c_0}e^{-\beta(\varepsilon_{node} - \mu_0)}
\end{equation}

\noindent In the final step, $k_{ij}$, a binding rate, is expressed as the concentration ($c$) of the molecule binding, multiplied by a bimolecular rate constant ($k_{\text{bi}}$), so that $k_{ij} = ck_{\text{bi}}$; and $k_{ij}$, a dissociation rate, is renamed $k_{\text{off-}node}$. Making these two substitutions and solving for $k_{\text{off-}node}$, gives: 

\begin{equation}
    k_{\text{off-}node} = c_0k_{\text{bi}} \, e^ {-\beta\Delta\varepsilon_{node}}
\end{equation}

\noindent where $\Delta\varepsilon_{node} = \mu_0 - \varepsilon_{node}$. Taking the log of both sides allows the node energy to be solved for in terms of the dissociation rate:

\begin{equation}
    \Delta\varepsilon_{node} = -k_BT\ln{\left(\frac{k_{\text{off-}node}}{c_0k_{\text{bi}}}\right)}
\end{equation}

\subsection{Behavior blocks} \label{si_find_paths}

Here we describe the different ways that substrate turnover takes place, using twelve flow charts ("blocks"). Each block contains multiple pathways by which turnover can take place. These twelve blocks are the twelve different categories of turnover behavior that were tracked by code, analyzed and graphed. All describe a set of unique futile behaviors, except the productive cycle block, which describes the desired non-futile behavior.

\subsubsection*{Reaction sequences that make the behavior blocks}

To generate the flow charts, we first defined a set of fourteen reaction sequences. Each of these sequences describes a different way that the composition of molecules bound to the enzyme changes. Eight sequences (ad2, ad2w/S, ad2*, sd, combo-I, combo-II, rd $\rightarrow$ \o\, $\rightarrow$ S, and rd $\rightarrow$ \o\, $\rightarrow$ L) describe eight different ways that P2 dissociates from the enzyme, accompanied by the binding of S, L, of a combination of S and L. Two sequences describe two variations of ligand displacement by susbtrate (ad1, ad1ws). One sequence describes how one substrate can be exchanged for another substrate (combo III). One describes cleavage. One describes rectification. And finally one describes idling:\\

\ctikzset{bipoles/length=.4cm}
\newcommand\esymbol[1]{\begin{circuitikz}
\draw (0,0) to [#1] (1,0); \end{circuitikz}}


\begin{enumerate}
    \item ad1 (allosteric displacement 1): 
    A single substrate binds and displaces ligand, such that when ligand dissociates, substrate is completely bound at all three nodes.
    \item ad1ws (allosteric displacement 1 weak start):
    A single substrate binds and displaces ligand, such that when ligand dissociates, substrate is NOT completely bound at all three nodes.
    \item ad2 (allosteric displacement 2):
    A single ligand binds and displaces P2, such that when P2 dissociates, ligand is completely bound at all three nodes.
    \item ad2w/S (allosteric displacement 2 with S):
    Same as ad2, except that when P2 dissociates, substrate happens to be bound as well, but bound without actively participating in the displacement of P2.
    \item ad2* (allosteric displacement 2*):
    Either a single ligand binds and displaces P2, such that the ligand is NOT completely bound when P2 dissociates; or, multiple ligand bind and displace P2, which requires that none of the ligands can be completely bound.
    \item sd (steric displacement):
    a single subtrate binds and displaces P2.
    \item combo I:
    Substrate and ligand displace P2 together, such that ligand is partially bound when P2 dissociates.
    \item combo II:
    Substrate and ligand displace P2 together, such that ligand is completely bound when P2 dissociates.
    \item combo III:
    The exchange of substrate before it is cleaved, mediated by another substrate alone, or another substrate and ligand working together. After exchange is completed, the new substrate will either displace the ligand, or the ligand will displace the new substrate.  If ligand wins, it will eventually be displaced by another substrate. Thus, we say that a combo III always ends with an ad1. 
    \item rd $\rightarrow$ \o\, $\rightarrow$ S:
    P2 spontaneously dissociates, followed by substrate binding to the empty enzyme.
    \item rd $\rightarrow$ \o\, $\rightarrow$ L:
    P2 spontaneously dissociates, followed by ligand binding to the empty enzyme.
    \item \protect\usym{2702} (cleavage):
    substrate is cleaved into P1 and P2
    \item \esymbol{diode} (rectification):
    P1 dissociates, which stops reversible catalysis
    \item idling:
    Ligand stays bound during the cleavage of substrate. Idling may of may not include ligand also staying bound through rectification.
\end{enumerate}

\begin{figure}[H]
\begin{center}
\includegraphics[width=1\columnwidth]{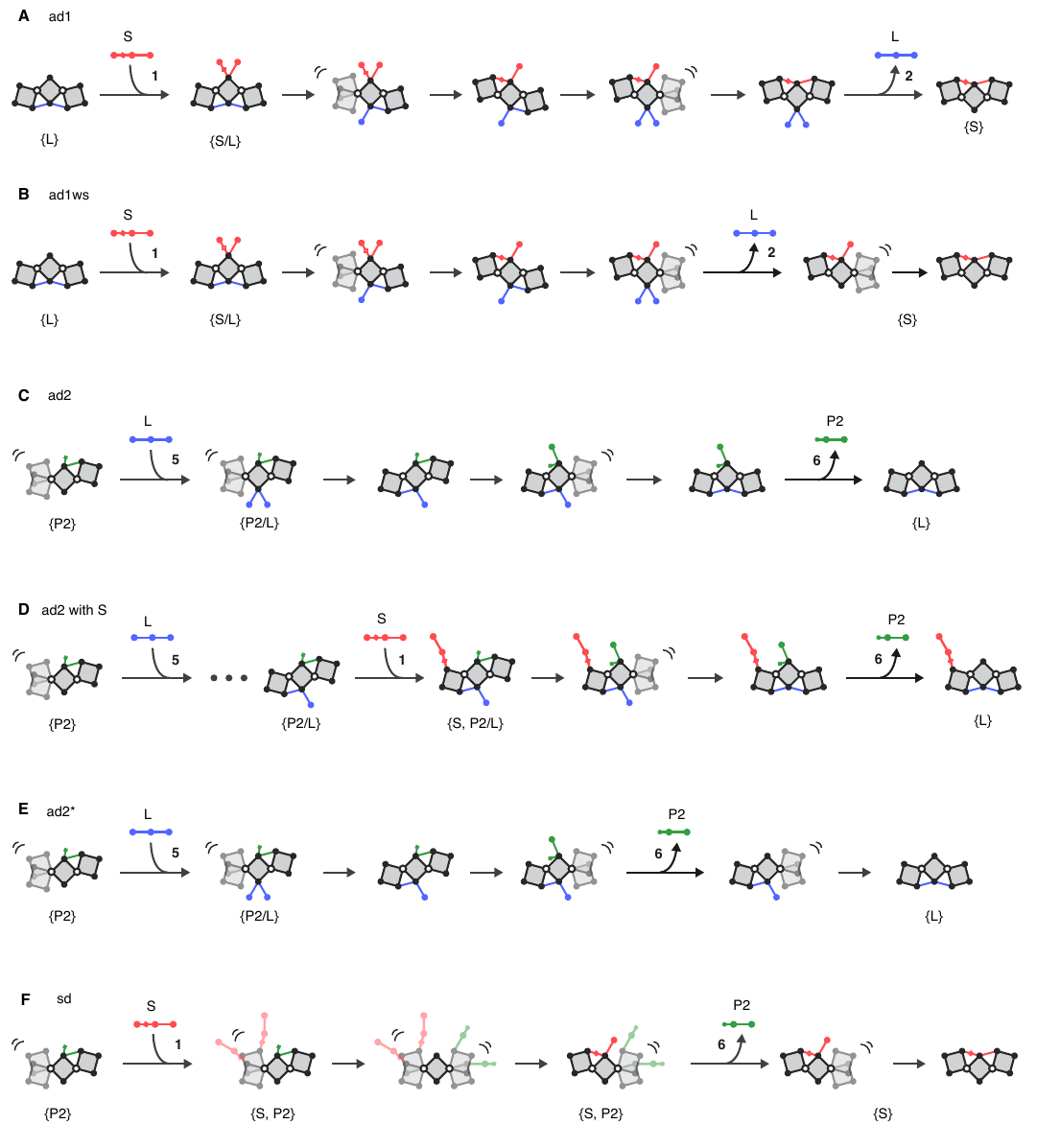}
\caption{\textbf{Reaction sequences.}}
\label{fig:seqs1}
\end{center}
\end{figure}

\begin{figure}[H]
\begin{center}
\includegraphics[width=1\columnwidth]{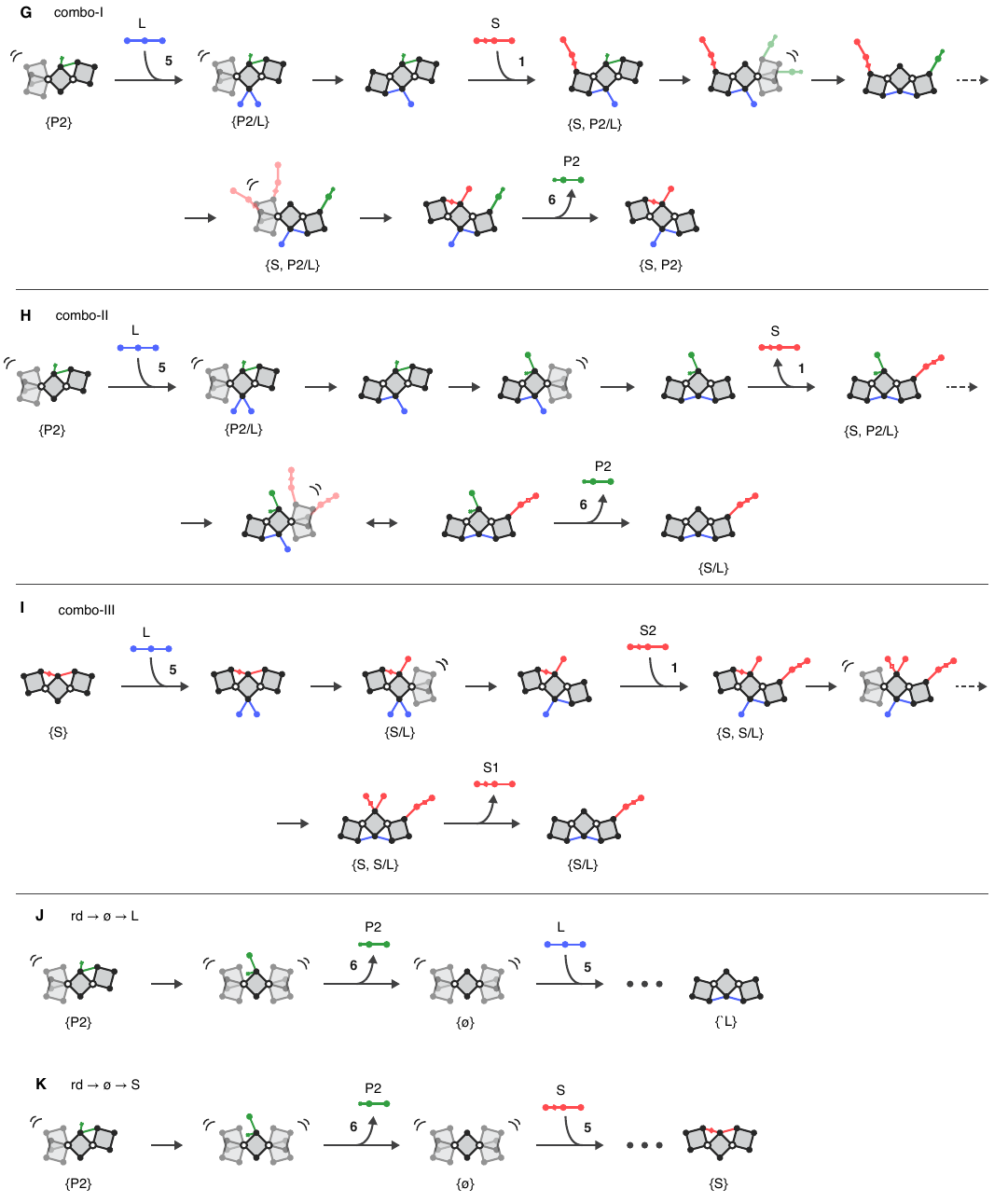}
\caption{\textbf{Reaction sequences.}}
\label{fig:seqs2}
\end{center}
\end{figure}

\begin{figure}[H]
\begin{center}
\includegraphics[width=1\columnwidth]{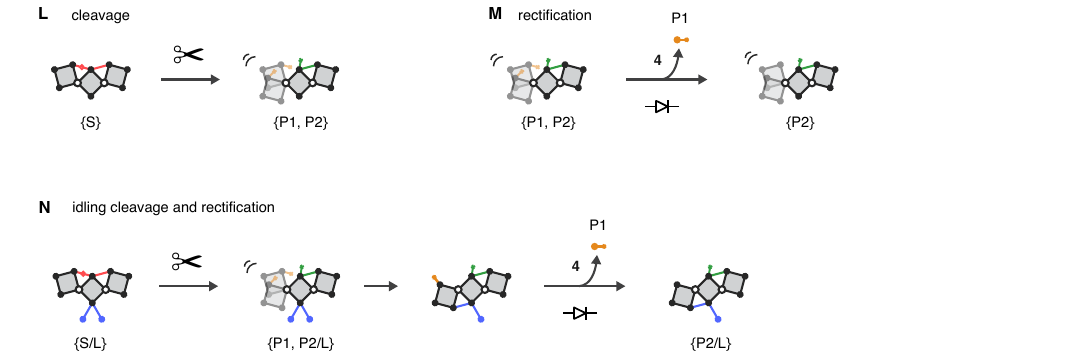}
\caption{\textbf{Reaction sequences.}}
\label{fig:seqs3}
\end{center}
\end{figure}

\subsubsection*{Behavior blocks}

\begin{figure}[H]
\begin{center}
\includegraphics[width=1\columnwidth]{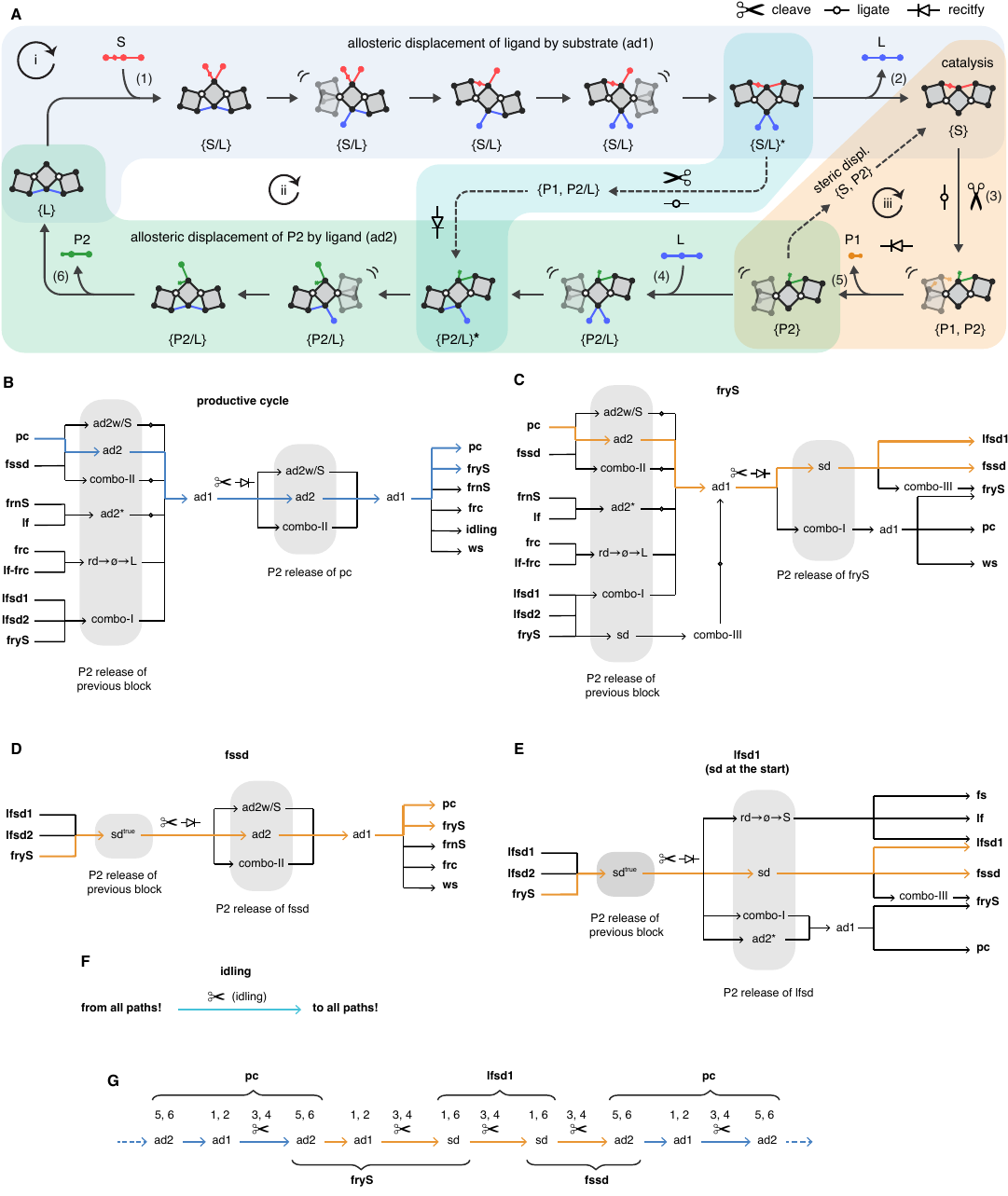}
\caption{\textbf{Behavior blocks.} \textbf{A,} \textbf{B,} \textbf{C,}  \textbf{D,}  \textbf{E,} }
\label{si_blocks1}
\end{center}
\end{figure}

\begin{figure}[H]
\begin{center}
\includegraphics[width=1\columnwidth]{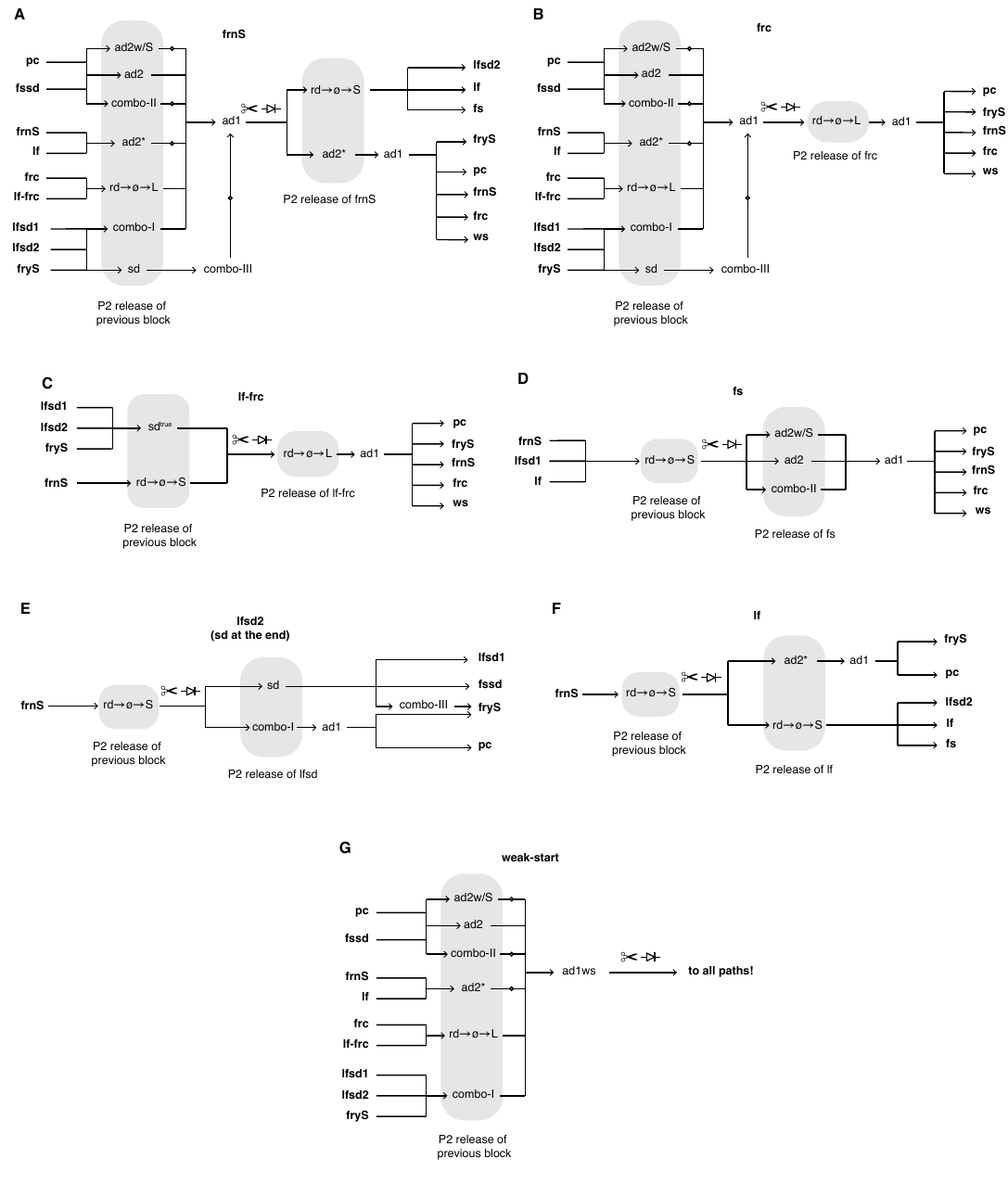}
\caption{\textbf{Behavior blocks continued.} \textbf{A,} \textbf{B,} \textbf{C,}  \textbf{D,}  \textbf{E,} }
\label{si_blocks2}
\end{center}
\end{figure}

There are twelve behavior blocks that are defined from the reaction sequences (list below; and SI Figs. \ref{si_blocks1} \& \ref{si_blocks2}). The first four paths on the list of blocks below are the most commonly occuring. Each path begins and ends with the dissociation of P2. In between the beginning an end a substrate binds and is cleaved. The first P2 that dissociates is the P2 born from the previous substrate that was bound and cleaved. The second P2 that dissociates is born from the substrate that binds and cleaved between the two P2 dissociation events. If the two P2 dissociations conform to the definitions of a productive cycle, then the cycle is a productive cycle. If they do not, the cycle is futile: \\

\begin{enumerate}
    \item pc (productive cycle):
    A pc begins and ends good.
    \item fryS (futile-reset-yes-S):
    A fryS begins good (unless it is preceded by a fryS or lfsd that end with a sd); and it ends badly. Its bad ending is either the steric displacement of P2 by S (sd), or the displacement of P2 by ligand and substrate together (combo I).
    \item lfsd1 (ligand-free-sd; sd at the beginning):
    A lfsd1 begins and ends badly. It begins with a sd, and ends in one of four futile ways (sd, combo-I, ad2*, or rd $\rightarrow$ \o\, $\rightarrow$ S). The `ligand-free' designation is accurate if it does not end with a combo-I or ad2*. Since a combo-I and ad2* occur rarely, more often then not, a lfsd is `ligand-free'.
    \item fssd (futile-start-sd):
    A fssd begins badly but ends good. It begins with a sd, and ends like a productive cycle.
    \item idling:
    Idling can begin and end good or bad. It is futile and unique in that cleavage takes place while ligand is bound.
    \item frnS (futile-reset-no-S):
    A frnS begins good (unless it is preceded by a fryS or lfsd that end with a sd, just like a fryS); and it ends badly. Its bad ending is either P2 spontaneously dissociating, followed substrate binding (rd $\rightarrow$ \o\, $\rightarrow$ S); or the dissociation of P2 when ligand is weakly bound (ad2*).
    \item frc (futile-reset-complete):
    A frc begins like a frnS and fryS, and ends badly, like frnS (rd $\rightarrow$ \o\, $\rightarrow$ S).
    \item lf-frc (ligand-free-futile-reset-complete):
    A lf-frc begins badly, with either a sd or rd $\rightarrow$ \o\, $\rightarrow$ S; and it ends badly, like a frc (rd $\rightarrow$ \o\, $\rightarrow$ S). The `ligand-free' designation is accurate if it does not ends with a combo-I.
    \item fs (futile-start):
    A futile-start begins bad, and ends good (like a pc). Its bad beginning is the random dissociation of P2, followed by the binding S to the empty enzyme (rd $\rightarrow$ \o\, $\rightarrow$ S).
    \item lfsd2 (ligand-free-sd; sd at the end):
    A lfsd2 begins bad (rd $\rightarrow$ \o\, $\rightarrow$ S), and ends bad (sd or combo-I). The "ligand-free" designation is accurate if ends with an sd.
    \item lf (ligand-free):
    A lf begins and ends bad. Its bad beginning is rd $\rightarrow$ \o\, $\rightarrow$ S, and its bad ending is either ad2* or rd $\rightarrow$ \o\, $\rightarrow$ S. The `ligand-free' designation is accurate if it ends with rd $\rightarrow$ \o\, $\rightarrow$ S.
    \item ws (weak-start):
    A ws almost begins good (like a pc), but is characterized by a displacement sequence in which ligand dissociates when substrate is partially bound.
\end{enumerate}

\subsubsection*{Identifying the behavior blocks}

\textbf{Step 1:} The first step was to go through each trajectory, line by line, and record relevant binding information about substrate, ligand and p2. \\

\noindent For each substrate that bound, we recorded:

\begin{enumerate}
    \item when it dissociated—if it dissociated as substrate
    \item when cleavage happened, if it happened
    \item when ligation happened, if it happened
    \item when P1 and P2 (created by cleavage) dissociated
\end{enumerate}

\noindent For each ligand that bound, we recorded:

\begin{enumerate}
    \item when it bound
    \item when it dissociated
    \item if cleavage events took place, while it was bound 
    \item if P2 dissociated, while it was bound
\end{enumerate}

\noindent For each P2 that dissociated, we recorded when it dissociated.\\

\noindent \textbf{Step 2:} In the second step, substrate turnover events were matched to specific behaviors that make them either part of a productive cylces, or part of a futile turnover. These behaviors are the initial definitions of the turnover blocks. The initial definitions are called "non-exclusive" (and labeled with nx superscript) because some turnover events satisfied conditions in more than one these "non-exclusive" blocks.\\

\textbf{Initial \& non-exclusive behavior blocks:}\\

\begin{enumerate}
    \item \textbf{fryS}$^{\mathbf{nx}}$:
    When P2 dissociates, substrate is bound, and ligand (if bound) is not fully bound. By default, ligand can be partially bound, though this is not explicitly tested for. The fryS$^{\text{nx}}$ P2 release conditions satisfy reaction sequences sd, or ad2*, at the end of a fryS, and lfsd1.
    \item \textbf{frnS}$^{\mathbf{nx}}$:
    When P2 dissociates, substrate is not bound at all, and ligand (if bound) is not fully bound. By default, ligand can be partially bound (in states \{11\} or \{12\}). In contrast to a fryS$^{\text{nx}}$, partially bound ligand is flagged here, for later use in defining other exclusive blocks. The frnS$^{\text{nx}}$ P2 release conditions satisfy reaction sequences rd $\rightarrow$ \o\, $\rightarrow$ S and ad2* at the ends of frnS, lf and lfsd1. Combined with the P2 release conditions for frc$^{\text{nx}}$, they satisfy reaction sequence, rd $\rightarrow$ \o\, $\rightarrow$ L, at the end of a frc and lf-frc.
    \item \textbf{frc}$^{\mathbf{nx}}$:
    When P2 dissociates, substrate is not bound at all, ligand is not fully bound, and following P2 dissociation, ligand binds to the empty enzyme. By default, ligand can be partially bound, though this is not explicitly tested for. Combined with frnS$^{\text{nx}}$, these P2 release conditions satisfy reaction sequence, rd $\rightarrow$ \o\, $\rightarrow$ L, at the end of a frc$^{\text{x}}$.
    \item \textbf{fs}$^{\mathbf{nx}}$:
    When substrate binds and is subsequently cleaved, ligand is not bound at all. If it is not the first substrate binding event (turnover event) of the simulation, the fs$^{\text{nx}}$ conditions satisfy reaction sequence, rd $\rightarrow$ \o\, $\rightarrow$ S, at the beginning of fs$^{\text{x}}$, lf$^{\text{x}}$ and lfsd1$^{\text{x}}$, where rd in the reaction sequence corresponds to the P2 release event of the preceding block. If it is the first substrate binding event of the simulation, they satisfy reaction sequence, \o\, $\rightarrow$ S, at their beginnings.  
    \item \textbf{fssd}$^{\mathbf{nx}}$:
    When substrate binds, P2 is bound, and when substrate is subsequently cleaved, ligand is not bound. These conditions eventually satisfy a sd at the beginning of a fssd, for the P2 release of the preceding block.
    \item \textbf{ws}$^{\mathbf{nx}}$:
    When P2 dissociates, ligand is partially bound (ligand is any state but \{10\}). This satisfies the ad1ws condition.
    \item \textbf{lf}$^{\mathbf{nx}}$ (logical combination of frnS and fs):
    lf is defined as a combination of the conditions set for frnS and fs. Here the flag set in frnS is used to rule out partial binding of ligand during P2 dissociation. Thes conditions satisfy the beginning and ending of lf as defined in Fig.
    \item \textbf{lfsd}$^{\mathbf{nx}}$ (logical combination of fryS, frnS, fs, fssd):
    lfsd is defined so that a sd takes place at the beginning, end, or both the beginning and end of a substrate turnover event. Three logical combination of fryS, frnS, fs and fssd achieve this goal:
    \begin{enumerate}[label=(\alph*)]
        \item fssd, fs, and fryS are true; frnS is false (lfsd1, sd at the beginning and end)
        \item fssd, fs and frnS are true; fryS is false (lfsd1, sd at the beginning only)
        \item fs and fryS are true; frnS and fssd are false (lfsd2, sd at the end only)
    \end{enumerate}
    \item \textbf{idling}$^{\mathbf{nx}}$:
    A futile event because ligand is bound during cleavage. Three cases are looked for, with each considering the possibility that substrate is cleaved (and thus ligated at least once) more that once before P1 dissociates, where category 2 and 3 are rare:
    \begin{enumerate}[label=(\alph*)]
        \item Ligand is bound at the first and last cleavage event
        \item Ligand is bound at the first cleavage event, but dissociates before the last cleavage. This case only applies when there are more than one cleavage events
        \item Ligand is NOT bound at the first cleavage event, but binds before the last cleavage. This case also only applies when there are more than one cleavage events.
    \end{enumerate}
    \item \textbf{pc}$^{\mathbf{nx}}$:
    The productive cycle has two main conditions tested for (steps i and ii below) to establish that the substrate turnover event in question is a productive cycle. If these two conditions are met, steps iii and iv determine further details of the two P2 displacements that place, where the P2 displacement tested for in step iii, is not required to occur, given how productive cycles are defined (e.g. they can start with rd $\rightarrow$ \o\, $\rightarrow$ L).
    \begin{enumerate}[label=(\roman*)]
        \item When P2 dissociates, ligand is fully bound (in \{10\}). This P2 is P2\_current, and this ligand is L2.
        \item The substrate from which P2\_current was cleaved is matched to a viable ligand (L1), if possible. Viable ligand are ligands that dissociated when a substrate was fully bound (in \{0\}). If a match is made, the substrate turnover event is a productive cycle. Three conditions are met for a successful match: 
        \begin{enumerate}
            \item substrate binds to the enzyme when a viable ligand is already bound
            \item substrate remains bound after the viable ligand dissociates
            \item substrate is cleaved after the viable ligand dissociates
        \end{enumerate}
        \item If L1 displaced a P2 (P2\_previous) when it bound, L1 is matched to this P2.
        \item P2\_current is finally matched to L2.
    \end{enumerate}
\end{enumerate}
\leavevmode\newline

\noindent \textbf{Step 3:} In the final step, a series of logical statements were used to construct mutually exclusive behavior blocks from the non-exclusive blocks, such that each exclusive block captured a unique set of pathways by which substrate was turned over, and all turnover events were accounted for. 

A tricky aspect of the non-exclusive blocks is that some of them are defined in such a way that they capture unexpected, or unintended behavior, and this behavior is not always consistent with their naming. For example, fssd$^{\text{nx}}$ captures steric displacements, but it also captures combo-I and combo-II displacements, because there is no restriction on the whether ligand is bound or not in the fssd$^{\text{nx}}$ definition. These combos can show up at the start of a productive cycle, and thus the "futile-start" and "sd" monikers for fssd$^{\text{nx}}$, do match the "productive cycle" moniker, but the behavior of fssd$^{nx}$ is consistent with the definition of a productive cycle.

Note that the starred exclusives (lfsd*, lf* and lf-frc*) are only defined as logical combinations of the non-exclusive blocks. However, in the code, lfsd$^{\text{nx}}$ and lf$^{\text{nx}}$ are already defined at step 2, before the final set of logical statements are used to make all the behavior blocks mutually exclusive from one another.

\begin{table}[H]
\centering
\begin{tabular}{r|ccccccccccc}  
 & \text{pc$^{\text{nx}}$} & \text{pci$^{\text{nx}}$} & \text{fryS$^{\text{nx}}$} & \text{frnS$^{\text{nx}}$} & \text{frc$^{\text{nx}}$} & \text{fssd$^{\text{nx}}$} & \text{fs$^{\text{nx}}$} & \text{ws$^{\text{nx}}$} & \text{idl$^{\text{nx}}$}\\
\hline
\multirow{2}{2.5em}{$\text{pc}$} & 1 & 0 & 0 & 0 & 0 & 0 & 0 & 0 & 0 \\
 & 1 & 0 & 0 & 0 & 0 & 1 & 0 & 0 & 0 \\
\hline
\multirow{2}{2.5em}{$\text{pci}$} & 0 & 1 & 0 & 0 & 0 & 0 & 0 & 0 & 0  \\
 & 0 & 1 & 0 & 0 & 0 & 1 & 0 & 0 & 0  \\
\hline
\multirow{2}{2.5em}{$\text{fryS}$} & 0 & 0 & 1 & 0 & 0 & 0 & 1 & 0 & 0 \\
 & 0 & 0 & 1 & 0 & 0 & 1 & 1 & 0 & 0 \\
 \hline
\multirow{2}{2.5em}{$\text{frnS}$} & 0 & 0 & 0 & 1 & 0 & 0 & 0 & 0 & 0 \\
 & 0 & 0 & 0 & 1 & 0 & 1 & 0 & 0 & 0 \\
 \hline
\multirow{1}{2.5em}{$\text{frc}$} & 0 & 0 & 0 & 1 & 1 & 0 & 0 & 0 & 0  \\
\hline
\multirow{3}{2.5em}{$\text{lfsd*}$} & 0 & 0 & 1 & 0 & 0 & 1 & 1 & 0 & 0  \\
& 0 & 0 & 0 & 1 & 0 & 1 & 1 & 0 & 0  \\
& 0 & 0 & 1 & 0 & 0 & 0 & 1 & 0 & 0  \\
\hline
\multirow{1}{2.5em}{$\text{fssd}$} & 0 & 0 & 0 & 0 & 0 & 1 & 1 & 0 & 0  \\
\hline
\multirow{1}{2.5em}{$\text{lf*}$} & 0 & 0 & 0 & 1 & 0 & 0 & 1 & 0 & 0  \\
\hline
\multirow{1}{2.5em}{$\text{lf-frc*}$} & 0 & 0 & 0 & 1 & 1 & 0 & 1 & 0 & 0  \\
\hline
\multirow{1}{2.5em}{$\text{fs}$} & 0 & 0 & 0 & 0 & 0 & 0 & 1 & 0 & 0  \\
\hline
\multirow{1}{2.5em}{$\text{ws}$} & 0 & 0 & 0 & 0 & 0 & 0 & 0 & 1 & 0  \\
\hline
\multirow{1}{2.5em}{$\text{idl}$} & 0 & 0 & 0 & 0 & 0 & 0 & 0 & 0 & 1  \\
\end{tabular}
\caption{\textbf{Exclusive behavior blocks from non-exclusive blocks}. This table shows the sets of non-exclusive behavior blocks (top row, `nx' subscript) that compose each exclusive behavior blocks (left column, no subscript). A `1' indicates that the non-exclusive block is an allowed element, and a `0' indicates that it is not an allowed element. The starred exclusive blocks, lfsd*, lf* and lf-frc*, do not have non-exclusive counterparts, as do the other blocks, and are only defined as compositions of the other non-exclusive blocks.}
\end{table}

\subsection{Movies} \label{si_movies}

(All animations were made by Christian Swinehart (drafting@samizdat.co).)\\

\begin{figure}[H]
\begin{center}
\includegraphics[width=1\columnwidth]{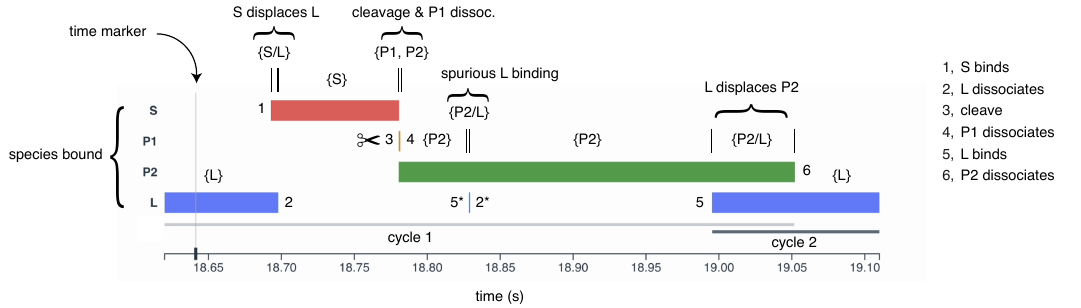}
\caption{\textbf{Movie 1 dwell legend.} This is a labeled version of the legend that appears in Movie 1. This dwell time plot shows a roughly a half second of the simulation, and the species that bound to the enzyme during this time. On the left, the row for each species is labeled (`species bound'), where S is shown in red, P1 in orange, P2 in green and L in blue. When multiple species are bound the bars of color overlap. For example, when S binds, close to 18.70 seconds, the red and blue bars overlap for a short section, during which the displacement of ligand by substrate takes place (see section labeled \{S, L\}). There are three more overlapping sections: \{P1, P2\}, where P1 and P2 are created just after cleavage; \{P2, L\}*, where short-lived spurious binding of L takes place; and \{P2, L\}, during which L displaces P2. From the beginning to roughly 19.05 seconds, one cycle (cycle 1) is completed. The beginning of cycle 1, where L first binds, takes place earlier in the simulation and is not shown. The end of cycle 1 overlaps with the beginning of cycle 2, where the overlap is the section where L displaces P2. The two cycles (cycle 1, grey; cycle 2, black) are demarcated at the bottom with horizontal lines.}
\label{fig:movie-1}
\end{center}
\end{figure}

\begin{figure}[H]
\begin{center}
\includegraphics[width=1\columnwidth]{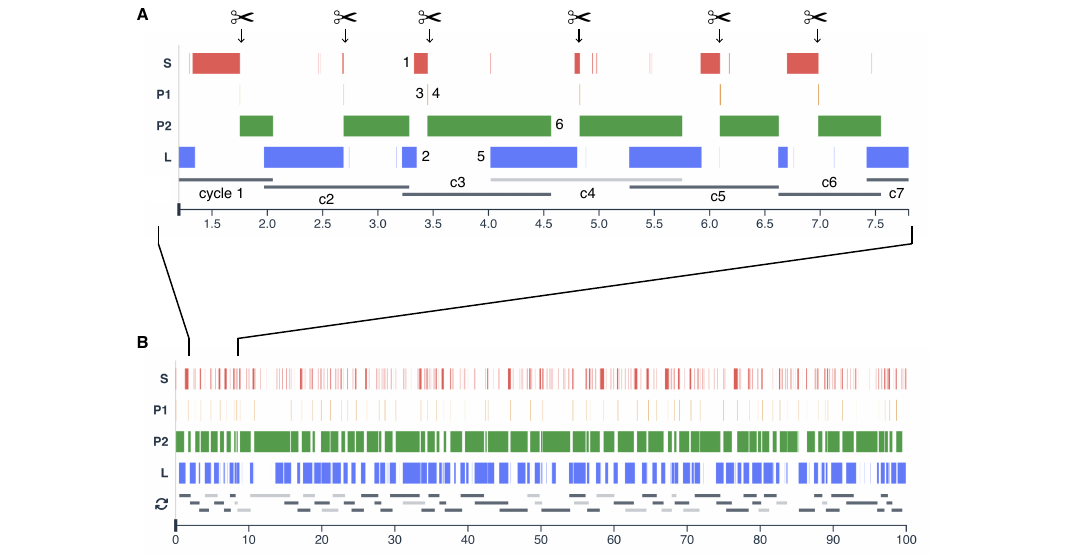}
\caption{\textbf{Movie 2 dwell legend.} \textbf{A.} This is the legend that appears in Movie 2. At the bottom, the seven cycles that take place (though cycles 1 and 7 are not complete) are demarcated. The six reactions are labeled by number for cycle 3. At the top the six cleavage events that take place are demarcated. \textbf{B.} This is the full 100 second simulation from which Movie 2 and Movie 1 were taken.}
\label{fig:movie-2}
\end{center}
\end{figure}

\noindent Each movie is accompanied below by a dwell time plot (see examples in SI Figs. \ref{fig:movie-1} and \ref{fig:movie-2}). In each dwell time plot, the time spent bound to the enzyme by each of the four reactants (dwell time) is represented by horizontal bars of color, with one color for each reactant: red for the substrate; orange for P1; green for P2; and blue for the ligand. The enzyme is represented the same as in the main text, except that all three squares are colored blue, and all the nodes are colored white when not bound. Instead, the reactants are depicted with colored nodes, although without any variation in shading. Thus, the substrate's nodes are all the same color red, P2's nodes are both the same green, and the ligand's nodes are the same blue. The final difference in representation is that the substrate is not depicted with a delta node, even though it is split in the same place. 

\subsubsection*{List of movies:} 

\begin{itemize}[label={}, leftmargin=2mm]
    \item Movie 1: single target cycle 
    \item Movie 2: seven consecutive target cycles
    \item Movie 3: side by side clips of a simulation started with substrate and ligand (left), and a simulation started with substrate only (right).
    \item Movie 4: steric displacement
    \item Movie 5: idling
    \item Movie 6: saturation
\end{itemize}

\subsection{Pathway data} \label{si_data}

Tables 7 and 8 give the total counts for each pathway. Each total count is a sum of ten counts, where each count is from the ten simulations performed at each of the twenty-four substrate concentrations. Each simulation was run for 100 seconds. Thus, for each simulation ($n$) the turnover rate ($v_n$), with units of $\text{no. P2 released}\,\times\,\text{s}^{-1}$, is the count for that simulation ($c_n$) divided by 100 seconds:

\begin{equation}
    v_n = \frac{c_n}{100}
\end{equation}

A mean turnover rate ($v$) was calculated for the ten simulations run:

\begin{equation}
    v = \frac{1}{10}\sum_{n=1}^{10}v_n
\end{equation}

and a standard deviation (sd) of the mean:

\begin{equation}
    \text{sd} = \sqrt{\frac{1}{10}\sum_{n=1}^{10}(v_n-v)^2}
\end{equation}

Each point on the plots shows the mean turnover rate for that pathway and its standard deviation.

\begin{table}[H]
\centering
\begin{tabular}{c|ccccccccccccc}  
[S] $\mu$M & total & idles & fryS & frnS & frc & fs & fssd & lf & lfsd & lf-frc & ws & pc & pci \\
\midrule
$1\times10^{-4}$&1&0&0&0&0&0&0&0&0&0&0&1&0\\
$2\times10^{-4}$&4&0&0&0&0&0&0&0&0&0&0&4&0\\
$5\times10^{-4}$&2&1&0&0&0&0&0&0&0&0&0&1&0\\
$1\times10^{-3}$&19&1&0&1&0&0&0&0&0&0&0&17&0\\
$2\times10^{-3}$&37&3&0&0&0&0&0&0&0&0&1&33&0\\
$5\times10^{-3}$&93&1&0&0&0&2&0&0&0&0&0&90&0\\
$1\times10^{-2}$&148&5&0&1&1&0&0&0&0&0&0&141&0\\
$2\times10^{-2}$&238&8&0&1&0&3&0&0&0&0&0&225&1\\
$5\times10^{-2}$&448&16&0&1&0&5&0&0&0&0&2&420&4\\
$1\times10^{-1}$&624&11&1&2&0&7&0&0&0&0&0&597&6\\
$2\times10^{-1}$&706&18&1&2&1&10&1&0&0&0&2&666&5\\
$5\times10^{-1}$&819&24&13&0&2&8&9&0&0&0&0&760&3\\
$1\times10^{0}$&875&25&33&4&0&13&28&0&1&0&1&765&5\\
$2\times10^{0}$&924&32&51&0&0&8&47&0&4&0&1&775&6\\
$5\times10^{0}$&984&30&104&3&0&11&92&0&13&0&1&724&6\\
$1\times10^{1}$&1061&36&157&2&0&6&133&0&35&0&1&686&5\\
$2\times10^{1}$&1193&41&225&0&0&7&194&0&77&0&2&639&8\\
$5\times10^{1}$&1327&32&308&1&0&9&268&0&133&0&1&566&9\\
$1\times10^{2}$&1476&41&331&0&0&5&303&0&203&0&6&587&0\\
$2\times10^{2}$&1484&37&349&0&0&7&334&0&242&0&11&499&5\\
$5\times10^{2}$&1236&35&283&0&0&7&254&0&196&0&12&444&5\\
$1\times10^{3}$&768&12&183&0&0&6&171&0&122&0&17&256&1\\
$2\times10^{3}$&358&8&86&0&0&7&79&0&49&0&11&118&0\\
$5\times10^{3}$&103&2&12&0&0&6&19&0&23&0&10&31&0
\end{tabular}
\caption{\textbf{Behavior block data for 'basic' simulations.}}
\end{table}

\begin{table}[H]
\centering
\begin{tabular}{c|ccccccccccccc}  
$\text{k}_{\text{cat}}$ $\text{s}^{-1}$ & total & idles & fryS & frnS & frc & fs & fssd & lf & lfsd & lf-frc & ws & pc & pci \\
\midrule
$1$&422&2&4&1&0&7&3&0&0&0&0&401&4\\
$10$&689&21&4&0&2&4&5&0&2&0&1&647&3\\
$100$&798&206&4&1&1&4&4&1&0&0&0&574&3\\
$1\times10^{3}$&1051&803&2&1&0&7&3&0&0&0&0&235&0\\
$1\times10^{4}$&1162&1118&0&0&0&9&0&0&0&0&0&35&0\\
$1\times10^{5}$&1164&1151&0&0&0&7&1&0&0&0&1&4&0\\
$1\times10^{6}$&1181&1170&0&0&0&7&4&0&0&0&0&0&0\\
$1\times10^{7}$&1124&1114&0&0&0&8&2&0&0&0&0&0&0\\
$1\times10^{8}$&1097&1091&0&1&0&5&1&0&0&0&0&0&0\\
$1\times10^{9}$&1109&1102&0&0&0&5&1&0&0&0&1&0&0
\end{tabular}
\caption{\textbf{kcat variation simulations.}}
\end{table}

\begin{table}[H]
\centering
\begin{tabular}{c|ccccccccccccc}  
$\text{k}_{\text{lig}}$ $\text{s}^{-1}$ & total & idles & fryS & frnS & frc & fs & fssd & lf & lfsd & lf-frc & ws & pc & pci \\
\midrule
$1$&700&22&4&3&0&5&4&0&0&0&1&655&6\\
$10$&689&21&4&0&2&4&5&0&2&0&1&647&3\\
$100$&706&27&7&2&2&9&4&0&0&0&0&649&6\\
$1\times10^{3}$&548&24&2&2&0&5&3&0&1&0&0&499&12\\
$1\times10^{4}$&227&10&2&0&0&4&0&0&0&0&0&180&31\\
$1\times10^{5}$&70&5&0&0&0&0&0&0&0&0&0&22&43\\
$1\times10^{6}$&18&4&0&0&0&0&0&0&0&0&0&1&13\\
$1\times10^{7}$&3&1&0&0&0&0&0&0&0&0&0&0&2\\
$1\times10^{8}$&0&0&0&0&0&0&0&0&0&0&0&0&0\\
$1\times10^{9}$&0&0&0&0&0&0&0&0&0&0&0&0&0
\end{tabular}
\caption{\textbf{klig variation simulations.}}
\end{table}

\subsection{Saturation} \label{si_saturation}

\begin{figure}[H]
\begin{center}
\includegraphics[width=1\columnwidth]{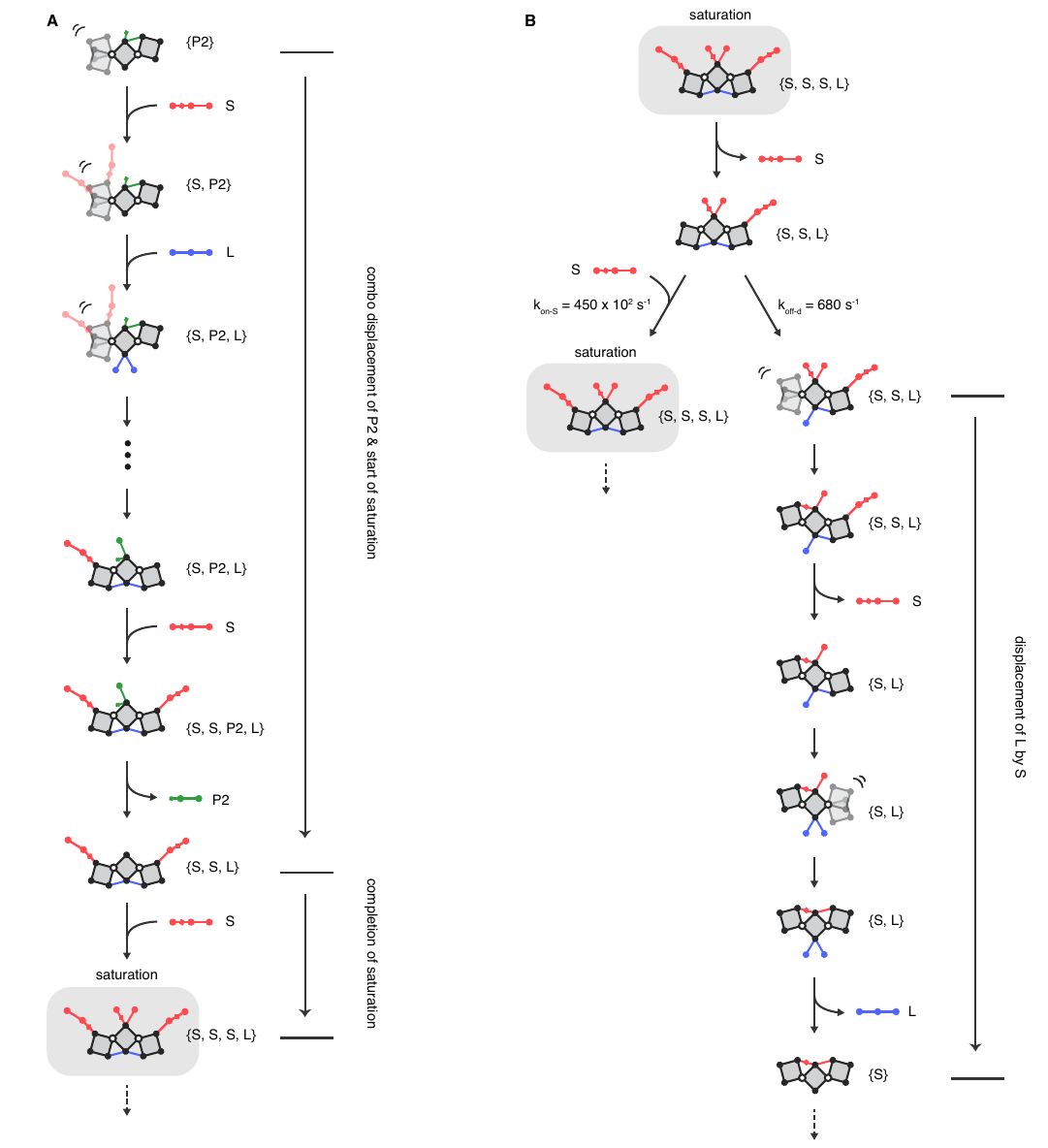}
\caption{\textbf{Saturation.} \textbf{A,} Pathway to saturation. Beginning with the P2-bound state (\{P2\}), substrate, at high concentration, binds at node a, followed by ligand. Trivalent ligand binding (abbreviated by `...') displaces P2, which exposes nodes a and b for substrate binding, and finally, substrate saturation. \textbf{B,} Persistence of the saturation state. The saturation state persists because any substrate that dissociates is likely replaced (left branch) before ligand dissociates from one of its nodes and opens up a pathway for the displacement of ligand by substrate (right branch). Substrate replacement beats ligand dissociation because at high [S], the binding rate surpasses the ligand dissociation by orders of magnitude. For example, at 5 mM [S], $k_{\text{on-S}} = 450\times10^2\,\text{s}^{-1}$, whereas $k_{\text{off-d}} = 680\,\text{s}^{-1}$. The figure shows the saturation state decaying by dissociation of substrate at node \textsf{a} (the second state from top; state \{4, 5, 10\} in basis form, see SI Fig. \ref{fig:twobasis}), but it can also decay by dissociation of substrate from node c (state \{3, 5, 10\} in basis form). }
\label{fig:saturation}
\end{center}
\end{figure}

Saturation (SI Fig. \ref{fig:saturation}) takes place at very high concentrations of substrate (e.g. $>$ 200 $\mu$M), where the rate at which substrate binds the enzyme surpasses the rate at which bound ligand dissociates from its nodes. At peak saturation, the substrate binding site is occupied by three substrate molecules, and the ligand binding site by one trivalently bound ligand (\{S, S, S, L\})(SI Fig. \ref{fig:saturation}A). The bound substrates cannot displace the ligand because the ligand controls the geometry, and each substrate sterically blocks the other from binding divalently or trivalently. The saturation state persists because if one substrate molecule dissociates, it will likely be replaced by another before the ligand dissociates from one of its nodes and allows a bound substrate to bind divalently.

The peak saturation state forms during the displacement of P2 by ligand (SI Fig. \ref{fig:saturation}B). During the displacement process, unoccupied nodes at the substrate binding site are rapidly bound by substrate. The first node to be occupied by substrate is a, which becomes available after P1 dissociates. The most likely second node to be occupied by substrate is c, which is a three-step process involving ligand and the displacement of P2 by ligand. 

To describe this process, the first thing to note is that spurious dissociates of P2 from c are much more likely than b, because the node c reaction is much weaker than that of node \textsf{b} ($k_{\text{off-c}} = 680\,\text{s}^{-1}$ vs. $k_{\text{off-b}} = 3\,\text{s}^{-1}$). During spurious dissociations of P2 from node c, ligand can bind trivalently by capturing node f and the right degree of freedom, thereby inhibiting P2 from rebinding c (SI Fig. \ref{fig:saturation}B, 4th state from top). In this state, where P2 is still bound at node \textsf{b} and ligand is bound trivalently, substrate will likely bind at node c before P2 dissociates from node \textsf{b} and is released into solution, as the rate at which substrate binds at high [S] is much greater than the rate at which P2 dissociates from b ($k_{\text{on-S}} = 4.5\times10^4\,\text{s}^{-1}$ at [S] = 5 mM\footnote{$k_{\text{on-S}}$, at 5 mM [S], is calculated by multiplying the bimolecular rate constant, $k_{\text{bi}}=9\times10^6\,\text{M}^{-1}\text{s}^{-1}$, by 5 mM} vs. $k_{\text{off-b}} = 3\,\text{s}^{-1}$.). Upon substrate binding to node c, the precursor state of saturation is reached, in which one P2, two substrates, and one ligand are bound to the enzyme (\{P2, S, S, L\})(SI Fig. \ref{fig:saturation}B, 5th state from top). When P2 dissociates from node \textsf{b}, rapid substrate binding to node \textsf{b} will complete formation of the peak saturation complex, in which three substrates and one ligand molecule are bound (\{S, S, S, L\}). Peak saturation will dynamically persist, as described above, until one substrate molecule manages to displace ligand.

\subsection{States of the system} \label{all_states}

\scriptsize

\{18\}, \{0\}, \{1\}, \{2\}, \{3\}, \{4\}, \{5\}, \{6\}, \{7\}, \{8\}, \{9\}, \{10\}, \{11\}, \{12\}, \{13\}, \{14\}, \{15\}, \{0, 13\}, \{0, 13, 14\}, \{0, 13, 14, 15\}, \{0, 13, 15\}, \{0, 14\}, \{0, 14, 15\}, \{0, 15\}, \{1, 4\}, \{1, 4, 11\}, \{1, 4, 11, 14\}, \{1, 4, 13\}, \{1, 4, 13, 14\}, \{1, 4, 13, 14, 15\}, \{1, 4, 13, 15\}, \{1, 4, 14\}, \{1, 4, 14, 15\}, \{1, 4, 15\}, \{1, 7\}, \{1, 7, 11\}, \{1, 7, 11, 14\}, \{1, 7, 13\}, \{1, 7, 13, 14\}, \{1, 7, 13, 14, 15\}, \{1, 7, 13, 15\}, \{1, 7, 14\}, \{1, 7, 14, 15\}, \{1, 7, 15\}, \{1, 11\}, \{1, 11, 14\}, \{1, 13\}, \{1, 13, 14\}, \{1, 13, 14, 15\}, \{1, 13, 15\}, \{1, 14\}, \{1, 14, 15\}, \{1, 15\}, \{2, 3\}, \{2, 3, 12\}, \{2, 3, 12, 13\}, \{2, 3, 13\}, \{2, 3, 13, 14\}, \{2, 3, 13, 14, 15\}, \{2, 3, 13, 15\}, \{2, 3, 14\}, \{2, 3, 14, 15\}, \{2, 3, 15\}, \{2, 9\}, \{2, 9, 12\}, \{2, 9, 12, 13\}, \{2, 9, 13\}, \{2, 9, 13, 14\}, \{2, 9, 13, 14, 15\}, \{2, 9, 13, 15\}, \{2, 9, 14\}, \{2, 9, 14, 15\}, \{2, 9, 15\}, \{2, 12\}, \{2, 12, 13\}, \{2, 13\}, \{2, 13, 14\}, \{2, 13, 14, 15\}, \{2, 13, 15\}, \{2, 14\}, \{2, 14, 15\}, \{2, 15\}, \{3, 4\}, \{3, 4, 5\}, \{3, 4, 5, 10\}, \{3, 4, 5, 11\}, \{3, 4, 5, 11, 14\}, \{3, 4, 5, 12\}, \{3, 4, 5, 12, 13\}, \{3, 4, 5, 13\}, \{3, 4, 5, 13, 14\}, \{3, 4, 5, 13, 14, 15\}, \{3, 4, 5, 13, 15\}, \{3, 4, 5, 14\}, \{3, 4, 5, 14, 15\}, \{3, 4, 5, 15\}, \{3, 4, 8\}, \{3, 4, 8, 10\}, \{3, 4, 8, 11\}, \{3, 4, 8, 11, 14\}, \{3, 4, 8, 12\}, \{3, 4, 8, 12, 13\}, \{3, 4, 8, 13\}, \{3, 4, 8, 13, 14\}, \{3, 4, 8, 13, 14, 15\}, \{3, 4, 8, 13, 15\}, \{3, 4, 8, 14\}, \{3, 4, 8, 14, 15\}, \{3, 4, 8, 15\}, \{3, 4, 10\}, \{3, 4, 11\}, \{3, 4, 11, 14\}, \{3, 4, 12\}, \{3, 4, 12, 13\}, \{3, 4, 13\}, \{3, 4, 13, 14\}, \{3, 4, 13, 14, 15\}, \{3, 4, 13, 15\}, \{3, 4, 14\}, \{3, 4, 14, 15\}, \{3, 4, 15\}, \{3, 5\}, \{3, 5, 7\}, \{3, 5, 7, 10\}, \{3, 5, 7, 11\}, \{3, 5, 7, 11, 14\}, \{3, 5, 7, 12\}, \{3, 5, 7, 12, 13\}, \{3, 5, 7, 13\}, \{3, 5, 7, 13, 14\}, \{3, 5, 7, 13, 14, 15\}, \{3, 5, 7, 13, 15\}, \{3, 5, 7, 14\}, \{3, 5, 7, 14, 15\}, \{3, 5, 7, 15\}, \{3, 5, 10\}, \{3, 5, 11\}, \{3, 5, 11, 14\}, \{3, 5, 12\}, \{3, 5, 12, 13\}, \{3, 5, 13\}, \{3, 5, 13, 14\}, \{3, 5, 13, 14, 15\}, \{3, 5, 13, 15\}, \{3, 5, 14\}, \{3, 5, 14, 15\}, \{3, 5, 15\}, \{3, 6\}, \{3, 6, 12\}, \{3, 6, 12, 13\}, \{3, 6, 13\}, \{3, 6, 13, 14\}, \{3, 6, 13, 14, 15\}, \{3, 6, 13, 15\}, \{3, 6, 14\}, \{3, 6, 14, 15\}, \{3, 6, 15\}, \{3, 7\}, \{3, 7, 8\}, \{3, 7, 8, 10\}, \{3, 7, 8, 11\}, \{3, 7, 8, 11, 14\}, \{3, 7, 8, 12\}, \{3, 7, 8, 12, 13\}, \{3, 7, 8, 13\}, \{3, 7, 8, 13, 14\}, \{3, 7, 8, 13, 14, 15\}, \{3, 7, 8, 13, 15\}, \{3, 7, 8, 14\}, \{3, 7, 8, 14, 15\}, \{3, 7, 8, 15\}, \{3, 7, 10\}, \{3, 7, 11\}, \{3, 7, 11, 14\}, \{3, 7, 12\}, \{3, 7, 12, 13\}, \{3, 7, 13\}, \{3, 7, 13, 14\}, \{3, 7, 13, 14, 15\}, \{3, 7, 13, 15\}, \{3, 7, 14\}, \{3, 7, 14, 15\}, \{3, 7, 15\}, \{3, 8\}, \{3, 8, 10\}, \{3, 8, 11\}, \{3, 8, 11, 14\}, \{3, 8, 12\}, \{3, 8, 12, 13\}, \{3, 8, 13\}, \{3, 8, 13, 14\}, \{3, 8, 13, 14, 15\}, \{3, 8, 13, 15\}, \{3, 8, 14\}, \{3, 8, 14, 15\}, \{3, 8, 15\}, \{3, 10\}, \{3, 11\}, \{3, 11, 14\}, \{3, 12\}, \{3, 12, 13\}, \{3, 13\}, \{3, 13, 14\}, \{3, 13, 14, 15\}, \{3, 13, 15\}, \{3, 14\}, \{3, 14, 15\}, \{3, 15\}, \{4, 5\}, \{4, 5, 9\}, \{4, 5, 9, 10\}, \{4, 5, 9, 11\}, \{4, 5, 9, 11, 14\}, \{4, 5, 9, 12\}, \{4, 5, 9, 12, 13\}, \{4, 5, 9, 13\}, \{4, 5, 9, 13, 14\}, \{4, 5, 9, 13, 14, 15\}, \{4, 5, 9, 13, 15\}, \{4, 5, 9, 14\}, \{4, 5, 9, 14, 15\}, \{4, 5, 9, 15\}, \{4, 5, 10\}, \{4, 5, 11\}, \{4, 5, 11, 14\}, \{4, 5, 12\}, \{4, 5, 12, 13\}, \{4, 5, 13\}, \{4, 5, 13, 14\}, \{4, 5, 13, 14, 15\}, \{4, 5, 13, 15\}, \{4, 5, 14\}, \{4, 5, 14, 15\}, \{4, 5, 15\}, \{4, 8\}, \{4, 8, 9\}, \{4, 8, 9, 10\}, \{4, 8, 9, 11\}, \{4, 8, 9, 11, 14\}, \{4, 8, 9, 12\}, \{4, 8, 9, 12, 13\}, \{4, 8, 9, 13\}, \{4, 8, 9, 13, 14\}, \{4, 8, 9, 13, 14, 15\}, \{4, 8, 9, 13, 15\}, \{4, 8, 9, 14\}, \{4, 8, 9, 14, 15\}, \{4, 8, 9, 15\}, \{4, 8, 10\}, \{4, 8, 11\}, \{4, 8, 11, 14\}, \{4, 8, 12\}, \{4, 8, 12, 13\}, \{4, 8, 13\}, \{4, 8, 13, 14\}, \{4, 8, 13, 14, 15\}, \{4, 8, 13, 15\}, \{4, 8, 14\}, \{4, 8, 14, 15\}, \{4, 8, 15\}, \{4, 9\}, \{4, 9, 10\}, \{4, 9, 11\}, \{4, 9, 11, 14\}, \{4, 9, 12\}, \{4, 9, 12, 13\}, \{4, 9, 13\}, \{4, 9, 13, 14\}, \{4, 9, 13, 14, 15\}, \{4, 9, 13, 15\}, \{4, 9, 14\}, \{4, 9, 14, 15\}, \{4, 9, 15\}, \{4, 10\}, \{4, 11\}, \{4, 11, 14\}, \{4, 12\}, \{4, 12, 13\}, \{4, 13\}, \{4, 13, 14\}, \{4, 13, 14, 15\}, \{4, 13, 15\}, \{4, 14\}, \{4, 14, 15\}, \{4, 15\}, \{5, 7\}, \{5, 7, 9\}, \{5, 7, 9, 10\}, \{5, 7, 9, 11\}, \{5, 7, 9, 11, 14\}, \{5, 7, 9, 12\}, \{5, 7, 9, 12, 13\}, \{5, 7, 9, 13\}, \{5, 7, 9, 13, 14\}, \{5, 7, 9, 13, 14, 15\}, \{5, 7, 9, 13, 15\}, \{5, 7, 9, 14\}, \{5, 7, 9, 14, 15\}, \{5, 7, 9, 15\}, \{5, 7, 10\}, \{5, 7, 11\}, \{5, 7, 11, 14\}, \{5, 7, 12\}, \{5, 7, 12, 13\}, \{5, 7, 13\}, \{5, 7, 13, 14\}, \{5, 7, 13, 14, 15\}, \{5, 7, 13, 15\}, \{5, 7, 14\}, \{5, 7, 14, 15\}, \{5, 7, 15\}, \{5, 9\}, \{5, 9, 10\}, \{5, 9, 11\}, \{5, 9, 11, 14\}, \{5, 9, 12\}, \{5, 9, 12, 13\}, \{5, 9, 13\}, \{5, 9, 13, 14\}, \{5, 9, 13, 14, 15\}, \{5, 9, 13, 15\}, \{5, 9, 14\}, \{5, 9, 14, 15\}, \{5, 9, 15\}, \{5, 10\}, \{5, 11\}, \{5, 11, 14\}, \{5, 12\}, \{5, 12, 13\}, \{5, 13\}, \{5, 13, 14\}, \{5, 13, 14, 15\}, \{5, 13, 15\}, \{5, 14\}, \{5, 14, 15\}, \{5, 15\}, \{6, 9\}, \{6, 9, 12\}, \{6, 9, 12, 13\}, \{6, 9, 13\}, \{6, 9, 13, 14\}, \{6, 9, 13, 14, 15\}, \{6, 9, 13, 15\}, \{6, 9, 14\}, \{6, 9, 14, 15\}, \{6, 9, 15\}, \{6, 12\}, \{6, 12, 13\}, \{6, 13\}, \{6, 13, 14\}, \{6, 13, 14, 15\}, \{6, 13, 15\}, \{6, 14\}, \{6, 14, 15\}, \{6, 15\}, \{7, 8\}, \{7, 8, 9\}, \{7, 8, 9, 10\}, \{7, 8, 9, 11\}, \{7, 8, 9, 11, 14\}, \{7, 8, 9, 12\}, \{7, 8, 9, 12, 13\}, \{7, 8, 9, 13\}, \{7, 8, 9, 13, 14\}, \{7, 8, 9, 13, 14, 15\}, \{7, 8, 9, 13, 15\}, \{7, 8, 9, 14\}, \{7, 8, 9, 14, 15\}, \{7, 8, 9, 15\}, \{7, 8, 10\}, \{7, 8, 11\}, \{7, 8, 11, 14\}, \{7, 8, 12\}, \{7, 8, 12, 13\}, \{7, 8, 13\}, \{7, 8, 13, 14\}, \{7, 8, 13, 14, 15\}, \{7, 8, 13, 15\}, \{7, 8, 14\}, \{7, 8, 14, 15\}, \{7, 8, 15\}, \{7, 9\}, \{7, 9, 10\}, \{7, 9, 11\}, \{7, 9, 11, 14\}, \{7, 9, 12\}, \{7, 9, 12, 13\}, \{7, 9, 13\}, \{7, 9, 13, 14\}, \{7, 9, 13, 14, 15\}, \{7, 9, 13, 15\}, \{7, 9, 14\}, \{7, 9, 14, 15\}, \{7, 9, 15\}, \{7, 10\}, \{7, 11\}, \{7, 11, 14\}, \{7, 12\}, \{7, 12, 13\}, \{7, 13\}, \{7, 13, 14\}, \{7, 13, 14, 15\}, \{7, 13, 15\}, \{7, 14\}, \{7, 14, 15\}, \{7, 15\}, \{8, 9\}, \{8, 9, 10\}, \{8, 9, 11\}, \{8, 9, 11, 14\}, \{8, 9, 12\}, \{8, 9, 12, 13\}, \{8, 9, 13\}, \{8, 9, 13, 14\}, \{8, 9, 13, 14, 15\}, \{8, 9, 13, 15\}, \{8, 9, 14\}, \{8, 9, 14, 15\}, \{8, 9, 15\}, \{8, 10\}, \{8, 11\}, \{8, 11, 14\}, \{8, 12\}, \{8, 12, 13\}, \{8, 13\}, \{8, 13, 14\}, \{8, 13, 14, 15\}, \{8, 13, 15\}, \{8, 14\}, \{8, 14, 15\}, \{8, 15\}, \{9, 10\}, \{9, 11\}, \{9, 11, 14\}, \{9, 12\}, \{9, 12, 13\}, \{9, 13\}, \{9, 13, 14\}, \{9, 13, 14, 15\}, \{9, 13, 15\}, \{9, 14\}, \{9, 14, 15\}, \{9, 15\}, \{11, 14\}, \{12, 13\}, \{13, 14\}, \{13, 14, 15\}, \{13, 15\}, \{14, 15\}

\subsection{Possible transitions out of each state} \label{all_transitions}

\scriptsize

\{18\} $\rightarrow$ \{3\}, \{4\}, \{5\}, \{7\}, \{8\}, \{9\}, \{13\}, \{14\}, \{15\}\\
\{0\} $\rightarrow$ \{1\}, \{2\}, \{0, 13\}, \{0, 14\}, \{0, 15\}, \{6, 9\}\\
\{1\} $\rightarrow$ \{0\}, \{3\}, \{5\}, \{1, 4\}, \{1, 7\}, \{1, 13\}, \{1, 14\}, \{1, 15\}\\
\{2\} $\rightarrow$ \{0\}, \{4\}, \{5\}, \{2, 3\}, \{2, 9\}, \{2, 13\}, \{2, 14\}, \{2, 15\}\\
\{3\} $\rightarrow$ \{1\}, \{18\}, \{3, 4\}, \{3, 5\}, \{3, 7\}, \{3, 8\}, \{3, 13\}, \{3, 14\}, \{3, 15\}\\
\{4\} $\rightarrow$ \{2\}, \{18\}, \{3, 4\}, \{4, 5\}, \{4, 8\}, \{4, 9\}, \{4, 13\}, \{4, 14\}, \{4, 15\}\\
\{5\} $\rightarrow$ \{1\}, \{2\}, \{18\}, \{3, 5\}, \{4, 5\}, \{5, 7\}, \{5, 9\}, \{5, 13\}, \{5, 14\}, \{5, 15\}\\
\{6\} $\rightarrow$ \{7\}, \{8\}, \{3, 6\}, \{6, 9\}, \{6, 13\}, \{6, 14\}, \{6, 15\}\\
\{7\} $\rightarrow$ \{6\}, \{18\}, \{3, 7\}, \{5, 7\}, \{7, 8\}, \{7, 9\}, \{7, 13\}, \{7, 14\}, \{7, 15\}\\
\{8\} $\rightarrow$ \{6\}, \{18\}, \{3, 8\}, \{4, 8\}, \{7, 8\}, \{8, 9\}, \{8, 13\}, \{8, 14\}, \{8, 15\}\\
\{9\} $\rightarrow$ \{18\}, \{4, 9\}, \{5, 9\}, \{7, 9\}, \{8, 9\}, \{9, 13\}, \{9, 14\}, \{9, 15\}\\
\{10\} $\rightarrow$ \{11\}, \{12\}, \{3, 10\}, \{4, 10\}, \{5, 10\}, \{7, 10\}, \{8, 10\}, \{9, 10\}\\
\{11\} $\rightarrow$ \{10\}, \{13\}, \{15\}, \{3, 11\}, \{4, 11\}, \{5, 11\}, \{7, 11\}, \{8, 11\}, \{9, 11\}, \{11, 14\}\\
\{12\} $\rightarrow$ \{10\}, \{14\}, \{15\}, \{3, 12\}, \{4, 12\}, \{5, 12\}, \{7, 12\}, \{8, 12\}, \{9, 12\}, \{12, 13\}\\
\{13\} $\rightarrow$ \{11\}, \{18\}, \{3, 13\}, \{4, 13\}, \{5, 13\}, \{7, 13\}, \{8, 13\}, \{9, 13\}, \{13, 14\}, \{13, 15\}\\
\{14\} $\rightarrow$ \{12\}, \{18\}, \{3, 14\}, \{4, 14\}, \{5, 14\}, \{7, 14\}, \{8, 14\}, \{9, 14\}, \{13, 14\}, \{14, 15\}\\
\{15\} $\rightarrow$ \{11\}, \{12\}, \{18\}, \{3, 15\}, \{4, 15\}, \{5, 15\}, \{7, 15\}, \{8, 15\}, \{9, 15\}, \{13, 15\}, \{14, 15\}\\
\{0, 13\} $\rightarrow$ \{1, 13\}, \{2, 13\}, \{0\}, \{0, 13, 14\}, \{0, 13, 15\}, \{6, 9, 13\}\\
\{0, 13, 14\} $\rightarrow$ \{1, 13, 14\}, \{2, 13, 14\}, \{0, 14\}, \{0, 13\}, \{0, 13, 14, 15\}, \{6, 9, 13, 14\}\\
\{0, 13, 14, 15\} $\rightarrow$ \{1, 13, 14, 15\}, \{2, 13, 14, 15\}, \{0, 14, 15\}, \{0, 13, 15\}, \{0, 13, 14\}, \{6, 9, 13, 14, 15\}\\
\{0, 13, 15\} $\rightarrow$ \{1, 13, 15\}, \{2, 13, 15\}, \{0, 15\}, \{0, 13\}, \{0, 13, 14, 15\}, \{6, 9, 13, 15\}\\
\{0, 14\} $\rightarrow$ \{1, 14\}, \{2, 14\}, \{0\}, \{0, 13, 14\}, \{0, 14, 15\}, \{6, 9, 14\}\\
\{0, 14, 15\} $\rightarrow$ \{1, 14, 15\}, \{2, 14, 15\}, \{0, 15\}, \{0, 14\}, \{0, 13, 14, 15\}, \{6, 9, 14, 15\}\\
\{0, 15\} $\rightarrow$ \{1, 15\}, \{2, 15\}, \{0\}, \{0, 13, 15\}, \{0, 14, 15\}, \{6, 9, 15\}\\
\{1, 4\} $\rightarrow$ \{3, 4\}, \{4, 5\}, \{1\}, \{1, 4, 13\}, \{1, 4, 14\}, \{1, 4, 15\}\\
\{1, 4, 11\} $\rightarrow$ \{3, 4, 11\}, \{4, 5, 11\}, \{1, 11\}, \{1, 4, 13\}, \{1, 4, 15\}, \{1, 4, 11, 14\}\\
\{1, 4, 11, 14\} $\rightarrow$ \{3, 4, 11, 14\}, \{4, 5, 11, 14\}, \{1, 11, 14\}, \{1, 4, 13, 14\}, \{1, 4, 14, 15\}, \{1, 4, 11\}\\
\{1, 4, 13\} $\rightarrow$ \{3, 4, 13\}, \{4, 5, 13\}, \{1, 13\}, \{1, 4, 11\}, \{1, 4\}, \{1, 4, 13, 14\}, \{1, 4, 13, 15\}\\
\{1, 4, 13, 14\} $\rightarrow$ \{3, 4, 13, 14\}, \{4, 5, 13, 14\}, \{1, 13, 14\}, \{1, 4, 11, 14\}, \{1, 4, 14\}, \{1, 4, 13\}, \{1, 4, 13, 14, 15\}\\
\{1, 4, 13, 14, 15\} $\rightarrow$ \{3, 4, 13, 14, 15\}, \{4, 5, 13, 14, 15\}, \{1, 13, 14, 15\}, \{1, 4, 14, 15\}, \{1, 4, 13, 15\}, \{1, 4, 13, 14\}\\
\{1, 4, 13, 15\} $\rightarrow$ \{3, 4, 13, 15\}, \{4, 5, 13, 15\}, \{1, 13, 15\}, \{1, 4, 15\}, \{1, 4, 13\}, \{1, 4, 13, 14, 15\}\\
\{1, 4, 14\} $\rightarrow$ \{3, 4, 14\}, \{4, 5, 14\}, \{1, 14\}, \{1, 4\}, \{1, 4, 13, 14\}, \{1, 4, 14, 15\}\\
\{1, 4, 14, 15\} $\rightarrow$ \{3, 4, 14, 15\}, \{4, 5, 14, 15\}, \{1, 14, 15\}, \{1, 4, 15\}, \{1, 4, 11, 14\}, \{1, 4, 14\}, \{1, 4, 13, 14, 15\}\\
\{1, 4, 15\} $\rightarrow$ \{3, 4, 15\}, \{4, 5, 15\}, \{1, 15\}, \{1, 4, 11\}, \{1, 4\}, \{1, 4, 13, 15\}, \{1, 4, 14, 15\}\\
\{1, 7\} $\rightarrow$ \{3, 7\}, \{5, 7\}, \{1\}, \{1, 7, 13\}, \{1, 7, 14\}, \{1, 7, 15\}\\
\{1, 7, 11\} $\rightarrow$ \{3, 7, 11\}, \{5, 7, 11\}, \{1, 11\}, \{1, 7, 13\}, \{1, 7, 15\}, \{1, 7, 11, 14\}\\
\{1, 7, 11, 14\} $\rightarrow$ \{3, 7, 11, 14\}, \{5, 7, 11, 14\}, \{1, 11, 14\}, \{1, 7, 13, 14\}, \{1, 7, 14, 15\}, \{1, 7, 11\}\\
\{1, 7, 13\} $\rightarrow$ \{3, 7, 13\}, \{5, 7, 13\}, \{1, 13\}, \{1, 7, 11\}, \{1, 7\}, \{1, 7, 13, 14\}, \{1, 7, 13, 15\}\\
\{1, 7, 13, 14\} $\rightarrow$ \{3, 7, 13, 14\}, \{5, 7, 13, 14\}, \{1, 13, 14\}, \{1, 7, 11, 14\}, \{1, 7, 14\}, \{1, 7, 13\}, \{1, 7, 13, 14, 15\}\\
\{1, 7, 13, 14, 15\} $\rightarrow$ \{3, 7, 13, 14, 15\}, \{5, 7, 13, 14, 15\}, \{1, 13, 14, 15\}, \{1, 7, 14, 15\}, \{1, 7, 13, 15\}, \{1, 7, 13, 14\}\\
\{1, 7, 13, 15\} $\rightarrow$ \{3, 7, 13, 15\}, \{5, 7, 13, 15\}, \{1, 13, 15\}, \{1, 7, 15\}, \{1, 7, 13\}, \{1, 7, 13, 14, 15\}\\
\{1, 7, 14\} $\rightarrow$ \{3, 7, 14\}, \{5, 7, 14\}, \{1, 14\}, \{1, 7\}, \{1, 7, 13, 14\}, \{1, 7, 14, 15\}\\
\{1, 7, 14, 15\} $\rightarrow$ \{3, 7, 14, 15\}, \{5, 7, 14, 15\}, \{1, 14, 15\}, \{1, 7, 15\}, \{1, 7, 11, 14\}, \{1, 7, 14\}, \{1, 7, 13, 14, 15\}\\
\{1, 7, 15\} $\rightarrow$ \{3, 7, 15\}, \{5, 7, 15\}, \{1, 15\}, \{1, 7, 11\}, \{1, 7\}, \{1, 7, 13, 15\}, \{1, 7, 14, 15\}\\
\{1, 11\} $\rightarrow$ \{3, 11\}, \{5, 11\}, \{1, 13\}, \{1, 15\}, \{1, 4, 11\}, \{1, 7, 11\}, \{1, 11, 14\}\\
\{1, 11, 14\} $\rightarrow$ \{3, 11, 14\}, \{5, 11, 14\}, \{1, 13, 14\}, \{1, 14, 15\}, \{1, 11\}, \{1, 4, 11, 14\}, \{1, 7, 11, 14\}\\
\{1, 13\} $\rightarrow$ \{0, 13\}, \{3, 13\}, \{5, 13\}, \{1, 11\}, \{1\}, \{1, 4, 13\}, \{1, 7, 13\}, \{1, 13, 14\}, \{1, 13, 15\}\\
\{1, 13, 14\} $\rightarrow$ \{0, 13, 14\}, \{3, 13, 14\}, \{5, 13, 14\}, \{1, 11, 14\}, \{1, 14\}, \{1, 13\}, \{1, 4, 13, 14\}, \{1, 7, 13, 14\}, \{1, 13, 14, 15\}\\
\{1, 13, 14, 15\} $\rightarrow$ \{0, 13, 14, 15\}, \{3, 13, 14, 15\}, \{5, 13, 14, 15\}, \{1, 14, 15\}, \{1, 13, 15\}, \{1, 13, 14\}, \{1, 4, 13, 14, 15\}, \{1, 7, 13, 14, 15\}\\
\{1, 13, 15\} $\rightarrow$ \{0, 13, 15\}, \{3, 13, 15\}, \{5, 13, 15\}, \{1, 15\}, \{1, 13\}, \{1, 4, 13, 15\}, \{1, 7, 13, 15\}, \{1, 13, 14, 15\}\\
\{1, 14\} $\rightarrow$ \{0, 14\}, \{3, 14\}, \{5, 14\}, \{1\}, \{1, 4, 14\}, \{1, 7, 14\}, \{1, 13, 14\}, \{1, 14, 15\}\\
\{1, 14, 15\} $\rightarrow$ \{0, 14, 15\}, \{3, 14, 15\}, \{5, 14, 15\}, \{1, 15\}, \{1, 11, 14\}, \{1, 14\}, \{1, 4, 14, 15\}, \{1, 7, 14, 15\}, \{1, 13, 14, 15\}\\
\{1, 15\} $\rightarrow$ \{0, 15\}, \{3, 15\}, \{5, 15\}, \{1, 11\}, \{1\}, \{1, 4, 15\}, \{1, 7, 15\}, \{1, 13, 15\}, \{1, 14, 15\}\\
\{2, 3\} $\rightarrow$ \{3, 4\}, \{3, 5\}, \{2\}, \{2, 3, 13\}, \{2, 3, 14\}, \{2, 3, 15\}\\
\{2, 3, 12\} $\rightarrow$ \{3, 4, 12\}, \{3, 5, 12\}, \{2, 12\}, \{2, 3, 14\}, \{2, 3, 15\}, \{2, 3, 12, 13\}\\
\{2, 3, 12, 13\} $\rightarrow$ \{3, 4, 12, 13\}, \{3, 5, 12, 13\}, \{2, 12, 13\}, \{2, 3, 13, 14\}, \{2, 3, 13, 15\}, \{2, 3, 12\}\\
\{2, 3, 13\} $\rightarrow$ \{3, 4, 13\}, \{3, 5, 13\}, \{2, 13\}, \{2, 3\}, \{2, 3, 13, 14\}, \{2, 3, 13, 15\}\\
\{2, 3, 13, 14\} $\rightarrow$ \{3, 4, 13, 14\}, \{3, 5, 13, 14\}, \{2, 13, 14\}, \{2, 3, 14\}, \{2, 3, 12, 13\}, \{2, 3, 13\}, \{2, 3, 13, 14, 15\}\\
\{2, 3, 13, 14, 15\} $\rightarrow$ \{3, 4, 13, 14, 15\}, \{3, 5, 13, 14, 15\}, \{2, 13, 14, 15\}, \{2, 3, 14, 15\}, \{2, 3, 13, 15\}, \{2, 3, 13, 14\}\\
\{2, 3, 13, 15\} $\rightarrow$ \{3, 4, 13, 15\}, \{3, 5, 13, 15\}, \{2, 13, 15\}, \{2, 3, 15\}, \{2, 3, 12, 13\}, \{2, 3, 13\}, \{2, 3, 13, 14, 15\}\\
\{2, 3, 14\} $\rightarrow$ \{3, 4, 14\}, \{3, 5, 14\}, \{2, 14\}, \{2, 3, 12\}, \{2, 3\}, \{2, 3, 13, 14\}, \{2, 3, 14, 15\}\\
\{2, 3, 14, 15\} $\rightarrow$ \{3, 4, 14, 15\}, \{3, 5, 14, 15\}, \{2, 14, 15\}, \{2, 3, 15\}, \{2, 3, 14\}, \{2, 3, 13, 14, 15\}\\
\{2, 3, 15\} $\rightarrow$ \{3, 4, 15\}, \{3, 5, 15\}, \{2, 15\}, \{2, 3, 12\}, \{2, 3\}, \{2, 3, 13, 15\}, \{2, 3, 14, 15\}\\
\{2, 9\} $\rightarrow$ \{4, 9\}, \{5, 9\}, \{2\}, \{2, 9, 13\}, \{2, 9, 14\}, \{2, 9, 15\}\\
\{2, 9, 12\} $\rightarrow$ \{4, 9, 12\}, \{5, 9, 12\}, \{2, 12\}, \{2, 9, 14\}, \{2, 9, 15\}, \{2, 9, 12, 13\}\\
\{2, 9, 12, 13\} $\rightarrow$ \{4, 9, 12, 13\}, \{5, 9, 12, 13\}, \{2, 12, 13\}, \{2, 9, 13, 14\}, \{2, 9, 13, 15\}, \{2, 9, 12\}\\
\{2, 9, 13\} $\rightarrow$ \{4, 9, 13\}, \{5, 9, 13\}, \{2, 13\}, \{2, 9\}, \{2, 9, 13, 14\}, \{2, 9, 13, 15\}\\
\{2, 9, 13, 14\} $\rightarrow$ \{4, 9, 13, 14\}, \{5, 9, 13, 14\}, \{2, 13, 14\}, \{2, 9, 14\}, \{2, 9, 12, 13\}, \{2, 9, 13\}, \{2, 9, 13, 14, 15\}\\
\{2, 9, 13, 14, 15\} $\rightarrow$ \{4, 9, 13, 14, 15\}, \{5, 9, 13, 14, 15\}, \{2, 13, 14, 15\}, \{2, 9, 14, 15\}, \{2, 9, 13, 15\}, \{2, 9, 13, 14\}\\
\{2, 9, 13, 15\} $\rightarrow$ \{4, 9, 13, 15\}, \{5, 9, 13, 15\}, \{2, 13, 15\}, \{2, 9, 15\}, \{2, 9, 12, 13\}, \{2, 9, 13\}, \{2, 9, 13, 14, 15\}\\
\{2, 9, 14\} $\rightarrow$ \{4, 9, 14\}, \{5, 9, 14\}, \{2, 14\}, \{2, 9, 12\}, \{2, 9\}, \{2, 9, 13, 14\}, \{2, 9, 14, 15\}\\
\{2, 9, 14, 15\} $\rightarrow$ \{4, 9, 14, 15\}, \{5, 9, 14, 15\}, \{2, 14, 15\}, \{2, 9, 15\}, \{2, 9, 14\}, \{2, 9, 13, 14, 15\}\\
\{2, 9, 15\} $\rightarrow$ \{4, 9, 15\}, \{5, 9, 15\}, \{2, 15\}, \{2, 9, 12\}, \{2, 9\}, \{2, 9, 13, 15\}, \{2, 9, 14, 15\}\\
\{2, 12\} $\rightarrow$ \{4, 12\}, \{5, 12\}, \{2, 14\}, \{2, 15\}, \{2, 3, 12\}, \{2, 9, 12\}, \{2, 12, 13\}\\
\{2, 12, 13\} $\rightarrow$ \{4, 12, 13\}, \{5, 12, 13\}, \{2, 13, 14\}, \{2, 13, 15\}, \{2, 12\}, \{2, 3, 12, 13\}, \{2, 9, 12, 13\}\\
\{2, 13\} $\rightarrow$ \{0, 13\}, \{4, 13\}, \{5, 13\}, \{2\}, \{2, 3, 13\}, \{2, 9, 13\}, \{2, 13, 14\}, \{2, 13, 15\}\\
\{2, 13, 14\} $\rightarrow$ \{0, 13, 14\}, \{4, 13, 14\}, \{5, 13, 14\}, \{2, 14\}, \{2, 12, 13\}, \{2, 13\}, \{2, 3, 13, 14\}, \{2, 9, 13, 14\}, \{2, 13, 14, 15\}\\
\{2, 13, 14, 15\} $\rightarrow$ \{0, 13, 14, 15\}, \{4, 13, 14, 15\}, \{5, 13, 14, 15\}, \{2, 14, 15\}, \{2, 13, 15\}, \{2, 13, 14\}, \{2, 3, 13, 14, 15\}, \{2, 9, 13, 14, 15\}\\
\{2, 13, 15\} $\rightarrow$ \{0, 13, 15\}, \{4, 13, 15\}, \{5, 13, 15\}, \{2, 15\}, \{2, 12, 13\}, \{2, 13\}, \{2, 3, 13, 15\}, \{2, 9, 13, 15\}, \{2, 13, 14, 15\}\\
\{2, 14\} $\rightarrow$ \{0, 14\}, \{4, 14\}, \{5, 14\}, \{2, 12\}, \{2\}, \{2, 3, 14\}, \{2, 9, 14\}, \{2, 13, 14\}, \{2, 14, 15\}\\
\{2, 14, 15\} $\rightarrow$ \{0, 14, 15\}, \{4, 14, 15\}, \{5, 14, 15\}, \{2, 15\}, \{2, 14\}, \{2, 3, 14, 15\}, \{2, 9, 14, 15\}, \{2, 13, 14, 15\}\\
\{2, 15\} $\rightarrow$ \{0, 15\}, \{4, 15\}, \{5, 15\}, \{2, 12\}, \{2\}, \{2, 3, 15\}, \{2, 9, 15\}, \{2, 13, 15\}, \{2, 14, 15\}\\
\{3, 4\} $\rightarrow$ \{1, 4\}, \{4\}, \{2, 3\}, \{3\}, \{3, 4, 5\}, \{3, 4, 8\}, \{3, 4, 13\}, \{3, 4, 14\}, \{3, 4, 15\}\\
\{3, 4, 5\} $\rightarrow$ \{4, 5\}, \{3, 5\}, \{3, 4\}, \{3, 4, 5, 13\}, \{3, 4, 5, 14\}, \{3, 4, 5, 15\}\\
\{3, 4, 5, 10\} $\rightarrow$ \{4, 5, 10\}, \{3, 5, 10\}, \{3, 4, 10\}, \{3, 4, 5, 11\}, \{3, 4, 5, 12\}\\
\{3, 4, 5, 11\} $\rightarrow$ \{4, 5, 11\}, \{3, 5, 11\}, \{3, 4, 11\}, \{3, 4, 5, 10\}, \{3, 4, 5, 13\}, \{3, 4, 5, 15\}, \{3, 4, 5, 11, 14\}\\
\{3, 4, 5, 11, 14\} $\rightarrow$ \{4, 5, 11, 14\}, \{3, 5, 11, 14\}, \{3, 4, 11, 14\}, \{3, 4, 5, 13, 14\}, \{3, 4, 5, 14, 15\}, \{3, 4, 5, 11\}\\
\{3, 4, 5, 12\} $\rightarrow$ \{4, 5, 12\}, \{3, 5, 12\}, \{3, 4, 12\}, \{3, 4, 5, 10\}, \{3, 4, 5, 14\}, \{3, 4, 5, 15\}, \{3, 4, 5, 12, 13\}\\
\{3, 4, 5, 12, 13\} $\rightarrow$ \{4, 5, 12, 13\}, \{3, 5, 12, 13\}, \{3, 4, 12, 13\}, \{3, 4, 5, 13, 14\}, \{3, 4, 5, 13, 15\}, \{3, 4, 5, 12\}\\
\{3, 4, 5, 13\} $\rightarrow$ \{4, 5, 13\}, \{3, 5, 13\}, \{3, 4, 13\}, \{3, 4, 5, 11\}, \{3, 4, 5\}, \{3, 4, 5, 13, 14\}, \{3, 4, 5, 13, 15\}\\
\{3, 4, 5, 13, 14\} $\rightarrow$ \{4, 5, 13, 14\}, \{3, 5, 13, 14\}, \{3, 4, 13, 14\}, \{3, 4, 5, 11, 14\}, \{3, 4, 5, 14\}, \{3, 4, 5, 12, 13\}, \{3, 4, 5, 13\}, \{3, 4, 5, 13, 14, 15\}\\
\{3, 4, 5, 13, 14, 15\} $\rightarrow$ \{4, 5, 13, 14, 15\}, \{3, 5, 13, 14, 15\}, \{3, 4, 13, 14, 15\}, \{3, 4, 5, 14, 15\}, \{3, 4, 5, 13, 15\}, \{3, 4, 5, 13, 14\}\\
\{3, 4, 5, 13, 15\} $\rightarrow$ \{4, 5, 13, 15\}, \{3, 5, 13, 15\}, \{3, 4, 13, 15\}, \{3, 4, 5, 15\}, \{3, 4, 5, 12, 13\}, \{3, 4, 5, 13\}, \{3, 4, 5, 13, 14, 15\}\\
\{3, 4, 5, 14\} $\rightarrow$ \{4, 5, 14\}, \{3, 5, 14\}, \{3, 4, 14\}, \{3, 4, 5, 12\}, \{3, 4, 5\}, \{3, 4, 5, 13, 14\}, \{3, 4, 5, 14, 15\}\\
\{3, 4, 5, 14, 15\} $\rightarrow$ \{4, 5, 14, 15\}, \{3, 5, 14, 15\}, \{3, 4, 14, 15\}, \{3, 4, 5, 15\}, \{3, 4, 5, 11, 14\}, \{3, 4, 5, 14\}, \{3, 4, 5, 13, 14, 15\}\\
\{3, 4, 5, 15\} $\rightarrow$ \{4, 5, 15\}, \{3, 5, 15\}, \{3, 4, 15\}, \{3, 4, 5, 11\}, \{3, 4, 5, 12\}, \{3, 4, 5\}, \{3, 4, 5, 13, 15\}, \{3, 4, 5, 14, 15\}\\
\{3, 4, 8\} $\rightarrow$ \{4, 8\}, \{3, 8\}, \{3, 4\}, \{3, 4, 8, 13\}, \{3, 4, 8, 14\}, \{3, 4, 8, 15\}\\
\{3, 4, 8, 10\} $\rightarrow$ \{4, 8, 10\}, \{3, 8, 10\}, \{3, 4, 10\}, \{3, 4, 8, 11\}, \{3, 4, 8, 12\}\\
\{3, 4, 8, 11\} $\rightarrow$ \{4, 8, 11\}, \{3, 8, 11\}, \{3, 4, 11\}, \{3, 4, 8, 10\}, \{3, 4, 8, 13\}, \{3, 4, 8, 15\}, \{3, 4, 8, 11, 14\}\\
\{3, 4, 8, 11, 14\} $\rightarrow$ \{4, 8, 11, 14\}, \{3, 8, 11, 14\}, \{3, 4, 11, 14\}, \{3, 4, 8, 13, 14\}, \{3, 4, 8, 14, 15\}, \{3, 4, 8, 11\}\\
\{3, 4, 8, 12\} $\rightarrow$ \{4, 8, 12\}, \{3, 8, 12\}, \{3, 4, 12\}, \{3, 4, 8, 10\}, \{3, 4, 8, 14\}, \{3, 4, 8, 15\}, \{3, 4, 8, 12, 13\}\\
\{3, 4, 8, 12, 13\} $\rightarrow$ \{4, 8, 12, 13\}, \{3, 8, 12, 13\}, \{3, 4, 12, 13\}, \{3, 4, 8, 13, 14\}, \{3, 4, 8, 13, 15\}, \{3, 4, 8, 12\}\\
\{3, 4, 8, 13\} $\rightarrow$ \{4, 8, 13\}, \{3, 8, 13\}, \{3, 4, 13\}, \{3, 4, 8, 11\}, \{3, 4, 8\}, \{3, 4, 8, 13, 14\}, \{3, 4, 8, 13, 15\}\\
\{3, 4, 8, 13, 14\} $\rightarrow$ \{4, 8, 13, 14\}, \{3, 8, 13, 14\}, \{3, 4, 13, 14\}, \{3, 4, 8, 11, 14\}, \{3, 4, 8, 14\}, \{3, 4, 8, 12, 13\}, \{3, 4, 8, 13\}, \{3, 4, 8, 13, 14, 15\}\\
\{3, 4, 8, 13, 14, 15\} $\rightarrow$ \{4, 8, 13, 14, 15\}, \{3, 8, 13, 14, 15\}, \{3, 4, 13, 14, 15\}, \{3, 4, 8, 14, 15\}, \{3, 4, 8, 13, 15\}, \{3, 4, 8, 13, 14\}\\
\{3, 4, 8, 13, 15\} $\rightarrow$ \{4, 8, 13, 15\}, \{3, 8, 13, 15\}, \{3, 4, 13, 15\}, \{3, 4, 8, 15\}, \{3, 4, 8, 12, 13\}, \{3, 4, 8, 13\}, \{3, 4, 8, 13, 14, 15\}\\
\{3, 4, 8, 14\} $\rightarrow$ \{4, 8, 14\}, \{3, 8, 14\}, \{3, 4, 14\}, \{3, 4, 8, 12\}, \{3, 4, 8\}, \{3, 4, 8, 13, 14\}, \{3, 4, 8, 14, 15\}\\
\{3, 4, 8, 14, 15\} $\rightarrow$ \{4, 8, 14, 15\}, \{3, 8, 14, 15\}, \{3, 4, 14, 15\}, \{3, 4, 8, 15\}, \{3, 4, 8, 11, 14\}, \{3, 4, 8, 14\}, \{3, 4, 8, 13, 14, 15\}\\
\{3, 4, 8, 15\} $\rightarrow$ \{4, 8, 15\}, \{3, 8, 15\}, \{3, 4, 15\}, \{3, 4, 8, 11\}, \{3, 4, 8, 12\}, \{3, 4, 8\}, \{3, 4, 8, 13, 15\}, \{3, 4, 8, 14, 15\}\\
\{3, 4, 10\} $\rightarrow$ \{4, 10\}, \{3, 10\}, \{3, 4, 11\}, \{3, 4, 12\}, \{3, 4, 5, 10\}, \{3, 4, 8, 10\}\\
\{3, 4, 11\} $\rightarrow$ \{1, 4, 11\}, \{4, 11\}, \{3, 11\}, \{3, 4, 10\}, \{3, 4, 13\}, \{3, 4, 15\}, \{3, 4, 5, 11\}, \{3, 4, 8, 11\}, \{3, 4, 11, 14\}\\
\{3, 4, 11, 14\} $\rightarrow$ \{1, 4, 11, 14\}, \{4, 11, 14\}, \{3, 11, 14\}, \{3, 4, 13, 14\}, \{3, 4, 14, 15\}, \{3, 4, 11\}, \{3, 4, 5, 11, 14\}, \{3, 4, 8, 11, 14\}\\
\{3, 4, 12\} $\rightarrow$ \{4, 12\}, \{2, 3, 12\}, \{3, 12\}, \{3, 4, 10\}, \{3, 4, 14\}, \{3, 4, 15\}, \{3, 4, 5, 12\}, \{3, 4, 8, 12\}, \{3, 4, 12, 13\}\\
\{3, 4, 12, 13\} $\rightarrow$ \{4, 12, 13\}, \{2, 3, 12, 13\}, \{3, 12, 13\}, \{3, 4, 13, 14\}, \{3, 4, 13, 15\}, \{3, 4, 12\}, \{3, 4, 5, 12, 13\}, \{3, 4, 8, 12, 13\}\\
\{3, 4, 13\} $\rightarrow$ \{1, 4, 13\}, \{4, 13\}, \{2, 3, 13\}, \{3, 13\}, \{3, 4, 11\}, \{3, 4\}, \{3, 4, 5, 13\}, \{3, 4, 8, 13\}, \{3, 4, 13, 14\}, \{3, 4, 13, 15\}\\
\{3, 4, 13, 14\} $\rightarrow$ \{1, 4, 13, 14\}, \{4, 13, 14\}, \{2, 3, 13, 14\}, \{3, 13, 14\}, \{3, 4, 11, 14\}, \{3, 4, 14\}, \{3, 4, 12, 13\}, \{3, 4, 13\}, \{3, 4, 5, 13, 14\}, \{3, 4, 8, 13, 14\}, \{3, 4, 13, 14, 15\}\\
\{3, 4, 13, 14, 15\} $\rightarrow$ \{1, 4, 13, 14, 15\}, \{4, 13, 14, 15\}, \{2, 3, 13, 14, 15\}, \{3, 13, 14, 15\}, \{3, 4, 14, 15\}, \{3, 4, 13, 15\}, \{3, 4, 13, 14\}, \{3, 4, 5, 13, 14, 15\}, \{3, 4, 8, 13, 14, 15\}\\
\{3, 4, 13, 15\} $\rightarrow$ \{1, 4, 13, 15\}, \{4, 13, 15\}, \{2, 3, 13, 15\}, \{3, 13, 15\}, \{3, 4, 15\}, \{3, 4, 12, 13\}, \{3, 4, 13\}, \{3, 4, 5, 13, 15\}, \{3, 4, 8, 13, 15\}, \{3, 4, 13, 14, 15\}\\
\{3, 4, 14\} $\rightarrow$ \{1, 4, 14\}, \{4, 14\}, \{2, 3, 14\}, \{3, 14\}, \{3, 4, 12\}, \{3, 4\}, \{3, 4, 5, 14\}, \{3, 4, 8, 14\}, \{3, 4, 13, 14\}, \{3, 4, 14, 15\}\\
\{3, 4, 14, 15\} $\rightarrow$ \{1, 4, 14, 15\}, \{4, 14, 15\}, \{2, 3, 14, 15\}, \{3, 14, 15\}, \{3, 4, 15\}, \{3, 4, 11, 14\}, \{3, 4, 14\}, \{3, 4, 5, 14, 15\}, \{3, 4, 8, 14, 15\}, \{3, 4, 13, 14, 15\}\\
\{3, 4, 15\} $\rightarrow$ \{1, 4, 15\}, \{4, 15\}, \{2, 3, 15\}, \{3, 15\}, \{3, 4, 11\}, \{3, 4, 12\}, \{3, 4\}, \{3, 4, 5, 15\}, \{3, 4, 8, 15\}, \{3, 4, 13, 15\}, \{3, 4, 14, 15\}\\
\{3, 5\} $\rightarrow$ \{5\}, \{2, 3\}, \{3\}, \{3, 4, 5\}, \{3, 5, 7\}, \{3, 5, 13\}, \{3, 5, 14\}, \{3, 5, 15\}\\
\{3, 5, 7\} $\rightarrow$ \{5, 7\}, \{3, 7\}, \{3, 5\}, \{3, 5, 7, 13\}, \{3, 5, 7, 14\}, \{3, 5, 7, 15\}\\
\{3, 5, 7, 10\} $\rightarrow$ \{5, 7, 10\}, \{3, 7, 10\}, \{3, 5, 10\}, \{3, 5, 7, 11\}, \{3, 5, 7, 12\}\\
\{3, 5, 7, 11\} $\rightarrow$ \{5, 7, 11\}, \{3, 7, 11\}, \{3, 5, 11\}, \{3, 5, 7, 10\}, \{3, 5, 7, 13\}, \{3, 5, 7, 15\}, \{3, 5, 7, 11, 14\}\\
\{3, 5, 7, 11, 14\} $\rightarrow$ \{5, 7, 11, 14\}, \{3, 7, 11, 14\}, \{3, 5, 11, 14\}, \{3, 5, 7, 13, 14\}, \{3, 5, 7, 14, 15\}, \{3, 5, 7, 11\}\\
\{3, 5, 7, 12\} $\rightarrow$ \{5, 7, 12\}, \{3, 7, 12\}, \{3, 5, 12\}, \{3, 5, 7, 10\}, \{3, 5, 7, 14\}, \{3, 5, 7, 15\}, \{3, 5, 7, 12, 13\}\\
\{3, 5, 7, 12, 13\} $\rightarrow$ \{5, 7, 12, 13\}, \{3, 7, 12, 13\}, \{3, 5, 12, 13\}, \{3, 5, 7, 13, 14\}, \{3, 5, 7, 13, 15\}, \{3, 5, 7, 12\}\\
\{3, 5, 7, 13\} $\rightarrow$ \{5, 7, 13\}, \{3, 7, 13\}, \{3, 5, 13\}, \{3, 5, 7, 11\}, \{3, 5, 7\}, \{3, 5, 7, 13, 14\}, \{3, 5, 7, 13, 15\}\\
\{3, 5, 7, 13, 14\} $\rightarrow$ \{5, 7, 13, 14\}, \{3, 7, 13, 14\}, \{3, 5, 13, 14\}, \{3, 5, 7, 11, 14\}, \{3, 5, 7, 14\}, \{3, 5, 7, 12, 13\}, \{3, 5, 7, 13\}, \{3, 5, 7, 13, 14, 15\}\\
\{3, 5, 7, 13, 14, 15\} $\rightarrow$ \{5, 7, 13, 14, 15\}, \{3, 7, 13, 14, 15\}, \{3, 5, 13, 14, 15\}, \{3, 5, 7, 14, 15\}, \{3, 5, 7, 13, 15\}, \{3, 5, 7, 13, 14\}\\
\{3, 5, 7, 13, 15\} $\rightarrow$ \{5, 7, 13, 15\}, \{3, 7, 13, 15\}, \{3, 5, 13, 15\}, \{3, 5, 7, 15\}, \{3, 5, 7, 12, 13\}, \{3, 5, 7, 13\}, \{3, 5, 7, 13, 14, 15\}\\
\{3, 5, 7, 14\} $\rightarrow$ \{5, 7, 14\}, \{3, 7, 14\}, \{3, 5, 14\}, \{3, 5, 7, 12\}, \{3, 5, 7\}, \{3, 5, 7, 13, 14\}, \{3, 5, 7, 14, 15\}\\
\{3, 5, 7, 14, 15\} $\rightarrow$ \{5, 7, 14, 15\}, \{3, 7, 14, 15\}, \{3, 5, 14, 15\}, \{3, 5, 7, 15\}, \{3, 5, 7, 11, 14\}, \{3, 5, 7, 14\}, \{3, 5, 7, 13, 14, 15\}\\
\{3, 5, 7, 15\} $\rightarrow$ \{5, 7, 15\}, \{3, 7, 15\}, \{3, 5, 15\}, \{3, 5, 7, 11\}, \{3, 5, 7, 12\}, \{3, 5, 7\}, \{3, 5, 7, 13, 15\}, \{3, 5, 7, 14, 15\}\\
\{3, 5, 10\} $\rightarrow$ \{5, 10\}, \{3, 10\}, \{3, 5, 11\}, \{3, 5, 12\}, \{3, 4, 5, 10\}, \{3, 5, 7, 10\}\\
\{3, 5, 11\} $\rightarrow$ \{5, 11\}, \{3, 11\}, \{3, 5, 10\}, \{3, 5, 13\}, \{3, 5, 15\}, \{3, 4, 5, 11\}, \{3, 5, 7, 11\}, \{3, 5, 11, 14\}\\
\{3, 5, 11, 14\} $\rightarrow$ \{5, 11, 14\}, \{3, 11, 14\}, \{3, 5, 13, 14\}, \{3, 5, 14, 15\}, \{3, 5, 11\}, \{3, 4, 5, 11, 14\}, \{3, 5, 7, 11, 14\}\\
\{3, 5, 12\} $\rightarrow$ \{5, 12\}, \{2, 3, 12\}, \{3, 12\}, \{3, 5, 10\}, \{3, 5, 14\}, \{3, 5, 15\}, \{3, 4, 5, 12\}, \{3, 5, 7, 12\}, \{3, 5, 12, 13\}\\
\{3, 5, 12, 13\} $\rightarrow$ \{5, 12, 13\}, \{2, 3, 12, 13\}, \{3, 12, 13\}, \{3, 5, 13, 14\}, \{3, 5, 13, 15\}, \{3, 5, 12\}, \{3, 4, 5, 12, 13\}, \{3, 5, 7, 12, 13\}\\
\{3, 5, 13\} $\rightarrow$ \{5, 13\}, \{2, 3, 13\}, \{3, 13\}, \{3, 5, 11\}, \{3, 5\}, \{3, 4, 5, 13\}, \{3, 5, 7, 13\}, \{3, 5, 13, 14\}, \{3, 5, 13, 15\}\\
\{3, 5, 13, 14\} $\rightarrow$ \{5, 13, 14\}, \{2, 3, 13, 14\}, \{3, 13, 14\}, \{3, 5, 11, 14\}, \{3, 5, 14\}, \{3, 5, 12, 13\}, \{3, 5, 13\}, \{3, 4, 5, 13, 14\}, \{3, 5, 7, 13, 14\}, \{3, 5, 13, 14, 15\}\\
\{3, 5, 13, 14, 15\} $\rightarrow$ \{5, 13, 14, 15\}, \{2, 3, 13, 14, 15\}, \{3, 13, 14, 15\}, \{3, 5, 14, 15\}, \{3, 5, 13, 15\}, \{3, 5, 13, 14\}, \{3, 4, 5, 13, 14, 15\}, \{3, 5, 7, 13, 14, 15\}\\
\{3, 5, 13, 15\} $\rightarrow$ \{5, 13, 15\}, \{2, 3, 13, 15\}, \{3, 13, 15\}, \{3, 5, 15\}, \{3, 5, 12, 13\}, \{3, 5, 13\}, \{3, 4, 5, 13, 15\}, \{3, 5, 7, 13, 15\}, \{3, 5, 13, 14, 15\}\\
\{3, 5, 14\} $\rightarrow$ \{5, 14\}, \{2, 3, 14\}, \{3, 14\}, \{3, 5, 12\}, \{3, 5\}, \{3, 4, 5, 14\}, \{3, 5, 7, 14\}, \{3, 5, 13, 14\}, \{3, 5, 14, 15\}\\
\{3, 5, 14, 15\} $\rightarrow$ \{5, 14, 15\}, \{2, 3, 14, 15\}, \{3, 14, 15\}, \{3, 5, 15\}, \{3, 5, 11, 14\}, \{3, 5, 14\}, \{3, 4, 5, 14, 15\}, \{3, 5, 7, 14, 15\}, \{3, 5, 13, 14, 15\}\\
\{3, 5, 15\} $\rightarrow$ \{5, 15\}, \{2, 3, 15\}, \{3, 15\}, \{3, 5, 11\}, \{3, 5, 12\}, \{3, 5\}, \{3, 4, 5, 15\}, \{3, 5, 7, 15\}, \{3, 5, 13, 15\}, \{3, 5, 14, 15\}\\
\{3, 6\} $\rightarrow$ \{6\}, \{3, 7\}, \{3, 8\}, \{3, 6, 13\}, \{3, 6, 14\}, \{3, 6, 15\}\\
\{3, 6, 12\} $\rightarrow$ \{6, 12\}, \{3, 7, 12\}, \{3, 8, 12\}, \{3, 6, 14\}, \{3, 6, 15\}, \{3, 6, 12, 13\}\\
\{3, 6, 12, 13\} $\rightarrow$ \{6, 12, 13\}, \{3, 7, 12, 13\}, \{3, 8, 12, 13\}, \{3, 6, 13, 14\}, \{3, 6, 13, 15\}, \{3, 6, 12\}\\
\{3, 6, 13\} $\rightarrow$ \{6, 13\}, \{3, 7, 13\}, \{3, 8, 13\}, \{3, 6\}, \{3, 6, 13, 14\}, \{3, 6, 13, 15\}\\
\{3, 6, 13, 14\} $\rightarrow$ \{6, 13, 14\}, \{3, 7, 13, 14\}, \{3, 8, 13, 14\}, \{3, 6, 14\}, \{3, 6, 12, 13\}, \{3, 6, 13\}, \{3, 6, 13, 14, 15\}\\
\{3, 6, 13, 14, 15\} $\rightarrow$ \{6, 13, 14, 15\}, \{3, 7, 13, 14, 15\}, \{3, 8, 13, 14, 15\}, \{3, 6, 14, 15\}, \{3, 6, 13, 15\}, \{3, 6, 13, 14\}\\
\{3, 6, 13, 15\} $\rightarrow$ \{6, 13, 15\}, \{3, 7, 13, 15\}, \{3, 8, 13, 15\}, \{3, 6, 15\}, \{3, 6, 12, 13\}, \{3, 6, 13\}, \{3, 6, 13, 14, 15\}\\
\{3, 6, 14\} $\rightarrow$ \{6, 14\}, \{3, 7, 14\}, \{3, 8, 14\}, \{3, 6, 12\}, \{3, 6\}, \{3, 6, 13, 14\}, \{3, 6, 14, 15\}\\
\{3, 6, 14, 15\} $\rightarrow$ \{6, 14, 15\}, \{3, 7, 14, 15\}, \{3, 8, 14, 15\}, \{3, 6, 15\}, \{3, 6, 14\}, \{3, 6, 13, 14, 15\}\\
\{3, 6, 15\} $\rightarrow$ \{6, 15\}, \{3, 7, 15\}, \{3, 8, 15\}, \{3, 6, 12\}, \{3, 6\}, \{3, 6, 13, 15\}, \{3, 6, 14, 15\}\\
\{3, 7\} $\rightarrow$ \{1, 7\}, \{7\}, \{3, 6\}, \{3\}, \{3, 5, 7\}, \{3, 7, 8\}, \{3, 7, 13\}, \{3, 7, 14\}, \{3, 7, 15\}\\
\{3, 7, 8\} $\rightarrow$ \{7, 8\}, \{3, 8\}, \{3, 7\}, \{3, 7, 8, 13\}, \{3, 7, 8, 14\}, \{3, 7, 8, 15\}\\
\{3, 7, 8, 10\} $\rightarrow$ \{7, 8, 10\}, \{3, 8, 10\}, \{3, 7, 10\}, \{3, 7, 8, 11\}, \{3, 7, 8, 12\}\\
\{3, 7, 8, 11\} $\rightarrow$ \{7, 8, 11\}, \{3, 8, 11\}, \{3, 7, 11\}, \{3, 7, 8, 10\}, \{3, 7, 8, 13\}, \{3, 7, 8, 15\}, \{3, 7, 8, 11, 14\}\\
\{3, 7, 8, 11, 14\} $\rightarrow$ \{7, 8, 11, 14\}, \{3, 8, 11, 14\}, \{3, 7, 11, 14\}, \{3, 7, 8, 13, 14\}, \{3, 7, 8, 14, 15\}, \{3, 7, 8, 11\}\\
\{3, 7, 8, 12\} $\rightarrow$ \{7, 8, 12\}, \{3, 8, 12\}, \{3, 7, 12\}, \{3, 7, 8, 10\}, \{3, 7, 8, 14\}, \{3, 7, 8, 15\}, \{3, 7, 8, 12, 13\}\\
\{3, 7, 8, 12, 13\} $\rightarrow$ \{7, 8, 12, 13\}, \{3, 8, 12, 13\}, \{3, 7, 12, 13\}, \{3, 7, 8, 13, 14\}, \{3, 7, 8, 13, 15\}, \{3, 7, 8, 12\}\\
\{3, 7, 8, 13\} $\rightarrow$ \{7, 8, 13\}, \{3, 8, 13\}, \{3, 7, 13\}, \{3, 7, 8, 11\}, \{3, 7, 8\}, \{3, 7, 8, 13, 14\}, \{3, 7, 8, 13, 15\}\\
\{3, 7, 8, 13, 14\} $\rightarrow$ \{7, 8, 13, 14\}, \{3, 8, 13, 14\}, \{3, 7, 13, 14\}, \{3, 7, 8, 11, 14\}, \{3, 7, 8, 14\}, \{3, 7, 8, 12, 13\}, \{3, 7, 8, 13\}, \{3, 7, 8, 13, 14, 15\}\\
\{3, 7, 8, 13, 14, 15\} $\rightarrow$ \{7, 8, 13, 14, 15\}, \{3, 8, 13, 14, 15\}, \{3, 7, 13, 14, 15\}, \{3, 7, 8, 14, 15\}, \{3, 7, 8, 13, 15\}, \{3, 7, 8, 13, 14\}\\
\{3, 7, 8, 13, 15\} $\rightarrow$ \{7, 8, 13, 15\}, \{3, 8, 13, 15\}, \{3, 7, 13, 15\}, \{3, 7, 8, 15\}, \{3, 7, 8, 12, 13\}, \{3, 7, 8, 13\}, \{3, 7, 8, 13, 14, 15\}\\
\{3, 7, 8, 14\} $\rightarrow$ \{7, 8, 14\}, \{3, 8, 14\}, \{3, 7, 14\}, \{3, 7, 8, 12\}, \{3, 7, 8\}, \{3, 7, 8, 13, 14\}, \{3, 7, 8, 14, 15\}\\
\{3, 7, 8, 14, 15\} $\rightarrow$ \{7, 8, 14, 15\}, \{3, 8, 14, 15\}, \{3, 7, 14, 15\}, \{3, 7, 8, 15\}, \{3, 7, 8, 11, 14\}, \{3, 7, 8, 14\}, \{3, 7, 8, 13, 14, 15\}\\
\{3, 7, 8, 15\} $\rightarrow$ \{7, 8, 15\}, \{3, 8, 15\}, \{3, 7, 15\}, \{3, 7, 8, 11\}, \{3, 7, 8, 12\}, \{3, 7, 8\}, \{3, 7, 8, 13, 15\}, \{3, 7, 8, 14, 15\}\\
\{3, 7, 10\} $\rightarrow$ \{7, 10\}, \{3, 10\}, \{3, 7, 11\}, \{3, 7, 12\}, \{3, 5, 7, 10\}, \{3, 7, 8, 10\}\\
\{3, 7, 11\} $\rightarrow$ \{1, 7, 11\}, \{7, 11\}, \{3, 11\}, \{3, 7, 10\}, \{3, 7, 13\}, \{3, 7, 15\}, \{3, 5, 7, 11\}, \{3, 7, 8, 11\}, \{3, 7, 11, 14\}\\
\{3, 7, 11, 14\} $\rightarrow$ \{1, 7, 11, 14\}, \{7, 11, 14\}, \{3, 11, 14\}, \{3, 7, 13, 14\}, \{3, 7, 14, 15\}, \{3, 7, 11\}, \{3, 5, 7, 11, 14\}, \{3, 7, 8, 11, 14\}\\
\{3, 7, 12\} $\rightarrow$ \{7, 12\}, \{3, 6, 12\}, \{3, 12\}, \{3, 7, 10\}, \{3, 7, 14\}, \{3, 7, 15\}, \{3, 5, 7, 12\}, \{3, 7, 8, 12\}, \{3, 7, 12, 13\}\\
\{3, 7, 12, 13\} $\rightarrow$ \{7, 12, 13\}, \{3, 6, 12, 13\}, \{3, 12, 13\}, \{3, 7, 13, 14\}, \{3, 7, 13, 15\}, \{3, 7, 12\}, \{3, 5, 7, 12, 13\}, \{3, 7, 8, 12, 13\}\\
\{3, 7, 13\} $\rightarrow$ \{1, 7, 13\}, \{7, 13\}, \{3, 6, 13\}, \{3, 13\}, \{3, 7, 11\}, \{3, 7\}, \{3, 5, 7, 13\}, \{3, 7, 8, 13\}, \{3, 7, 13, 14\}, \{3, 7, 13, 15\}\\
\{3, 7, 13, 14\} $\rightarrow$ \{1, 7, 13, 14\}, \{7, 13, 14\}, \{3, 6, 13, 14\}, \{3, 13, 14\}, \{3, 7, 11, 14\}, \{3, 7, 14\}, \{3, 7, 12, 13\}, \{3, 7, 13\}, \{3, 5, 7, 13, 14\}, \{3, 7, 8, 13, 14\}, \{3, 7, 13, 14, 15\}\\
\{3, 7, 13, 14, 15\} $\rightarrow$ \{1, 7, 13, 14, 15\}, \{7, 13, 14, 15\}, \{3, 6, 13, 14, 15\}, \{3, 13, 14, 15\}, \{3, 7, 14, 15\}, \{3, 7, 13, 15\}, \{3, 7, 13, 14\}, \{3, 5, 7, 13, 14, 15\}, \{3, 7, 8, 13, 14, 15\}\\
\{3, 7, 13, 15\} $\rightarrow$ \{1, 7, 13, 15\}, \{7, 13, 15\}, \{3, 6, 13, 15\}, \{3, 13, 15\}, \{3, 7, 15\}, \{3, 7, 12, 13\}, \{3, 7, 13\}, \{3, 5, 7, 13, 15\}, \{3, 7, 8, 13, 15\}, \{3, 7, 13, 14, 15\}\\
\{3, 7, 14\} $\rightarrow$ \{1, 7, 14\}, \{7, 14\}, \{3, 6, 14\}, \{3, 14\}, \{3, 7, 12\}, \{3, 7\}, \{3, 5, 7, 14\}, \{3, 7, 8, 14\}, \{3, 7, 13, 14\}, \{3, 7, 14, 15\}\\
\{3, 7, 14, 15\} $\rightarrow$ \{1, 7, 14, 15\}, \{7, 14, 15\}, \{3, 6, 14, 15\}, \{3, 14, 15\}, \{3, 7, 15\}, \{3, 7, 11, 14\}, \{3, 7, 14\}, \{3, 5, 7, 14, 15\}, \{3, 7, 8, 14, 15\}, \{3, 7, 13, 14, 15\}\\
\{3, 7, 15\} $\rightarrow$ \{1, 7, 15\}, \{7, 15\}, \{3, 6, 15\}, \{3, 15\}, \{3, 7, 11\}, \{3, 7, 12\}, \{3, 7\}, \{3, 5, 7, 15\}, \{3, 7, 8, 15\}, \{3, 7, 13, 15\}, \{3, 7, 14, 15\}\\
\{3, 8\} $\rightarrow$ \{8\}, \{3, 6\}, \{3\}, \{3, 4, 8\}, \{3, 7, 8\}, \{3, 8, 13\}, \{3, 8, 14\}, \{3, 8, 15\}\\
\{3, 8, 10\} $\rightarrow$ \{8, 10\}, \{3, 10\}, \{3, 8, 11\}, \{3, 8, 12\}, \{3, 4, 8, 10\}, \{3, 7, 8, 10\}\\
\{3, 8, 11\} $\rightarrow$ \{8, 11\}, \{3, 11\}, \{3, 8, 10\}, \{3, 8, 13\}, \{3, 8, 15\}, \{3, 4, 8, 11\}, \{3, 7, 8, 11\}, \{3, 8, 11, 14\}\\
\{3, 8, 11, 14\} $\rightarrow$ \{8, 11, 14\}, \{3, 11, 14\}, \{3, 8, 13, 14\}, \{3, 8, 14, 15\}, \{3, 8, 11\}, \{3, 4, 8, 11, 14\}, \{3, 7, 8, 11, 14\}\\
\{3, 8, 12\} $\rightarrow$ \{8, 12\}, \{3, 6, 12\}, \{3, 12\}, \{3, 8, 10\}, \{3, 8, 14\}, \{3, 8, 15\}, \{3, 4, 8, 12\}, \{3, 7, 8, 12\}, \{3, 8, 12, 13\}\\
\{3, 8, 12, 13\} $\rightarrow$ \{8, 12, 13\}, \{3, 6, 12, 13\}, \{3, 12, 13\}, \{3, 8, 13, 14\}, \{3, 8, 13, 15\}, \{3, 8, 12\}, \{3, 4, 8, 12, 13\}, \{3, 7, 8, 12, 13\}\\
\{3, 8, 13\} $\rightarrow$ \{8, 13\}, \{3, 6, 13\}, \{3, 13\}, \{3, 8, 11\}, \{3, 8\}, \{3, 4, 8, 13\}, \{3, 7, 8, 13\}, \{3, 8, 13, 14\}, \{3, 8, 13, 15\}\\
\{3, 8, 13, 14\} $\rightarrow$ \{8, 13, 14\}, \{3, 6, 13, 14\}, \{3, 13, 14\}, \{3, 8, 11, 14\}, \{3, 8, 14\}, \{3, 8, 12, 13\}, \{3, 8, 13\}, \{3, 4, 8, 13, 14\}, \{3, 7, 8, 13, 14\}, \{3, 8, 13, 14, 15\}\\
\{3, 8, 13, 14, 15\} $\rightarrow$ \{8, 13, 14, 15\}, \{3, 6, 13, 14, 15\}, \{3, 13, 14, 15\}, \{3, 8, 14, 15\}, \{3, 8, 13, 15\}, \{3, 8, 13, 14\}, \{3, 4, 8, 13, 14, 15\}, \{3, 7, 8, 13, 14, 15\}\\
\{3, 8, 13, 15\} $\rightarrow$ \{8, 13, 15\}, \{3, 6, 13, 15\}, \{3, 13, 15\}, \{3, 8, 15\}, \{3, 8, 12, 13\}, \{3, 8, 13\}, \{3, 4, 8, 13, 15\}, \{3, 7, 8, 13, 15\}, \{3, 8, 13, 14, 15\}\\
\{3, 8, 14\} $\rightarrow$ \{8, 14\}, \{3, 6, 14\}, \{3, 14\}, \{3, 8, 12\}, \{3, 8\}, \{3, 4, 8, 14\}, \{3, 7, 8, 14\}, \{3, 8, 13, 14\}, \{3, 8, 14, 15\}\\
\{3, 8, 14, 15\} $\rightarrow$ \{8, 14, 15\}, \{3, 6, 14, 15\}, \{3, 14, 15\}, \{3, 8, 15\}, \{3, 8, 11, 14\}, \{3, 8, 14\}, \{3, 4, 8, 14, 15\}, \{3, 7, 8, 14, 15\}, \{3, 8, 13, 14, 15\}\\
\{3, 8, 15\} $\rightarrow$ \{8, 15\}, \{3, 6, 15\}, \{3, 15\}, \{3, 8, 11\}, \{3, 8, 12\}, \{3, 8\}, \{3, 4, 8, 15\}, \{3, 7, 8, 15\}, \{3, 8, 13, 15\}, \{3, 8, 14, 15\}\\
\{3, 10\} $\rightarrow$ \{10\}, \{3, 11\}, \{3, 12\}, \{3, 4, 10\}, \{3, 5, 10\}, \{3, 7, 10\}, \{3, 8, 10\}\\
\{3, 11\} $\rightarrow$ \{1, 11\}, \{11\}, \{3, 10\}, \{3, 13\}, \{3, 15\}, \{3, 4, 11\}, \{3, 5, 11\}, \{3, 7, 11\}, \{3, 8, 11\}, \{3, 11, 14\}\\
\{3, 11, 14\} $\rightarrow$ \{1, 11, 14\}, \{11, 14\}, \{3, 13, 14\}, \{3, 14, 15\}, \{3, 11\}, \{3, 4, 11, 14\}, \{3, 5, 11, 14\}, \{3, 7, 11, 14\}, \{3, 8, 11, 14\}\\
\{3, 12\} $\rightarrow$ \{12\}, \{3, 10\}, \{3, 14\}, \{3, 15\}, \{3, 4, 12\}, \{3, 5, 12\}, \{3, 7, 12\}, \{3, 8, 12\}, \{3, 12, 13\}\\
\{3, 12, 13\} $\rightarrow$ \{12, 13\}, \{3, 13, 14\}, \{3, 13, 15\}, \{3, 12\}, \{3, 4, 12, 13\}, \{3, 5, 12, 13\}, \{3, 7, 12, 13\}, \{3, 8, 12, 13\}\\
\{3, 13\} $\rightarrow$ \{1, 13\}, \{13\}, \{3, 11\}, \{3\}, \{3, 4, 13\}, \{3, 5, 13\}, \{3, 7, 13\}, \{3, 8, 13\}, \{3, 13, 14\}, \{3, 13, 15\}\\
\{3, 13, 14\} $\rightarrow$ \{1, 13, 14\}, \{13, 14\}, \{3, 11, 14\}, \{3, 14\}, \{3, 12, 13\}, \{3, 13\}, \{3, 4, 13, 14\}, \{3, 5, 13, 14\}, \{3, 7, 13, 14\}, \{3, 8, 13, 14\}, \{3, 13, 14, 15\}\\
\{3, 13, 14, 15\} $\rightarrow$ \{1, 13, 14, 15\}, \{13, 14, 15\}, \{3, 14, 15\}, \{3, 13, 15\}, \{3, 13, 14\}, \{3, 4, 13, 14, 15\}, \{3, 5, 13, 14, 15\}, \{3, 7, 13, 14, 15\}, \{3, 8, 13, 14, 15\}\\
\{3, 13, 15\} $\rightarrow$ \{1, 13, 15\}, \{13, 15\}, \{3, 15\}, \{3, 12, 13\}, \{3, 13\}, \{3, 4, 13, 15\}, \{3, 5, 13, 15\}, \{3, 7, 13, 15\}, \{3, 8, 13, 15\}, \{3, 13, 14, 15\}\\
\{3, 14\} $\rightarrow$ \{1, 14\}, \{14\}, \{3, 12\}, \{3\}, \{3, 4, 14\}, \{3, 5, 14\}, \{3, 7, 14\}, \{3, 8, 14\}, \{3, 13, 14\}, \{3, 14, 15\}\\
\{3, 14, 15\} $\rightarrow$ \{1, 14, 15\}, \{14, 15\}, \{3, 15\}, \{3, 11, 14\}, \{3, 14\}, \{3, 4, 14, 15\}, \{3, 5, 14, 15\}, \{3, 7, 14, 15\}, \{3, 8, 14, 15\}, \{3, 13, 14, 15\}\\
\{3, 15\} $\rightarrow$ \{1, 15\}, \{15\}, \{3, 11\}, \{3, 12\}, \{3\}, \{3, 4, 15\}, \{3, 5, 15\}, \{3, 7, 15\}, \{3, 8, 15\}, \{3, 13, 15\}, \{3, 14, 15\}\\
\{4, 5\} $\rightarrow$ \{5\}, \{1, 4\}, \{4\}, \{3, 4, 5\}, \{4, 5, 9\}, \{4, 5, 13\}, \{4, 5, 14\}, \{4, 5, 15\}\\
\{4, 5, 9\} $\rightarrow$ \{5, 9\}, \{4, 9\}, \{4, 5\}, \{4, 5, 9, 13\}, \{4, 5, 9, 14\}, \{4, 5, 9, 15\}\\
\{4, 5, 9, 10\} $\rightarrow$ \{5, 9, 10\}, \{4, 9, 10\}, \{4, 5, 10\}, \{4, 5, 9, 11\}, \{4, 5, 9, 12\}\\
\{4, 5, 9, 11\} $\rightarrow$ \{5, 9, 11\}, \{4, 9, 11\}, \{4, 5, 11\}, \{4, 5, 9, 10\}, \{4, 5, 9, 13\}, \{4, 5, 9, 15\}, \{4, 5, 9, 11, 14\}\\
\{4, 5, 9, 11, 14\} $\rightarrow$ \{5, 9, 11, 14\}, \{4, 9, 11, 14\}, \{4, 5, 11, 14\}, \{4, 5, 9, 13, 14\}, \{4, 5, 9, 14, 15\}, \{4, 5, 9, 11\}\\
\{4, 5, 9, 12\} $\rightarrow$ \{5, 9, 12\}, \{4, 9, 12\}, \{4, 5, 12\}, \{4, 5, 9, 10\}, \{4, 5, 9, 14\}, \{4, 5, 9, 15\}, \{4, 5, 9, 12, 13\}\\
\{4, 5, 9, 12, 13\} $\rightarrow$ \{5, 9, 12, 13\}, \{4, 9, 12, 13\}, \{4, 5, 12, 13\}, \{4, 5, 9, 13, 14\}, \{4, 5, 9, 13, 15\}, \{4, 5, 9, 12\}\\
\{4, 5, 9, 13\} $\rightarrow$ \{5, 9, 13\}, \{4, 9, 13\}, \{4, 5, 13\}, \{4, 5, 9, 11\}, \{4, 5, 9\}, \{4, 5, 9, 13, 14\}, \{4, 5, 9, 13, 15\}\\
\{4, 5, 9, 13, 14\} $\rightarrow$ \{5, 9, 13, 14\}, \{4, 9, 13, 14\}, \{4, 5, 13, 14\}, \{4, 5, 9, 11, 14\}, \{4, 5, 9, 14\}, \{4, 5, 9, 12, 13\}, \{4, 5, 9, 13\}, \{4, 5, 9, 13, 14, 15\}\\
\{4, 5, 9, 13, 14, 15\} $\rightarrow$ \{5, 9, 13, 14, 15\}, \{4, 9, 13, 14, 15\}, \{4, 5, 13, 14, 15\}, \{4, 5, 9, 14, 15\}, \{4, 5, 9, 13, 15\}, \{4, 5, 9, 13, 14\}\\
\{4, 5, 9, 13, 15\} $\rightarrow$ \{5, 9, 13, 15\}, \{4, 9, 13, 15\}, \{4, 5, 13, 15\}, \{4, 5, 9, 15\}, \{4, 5, 9, 12, 13\}, \{4, 5, 9, 13\}, \{4, 5, 9, 13, 14, 15\}\\
\{4, 5, 9, 14\} $\rightarrow$ \{5, 9, 14\}, \{4, 9, 14\}, \{4, 5, 14\}, \{4, 5, 9, 12\}, \{4, 5, 9\}, \{4, 5, 9, 13, 14\}, \{4, 5, 9, 14, 15\}\\
\{4, 5, 9, 14, 15\} $\rightarrow$ \{5, 9, 14, 15\}, \{4, 9, 14, 15\}, \{4, 5, 14, 15\}, \{4, 5, 9, 15\}, \{4, 5, 9, 11, 14\}, \{4, 5, 9, 14\}, \{4, 5, 9, 13, 14, 15\}\\
\{4, 5, 9, 15\} $\rightarrow$ \{5, 9, 15\}, \{4, 9, 15\}, \{4, 5, 15\}, \{4, 5, 9, 11\}, \{4, 5, 9, 12\}, \{4, 5, 9\}, \{4, 5, 9, 13, 15\}, \{4, 5, 9, 14, 15\}\\
\{4, 5, 10\} $\rightarrow$ \{5, 10\}, \{4, 10\}, \{4, 5, 11\}, \{4, 5, 12\}, \{3, 4, 5, 10\}, \{4, 5, 9, 10\}\\
\{4, 5, 11\} $\rightarrow$ \{5, 11\}, \{1, 4, 11\}, \{4, 11\}, \{4, 5, 10\}, \{4, 5, 13\}, \{4, 5, 15\}, \{3, 4, 5, 11\}, \{4, 5, 9, 11\}, \{4, 5, 11, 14\}\\
\{4, 5, 11, 14\} $\rightarrow$ \{5, 11, 14\}, \{1, 4, 11, 14\}, \{4, 11, 14\}, \{4, 5, 13, 14\}, \{4, 5, 14, 15\}, \{4, 5, 11\}, \{3, 4, 5, 11, 14\}, \{4, 5, 9, 11, 14\}\\
\{4, 5, 12\} $\rightarrow$ \{5, 12\}, \{4, 12\}, \{4, 5, 10\}, \{4, 5, 14\}, \{4, 5, 15\}, \{3, 4, 5, 12\}, \{4, 5, 9, 12\}, \{4, 5, 12, 13\}\\
\{4, 5, 12, 13\} $\rightarrow$ \{5, 12, 13\}, \{4, 12, 13\}, \{4, 5, 13, 14\}, \{4, 5, 13, 15\}, \{4, 5, 12\}, \{3, 4, 5, 12, 13\}, \{4, 5, 9, 12, 13\}\\
\{4, 5, 13\} $\rightarrow$ \{5, 13\}, \{1, 4, 13\}, \{4, 13\}, \{4, 5, 11\}, \{4, 5\}, \{3, 4, 5, 13\}, \{4, 5, 9, 13\}, \{4, 5, 13, 14\}, \{4, 5, 13, 15\}\\
\{4, 5, 13, 14\} $\rightarrow$ \{5, 13, 14\}, \{1, 4, 13, 14\}, \{4, 13, 14\}, \{4, 5, 11, 14\}, \{4, 5, 14\}, \{4, 5, 12, 13\}, \{4, 5, 13\}, \{3, 4, 5, 13, 14\}, \{4, 5, 9, 13, 14\}, \{4, 5, 13, 14, 15\}\\
\{4, 5, 13, 14, 15\} $\rightarrow$ \{5, 13, 14, 15\}, \{1, 4, 13, 14, 15\}, \{4, 13, 14, 15\}, \{4, 5, 14, 15\}, \{4, 5, 13, 15\}, \{4, 5, 13, 14\}, \{3, 4, 5, 13, 14, 15\}, \{4, 5, 9, 13, 14, 15\}\\
\{4, 5, 13, 15\} $\rightarrow$ \{5, 13, 15\}, \{1, 4, 13, 15\}, \{4, 13, 15\}, \{4, 5, 15\}, \{4, 5, 12, 13\}, \{4, 5, 13\}, \{3, 4, 5, 13, 15\}, \{4, 5, 9, 13, 15\}, \{4, 5, 13, 14, 15\}\\
\{4, 5, 14\} $\rightarrow$ \{5, 14\}, \{1, 4, 14\}, \{4, 14\}, \{4, 5, 12\}, \{4, 5\}, \{3, 4, 5, 14\}, \{4, 5, 9, 14\}, \{4, 5, 13, 14\}, \{4, 5, 14, 15\}\\
\{4, 5, 14, 15\} $\rightarrow$ \{5, 14, 15\}, \{1, 4, 14, 15\}, \{4, 14, 15\}, \{4, 5, 15\}, \{4, 5, 11, 14\}, \{4, 5, 14\}, \{3, 4, 5, 14, 15\}, \{4, 5, 9, 14, 15\}, \{4, 5, 13, 14, 15\}\\
\{4, 5, 15\} $\rightarrow$ \{5, 15\}, \{1, 4, 15\}, \{4, 15\}, \{4, 5, 11\}, \{4, 5, 12\}, \{4, 5\}, \{3, 4, 5, 15\}, \{4, 5, 9, 15\}, \{4, 5, 13, 15\}, \{4, 5, 14, 15\}\\
\{4, 8\} $\rightarrow$ \{8\}, \{4\}, \{3, 4, 8\}, \{4, 8, 9\}, \{4, 8, 13\}, \{4, 8, 14\}, \{4, 8, 15\}\\
\{4, 8, 9\} $\rightarrow$ \{8, 9\}, \{4, 9\}, \{4, 8\}, \{4, 8, 9, 13\}, \{4, 8, 9, 14\}, \{4, 8, 9, 15\}\\
\{4, 8, 9, 10\} $\rightarrow$ \{8, 9, 10\}, \{4, 9, 10\}, \{4, 8, 10\}, \{4, 8, 9, 11\}, \{4, 8, 9, 12\}\\
\{4, 8, 9, 11\} $\rightarrow$ \{8, 9, 11\}, \{4, 9, 11\}, \{4, 8, 11\}, \{4, 8, 9, 10\}, \{4, 8, 9, 13\}, \{4, 8, 9, 15\}, \{4, 8, 9, 11, 14\}\\
\{4, 8, 9, 11, 14\} $\rightarrow$ \{8, 9, 11, 14\}, \{4, 9, 11, 14\}, \{4, 8, 11, 14\}, \{4, 8, 9, 13, 14\}, \{4, 8, 9, 14, 15\}, \{4, 8, 9, 11\}\\
\{4, 8, 9, 12\} $\rightarrow$ \{8, 9, 12\}, \{4, 9, 12\}, \{4, 8, 12\}, \{4, 8, 9, 10\}, \{4, 8, 9, 14\}, \{4, 8, 9, 15\}, \{4, 8, 9, 12, 13\}\\
\{4, 8, 9, 12, 13\} $\rightarrow$ \{8, 9, 12, 13\}, \{4, 9, 12, 13\}, \{4, 8, 12, 13\}, \{4, 8, 9, 13, 14\}, \{4, 8, 9, 13, 15\}, \{4, 8, 9, 12\}\\
\{4, 8, 9, 13\} $\rightarrow$ \{8, 9, 13\}, \{4, 9, 13\}, \{4, 8, 13\}, \{4, 8, 9, 11\}, \{4, 8, 9\}, \{4, 8, 9, 13, 14\}, \{4, 8, 9, 13, 15\}\\
\{4, 8, 9, 13, 14\} $\rightarrow$ \{8, 9, 13, 14\}, \{4, 9, 13, 14\}, \{4, 8, 13, 14\}, \{4, 8, 9, 11, 14\}, \{4, 8, 9, 14\}, \{4, 8, 9, 12, 13\}, \{4, 8, 9, 13\}, \{4, 8, 9, 13, 14, 15\}\\
\{4, 8, 9, 13, 14, 15\} $\rightarrow$ \{8, 9, 13, 14, 15\}, \{4, 9, 13, 14, 15\}, \{4, 8, 13, 14, 15\}, \{4, 8, 9, 14, 15\}, \{4, 8, 9, 13, 15\}, \{4, 8, 9, 13, 14\}\\
\{4, 8, 9, 13, 15\} $\rightarrow$ \{8, 9, 13, 15\}, \{4, 9, 13, 15\}, \{4, 8, 13, 15\}, \{4, 8, 9, 15\}, \{4, 8, 9, 12, 13\}, \{4, 8, 9, 13\}, \{4, 8, 9, 13, 14, 15\}\\
\{4, 8, 9, 14\} $\rightarrow$ \{8, 9, 14\}, \{4, 9, 14\}, \{4, 8, 14\}, \{4, 8, 9, 12\}, \{4, 8, 9\}, \{4, 8, 9, 13, 14\}, \{4, 8, 9, 14, 15\}\\
\{4, 8, 9, 14, 15\} $\rightarrow$ \{8, 9, 14, 15\}, \{4, 9, 14, 15\}, \{4, 8, 14, 15\}, \{4, 8, 9, 15\}, \{4, 8, 9, 11, 14\}, \{4, 8, 9, 14\}, \{4, 8, 9, 13, 14, 15\}\\
\{4, 8, 9, 15\} $\rightarrow$ \{8, 9, 15\}, \{4, 9, 15\}, \{4, 8, 15\}, \{4, 8, 9, 11\}, \{4, 8, 9, 12\}, \{4, 8, 9\}, \{4, 8, 9, 13, 15\}, \{4, 8, 9, 14, 15\}\\
\{4, 8, 10\} $\rightarrow$ \{8, 10\}, \{4, 10\}, \{4, 8, 11\}, \{4, 8, 12\}, \{3, 4, 8, 10\}, \{4, 8, 9, 10\}\\
\{4, 8, 11\} $\rightarrow$ \{8, 11\}, \{4, 11\}, \{4, 8, 10\}, \{4, 8, 13\}, \{4, 8, 15\}, \{3, 4, 8, 11\}, \{4, 8, 9, 11\}, \{4, 8, 11, 14\}\\
\{4, 8, 11, 14\} $\rightarrow$ \{8, 11, 14\}, \{4, 11, 14\}, \{4, 8, 13, 14\}, \{4, 8, 14, 15\}, \{4, 8, 11\}, \{3, 4, 8, 11, 14\}, \{4, 8, 9, 11, 14\}\\
\{4, 8, 12\} $\rightarrow$ \{8, 12\}, \{4, 12\}, \{4, 8, 10\}, \{4, 8, 14\}, \{4, 8, 15\}, \{3, 4, 8, 12\}, \{4, 8, 9, 12\}, \{4, 8, 12, 13\}\\
\{4, 8, 12, 13\} $\rightarrow$ \{8, 12, 13\}, \{4, 12, 13\}, \{4, 8, 13, 14\}, \{4, 8, 13, 15\}, \{4, 8, 12\}, \{3, 4, 8, 12, 13\}, \{4, 8, 9, 12, 13\}\\
\{4, 8, 13\} $\rightarrow$ \{8, 13\}, \{4, 13\}, \{4, 8, 11\}, \{4, 8\}, \{3, 4, 8, 13\}, \{4, 8, 9, 13\}, \{4, 8, 13, 14\}, \{4, 8, 13, 15\}\\
\{4, 8, 13, 14\} $\rightarrow$ \{8, 13, 14\}, \{4, 13, 14\}, \{4, 8, 11, 14\}, \{4, 8, 14\}, \{4, 8, 12, 13\}, \{4, 8, 13\}, \{3, 4, 8, 13, 14\}, \{4, 8, 9, 13, 14\}, \{4, 8, 13, 14, 15\}\\
\{4, 8, 13, 14, 15\} $\rightarrow$ \{8, 13, 14, 15\}, \{4, 13, 14, 15\}, \{4, 8, 14, 15\}, \{4, 8, 13, 15\}, \{4, 8, 13, 14\}, \{3, 4, 8, 13, 14, 15\}, \{4, 8, 9, 13, 14, 15\}\\
\{4, 8, 13, 15\} $\rightarrow$ \{8, 13, 15\}, \{4, 13, 15\}, \{4, 8, 15\}, \{4, 8, 12, 13\}, \{4, 8, 13\}, \{3, 4, 8, 13, 15\}, \{4, 8, 9, 13, 15\}, \{4, 8, 13, 14, 15\}\\
\{4, 8, 14\} $\rightarrow$ \{8, 14\}, \{4, 14\}, \{4, 8, 12\}, \{4, 8\}, \{3, 4, 8, 14\}, \{4, 8, 9, 14\}, \{4, 8, 13, 14\}, \{4, 8, 14, 15\}\\
\{4, 8, 14, 15\} $\rightarrow$ \{8, 14, 15\}, \{4, 14, 15\}, \{4, 8, 15\}, \{4, 8, 11, 14\}, \{4, 8, 14\}, \{3, 4, 8, 14, 15\}, \{4, 8, 9, 14, 15\}, \{4, 8, 13, 14, 15\}\\
\{4, 8, 15\} $\rightarrow$ \{8, 15\}, \{4, 15\}, \{4, 8, 11\}, \{4, 8, 12\}, \{4, 8\}, \{3, 4, 8, 15\}, \{4, 8, 9, 15\}, \{4, 8, 13, 15\}, \{4, 8, 14, 15\}\\
\{4, 9\} $\rightarrow$ \{2, 9\}, \{9\}, \{4\}, \{4, 5, 9\}, \{4, 8, 9\}, \{4, 9, 13\}, \{4, 9, 14\}, \{4, 9, 15\}\\
\{4, 9, 10\} $\rightarrow$ \{9, 10\}, \{4, 10\}, \{4, 9, 11\}, \{4, 9, 12\}, \{4, 5, 9, 10\}, \{4, 8, 9, 10\}\\
\{4, 9, 11\} $\rightarrow$ \{9, 11\}, \{4, 11\}, \{4, 9, 10\}, \{4, 9, 13\}, \{4, 9, 15\}, \{4, 5, 9, 11\}, \{4, 8, 9, 11\}, \{4, 9, 11, 14\}\\
\{4, 9, 11, 14\} $\rightarrow$ \{9, 11, 14\}, \{4, 11, 14\}, \{4, 9, 13, 14\}, \{4, 9, 14, 15\}, \{4, 9, 11\}, \{4, 5, 9, 11, 14\}, \{4, 8, 9, 11, 14\}\\
\{4, 9, 12\} $\rightarrow$ \{2, 9, 12\}, \{9, 12\}, \{4, 12\}, \{4, 9, 10\}, \{4, 9, 14\}, \{4, 9, 15\}, \{4, 5, 9, 12\}, \{4, 8, 9, 12\}, \{4, 9, 12, 13\}\\
\{4, 9, 12, 13\} $\rightarrow$ \{2, 9, 12, 13\}, \{9, 12, 13\}, \{4, 12, 13\}, \{4, 9, 13, 14\}, \{4, 9, 13, 15\}, \{4, 9, 12\}, \{4, 5, 9, 12, 13\}, \{4, 8, 9, 12, 13\}\\
\{4, 9, 13\} $\rightarrow$ \{2, 9, 13\}, \{9, 13\}, \{4, 13\}, \{4, 9, 11\}, \{4, 9\}, \{4, 5, 9, 13\}, \{4, 8, 9, 13\}, \{4, 9, 13, 14\}, \{4, 9, 13, 15\}\\
\{4, 9, 13, 14\} $\rightarrow$ \{2, 9, 13, 14\}, \{9, 13, 14\}, \{4, 13, 14\}, \{4, 9, 11, 14\}, \{4, 9, 14\}, \{4, 9, 12, 13\}, \{4, 9, 13\}, \{4, 5, 9, 13, 14\}, \{4, 8, 9, 13, 14\}, \{4, 9, 13, 14, 15\}\\
\{4, 9, 13, 14, 15\} $\rightarrow$ \{2, 9, 13, 14, 15\}, \{9, 13, 14, 15\}, \{4, 13, 14, 15\}, \{4, 9, 14, 15\}, \{4, 9, 13, 15\}, \{4, 9, 13, 14\}, \{4, 5, 9, 13, 14, 15\}, \{4, 8, 9, 13, 14, 15\}\\
\{4, 9, 13, 15\} $\rightarrow$ \{2, 9, 13, 15\}, \{9, 13, 15\}, \{4, 13, 15\}, \{4, 9, 15\}, \{4, 9, 12, 13\}, \{4, 9, 13\}, \{4, 5, 9, 13, 15\}, \{4, 8, 9, 13, 15\}, \{4, 9, 13, 14, 15\}\\
\{4, 9, 14\} $\rightarrow$ \{2, 9, 14\}, \{9, 14\}, \{4, 14\}, \{4, 9, 12\}, \{4, 9\}, \{4, 5, 9, 14\}, \{4, 8, 9, 14\}, \{4, 9, 13, 14\}, \{4, 9, 14, 15\}\\
\{4, 9, 14, 15\} $\rightarrow$ \{2, 9, 14, 15\}, \{9, 14, 15\}, \{4, 14, 15\}, \{4, 9, 15\}, \{4, 9, 11, 14\}, \{4, 9, 14\}, \{4, 5, 9, 14, 15\}, \{4, 8, 9, 14, 15\}, \{4, 9, 13, 14, 15\}\\
\{4, 9, 15\} $\rightarrow$ \{2, 9, 15\}, \{9, 15\}, \{4, 15\}, \{4, 9, 11\}, \{4, 9, 12\}, \{4, 9\}, \{4, 5, 9, 15\}, \{4, 8, 9, 15\}, \{4, 9, 13, 15\}, \{4, 9, 14, 15\}\\
\{4, 10\} $\rightarrow$ \{10\}, \{4, 11\}, \{4, 12\}, \{3, 4, 10\}, \{4, 5, 10\}, \{4, 8, 10\}, \{4, 9, 10\}\\
\{4, 11\} $\rightarrow$ \{11\}, \{4, 10\}, \{4, 13\}, \{4, 15\}, \{3, 4, 11\}, \{4, 5, 11\}, \{4, 8, 11\}, \{4, 9, 11\}, \{4, 11, 14\}\\
\{4, 11, 14\} $\rightarrow$ \{11, 14\}, \{4, 13, 14\}, \{4, 14, 15\}, \{4, 11\}, \{3, 4, 11, 14\}, \{4, 5, 11, 14\}, \{4, 8, 11, 14\}, \{4, 9, 11, 14\}\\
\{4, 12\} $\rightarrow$ \{2, 12\}, \{12\}, \{4, 10\}, \{4, 14\}, \{4, 15\}, \{3, 4, 12\}, \{4, 5, 12\}, \{4, 8, 12\}, \{4, 9, 12\}, \{4, 12, 13\}\\
\{4, 12, 13\} $\rightarrow$ \{2, 12, 13\}, \{12, 13\}, \{4, 13, 14\}, \{4, 13, 15\}, \{4, 12\}, \{3, 4, 12, 13\}, \{4, 5, 12, 13\}, \{4, 8, 12, 13\}, \{4, 9, 12, 13\}\\
\{4, 13\} $\rightarrow$ \{2, 13\}, \{13\}, \{4, 11\}, \{4\}, \{3, 4, 13\}, \{4, 5, 13\}, \{4, 8, 13\}, \{4, 9, 13\}, \{4, 13, 14\}, \{4, 13, 15\}\\
\{4, 13, 14\} $\rightarrow$ \{2, 13, 14\}, \{13, 14\}, \{4, 11, 14\}, \{4, 14\}, \{4, 12, 13\}, \{4, 13\}, \{3, 4, 13, 14\}, \{4, 5, 13, 14\}, \{4, 8, 13, 14\}, \{4, 9, 13, 14\}, \{4, 13, 14, 15\}\\
\{4, 13, 14, 15\} $\rightarrow$ \{2, 13, 14, 15\}, \{13, 14, 15\}, \{4, 14, 15\}, \{4, 13, 15\}, \{4, 13, 14\}, \{3, 4, 13, 14, 15\}, \{4, 5, 13, 14, 15\}, \{4, 8, 13, 14, 15\}, \{4, 9, 13, 14, 15\}\\
\{4, 13, 15\} $\rightarrow$ \{2, 13, 15\}, \{13, 15\}, \{4, 15\}, \{4, 12, 13\}, \{4, 13\}, \{3, 4, 13, 15\}, \{4, 5, 13, 15\}, \{4, 8, 13, 15\}, \{4, 9, 13, 15\}, \{4, 13, 14, 15\}\\
\{4, 14\} $\rightarrow$ \{2, 14\}, \{14\}, \{4, 12\}, \{4\}, \{3, 4, 14\}, \{4, 5, 14\}, \{4, 8, 14\}, \{4, 9, 14\}, \{4, 13, 14\}, \{4, 14, 15\}\\
\{4, 14, 15\} $\rightarrow$ \{2, 14, 15\}, \{14, 15\}, \{4, 15\}, \{4, 11, 14\}, \{4, 14\}, \{3, 4, 14, 15\}, \{4, 5, 14, 15\}, \{4, 8, 14, 15\}, \{4, 9, 14, 15\}, \{4, 13, 14, 15\}\\
\{4, 15\} $\rightarrow$ \{2, 15\}, \{15\}, \{4, 11\}, \{4, 12\}, \{4\}, \{3, 4, 15\}, \{4, 5, 15\}, \{4, 8, 15\}, \{4, 9, 15\}, \{4, 13, 15\}, \{4, 14, 15\}\\
\{5, 7\} $\rightarrow$ \{1, 7\}, \{7\}, \{5\}, \{3, 5, 7\}, \{5, 7, 9\}, \{5, 7, 13\}, \{5, 7, 14\}, \{5, 7, 15\}\\
\{5, 7, 9\} $\rightarrow$ \{7, 9\}, \{5, 9\}, \{5, 7\}, \{5, 7, 9, 13\}, \{5, 7, 9, 14\}, \{5, 7, 9, 15\}\\
\{5, 7, 9, 10\} $\rightarrow$ \{7, 9, 10\}, \{5, 9, 10\}, \{5, 7, 10\}, \{5, 7, 9, 11\}, \{5, 7, 9, 12\}\\
\{5, 7, 9, 11\} $\rightarrow$ \{7, 9, 11\}, \{5, 9, 11\}, \{5, 7, 11\}, \{5, 7, 9, 10\}, \{5, 7, 9, 13\}, \{5, 7, 9, 15\}, \{5, 7, 9, 11, 14\}\\
\{5, 7, 9, 11, 14\} $\rightarrow$ \{7, 9, 11, 14\}, \{5, 9, 11, 14\}, \{5, 7, 11, 14\}, \{5, 7, 9, 13, 14\}, \{5, 7, 9, 14, 15\}, \{5, 7, 9, 11\}\\
\{5, 7, 9, 12\} $\rightarrow$ \{7, 9, 12\}, \{5, 9, 12\}, \{5, 7, 12\}, \{5, 7, 9, 10\}, \{5, 7, 9, 14\}, \{5, 7, 9, 15\}, \{5, 7, 9, 12, 13\}\\
\{5, 7, 9, 12, 13\} $\rightarrow$ \{7, 9, 12, 13\}, \{5, 9, 12, 13\}, \{5, 7, 12, 13\}, \{5, 7, 9, 13, 14\}, \{5, 7, 9, 13, 15\}, \{5, 7, 9, 12\}\\
\{5, 7, 9, 13\} $\rightarrow$ \{7, 9, 13\}, \{5, 9, 13\}, \{5, 7, 13\}, \{5, 7, 9, 11\}, \{5, 7, 9\}, \{5, 7, 9, 13, 14\}, \{5, 7, 9, 13, 15\}\\
\{5, 7, 9, 13, 14\} $\rightarrow$ \{7, 9, 13, 14\}, \{5, 9, 13, 14\}, \{5, 7, 13, 14\}, \{5, 7, 9, 11, 14\}, \{5, 7, 9, 14\}, \{5, 7, 9, 12, 13\}, \{5, 7, 9, 13\}, \{5, 7, 9, 13, 14, 15\}\\
\{5, 7, 9, 13, 14, 15\} $\rightarrow$ \{7, 9, 13, 14, 15\}, \{5, 9, 13, 14, 15\}, \{5, 7, 13, 14, 15\}, \{5, 7, 9, 14, 15\}, \{5, 7, 9, 13, 15\}, \{5, 7, 9, 13, 14\}\\
\{5, 7, 9, 13, 15\} $\rightarrow$ \{7, 9, 13, 15\}, \{5, 9, 13, 15\}, \{5, 7, 13, 15\}, \{5, 7, 9, 15\}, \{5, 7, 9, 12, 13\}, \{5, 7, 9, 13\}, \{5, 7, 9, 13, 14, 15\}\\
\{5, 7, 9, 14\} $\rightarrow$ \{7, 9, 14\}, \{5, 9, 14\}, \{5, 7, 14\}, \{5, 7, 9, 12\}, \{5, 7, 9\}, \{5, 7, 9, 13, 14\}, \{5, 7, 9, 14, 15\}\\
\{5, 7, 9, 14, 15\} $\rightarrow$ \{7, 9, 14, 15\}, \{5, 9, 14, 15\}, \{5, 7, 14, 15\}, \{5, 7, 9, 15\}, \{5, 7, 9, 11, 14\}, \{5, 7, 9, 14\}, \{5, 7, 9, 13, 14, 15\}\\
\{5, 7, 9, 15\} $\rightarrow$ \{7, 9, 15\}, \{5, 9, 15\}, \{5, 7, 15\}, \{5, 7, 9, 11\}, \{5, 7, 9, 12\}, \{5, 7, 9\}, \{5, 7, 9, 13, 15\}, \{5, 7, 9, 14, 15\}\\
\{5, 7, 10\} $\rightarrow$ \{7, 10\}, \{5, 10\}, \{5, 7, 11\}, \{5, 7, 12\}, \{3, 5, 7, 10\}, \{5, 7, 9, 10\}\\
\{5, 7, 11\} $\rightarrow$ \{1, 7, 11\}, \{7, 11\}, \{5, 11\}, \{5, 7, 10\}, \{5, 7, 13\}, \{5, 7, 15\}, \{3, 5, 7, 11\}, \{5, 7, 9, 11\}, \{5, 7, 11, 14\}\\
\{5, 7, 11, 14\} $\rightarrow$ \{1, 7, 11, 14\}, \{7, 11, 14\}, \{5, 11, 14\}, \{5, 7, 13, 14\}, \{5, 7, 14, 15\}, \{5, 7, 11\}, \{3, 5, 7, 11, 14\}, \{5, 7, 9, 11, 14\}\\
\{5, 7, 12\} $\rightarrow$ \{7, 12\}, \{5, 12\}, \{5, 7, 10\}, \{5, 7, 14\}, \{5, 7, 15\}, \{3, 5, 7, 12\}, \{5, 7, 9, 12\}, \{5, 7, 12, 13\}\\
\{5, 7, 12, 13\} $\rightarrow$ \{7, 12, 13\}, \{5, 12, 13\}, \{5, 7, 13, 14\}, \{5, 7, 13, 15\}, \{5, 7, 12\}, \{3, 5, 7, 12, 13\}, \{5, 7, 9, 12, 13\}\\
\{5, 7, 13\} $\rightarrow$ \{1, 7, 13\}, \{7, 13\}, \{5, 13\}, \{5, 7, 11\}, \{5, 7\}, \{3, 5, 7, 13\}, \{5, 7, 9, 13\}, \{5, 7, 13, 14\}, \{5, 7, 13, 15\}\\
\{5, 7, 13, 14\} $\rightarrow$ \{1, 7, 13, 14\}, \{7, 13, 14\}, \{5, 13, 14\}, \{5, 7, 11, 14\}, \{5, 7, 14\}, \{5, 7, 12, 13\}, \{5, 7, 13\}, \{3, 5, 7, 13, 14\}, \{5, 7, 9, 13, 14\}, \{5, 7, 13, 14, 15\}\\
\{5, 7, 13, 14, 15\} $\rightarrow$ \{1, 7, 13, 14, 15\}, \{7, 13, 14, 15\}, \{5, 13, 14, 15\}, \{5, 7, 14, 15\}, \{5, 7, 13, 15\}, \{5, 7, 13, 14\}, \{3, 5, 7, 13, 14, 15\}, \{5, 7, 9, 13, 14, 15\}\\
\{5, 7, 13, 15\} $\rightarrow$ \{1, 7, 13, 15\}, \{7, 13, 15\}, \{5, 13, 15\}, \{5, 7, 15\}, \{5, 7, 12, 13\}, \{5, 7, 13\}, \{3, 5, 7, 13, 15\}, \{5, 7, 9, 13, 15\}, \{5, 7, 13, 14, 15\}\\
\{5, 7, 14\} $\rightarrow$ \{1, 7, 14\}, \{7, 14\}, \{5, 14\}, \{5, 7, 12\}, \{5, 7\}, \{3, 5, 7, 14\}, \{5, 7, 9, 14\}, \{5, 7, 13, 14\}, \{5, 7, 14, 15\}\\
\{5, 7, 14, 15\} $\rightarrow$ \{1, 7, 14, 15\}, \{7, 14, 15\}, \{5, 14, 15\}, \{5, 7, 15\}, \{5, 7, 11, 14\}, \{5, 7, 14\}, \{3, 5, 7, 14, 15\}, \{5, 7, 9, 14, 15\}, \{5, 7, 13, 14, 15\}\\
\{5, 7, 15\} $\rightarrow$ \{1, 7, 15\}, \{7, 15\}, \{5, 15\}, \{5, 7, 11\}, \{5, 7, 12\}, \{5, 7\}, \{3, 5, 7, 15\}, \{5, 7, 9, 15\}, \{5, 7, 13, 15\}, \{5, 7, 14, 15\}\\
\{5, 9\} $\rightarrow$ \{2, 9\}, \{9\}, \{5\}, \{4, 5, 9\}, \{5, 7, 9\}, \{5, 9, 13\}, \{5, 9, 14\}, \{5, 9, 15\}\\
\{5, 9, 10\} $\rightarrow$ \{9, 10\}, \{5, 10\}, \{5, 9, 11\}, \{5, 9, 12\}, \{4, 5, 9, 10\}, \{5, 7, 9, 10\}\\
\{5, 9, 11\} $\rightarrow$ \{9, 11\}, \{5, 11\}, \{5, 9, 10\}, \{5, 9, 13\}, \{5, 9, 15\}, \{4, 5, 9, 11\}, \{5, 7, 9, 11\}, \{5, 9, 11, 14\}\\
\{5, 9, 11, 14\} $\rightarrow$ \{9, 11, 14\}, \{5, 11, 14\}, \{5, 9, 13, 14\}, \{5, 9, 14, 15\}, \{5, 9, 11\}, \{4, 5, 9, 11, 14\}, \{5, 7, 9, 11, 14\}\\
\{5, 9, 12\} $\rightarrow$ \{2, 9, 12\}, \{9, 12\}, \{5, 12\}, \{5, 9, 10\}, \{5, 9, 14\}, \{5, 9, 15\}, \{4, 5, 9, 12\}, \{5, 7, 9, 12\}, \{5, 9, 12, 13\}\\
\{5, 9, 12, 13\} $\rightarrow$ \{2, 9, 12, 13\}, \{9, 12, 13\}, \{5, 12, 13\}, \{5, 9, 13, 14\}, \{5, 9, 13, 15\}, \{5, 9, 12\}, \{4, 5, 9, 12, 13\}, \{5, 7, 9, 12, 13\}\\
\{5, 9, 13\} $\rightarrow$ \{2, 9, 13\}, \{9, 13\}, \{5, 13\}, \{5, 9, 11\}, \{5, 9\}, \{4, 5, 9, 13\}, \{5, 7, 9, 13\}, \{5, 9, 13, 14\}, \{5, 9, 13, 15\}\\
\{5, 9, 13, 14\} $\rightarrow$ \{2, 9, 13, 14\}, \{9, 13, 14\}, \{5, 13, 14\}, \{5, 9, 11, 14\}, \{5, 9, 14\}, \{5, 9, 12, 13\}, \{5, 9, 13\}, \{4, 5, 9, 13, 14\}, \{5, 7, 9, 13, 14\}, \{5, 9, 13, 14, 15\}\\
\{5, 9, 13, 14, 15\} $\rightarrow$ \{2, 9, 13, 14, 15\}, \{9, 13, 14, 15\}, \{5, 13, 14, 15\}, \{5, 9, 14, 15\}, \{5, 9, 13, 15\}, \{5, 9, 13, 14\}, \{4, 5, 9, 13, 14, 15\}, \{5, 7, 9, 13, 14, 15\}\\
\{5, 9, 13, 15\} $\rightarrow$ \{2, 9, 13, 15\}, \{9, 13, 15\}, \{5, 13, 15\}, \{5, 9, 15\}, \{5, 9, 12, 13\}, \{5, 9, 13\}, \{4, 5, 9, 13, 15\}, \{5, 7, 9, 13, 15\}, \{5, 9, 13, 14, 15\}\\
\{5, 9, 14\} $\rightarrow$ \{2, 9, 14\}, \{9, 14\}, \{5, 14\}, \{5, 9, 12\}, \{5, 9\}, \{4, 5, 9, 14\}, \{5, 7, 9, 14\}, \{5, 9, 13, 14\}, \{5, 9, 14, 15\}\\
\{5, 9, 14, 15\} $\rightarrow$ \{2, 9, 14, 15\}, \{9, 14, 15\}, \{5, 14, 15\}, \{5, 9, 15\}, \{5, 9, 11, 14\}, \{5, 9, 14\}, \{4, 5, 9, 14, 15\}, \{5, 7, 9, 14, 15\}, \{5, 9, 13, 14, 15\}\\
\{5, 9, 15\} $\rightarrow$ \{2, 9, 15\}, \{9, 15\}, \{5, 15\}, \{5, 9, 11\}, \{5, 9, 12\}, \{5, 9\}, \{4, 5, 9, 15\}, \{5, 7, 9, 15\}, \{5, 9, 13, 15\}, \{5, 9, 14, 15\}\\
\{5, 10\} $\rightarrow$ \{10\}, \{5, 11\}, \{5, 12\}, \{3, 5, 10\}, \{4, 5, 10\}, \{5, 7, 10\}, \{5, 9, 10\}\\
\{5, 11\} $\rightarrow$ \{1, 11\}, \{11\}, \{5, 10\}, \{5, 13\}, \{5, 15\}, \{3, 5, 11\}, \{4, 5, 11\}, \{5, 7, 11\}, \{5, 9, 11\}, \{5, 11, 14\}\\
\{5, 11, 14\} $\rightarrow$ \{1, 11, 14\}, \{11, 14\}, \{5, 13, 14\}, \{5, 14, 15\}, \{5, 11\}, \{3, 5, 11, 14\}, \{4, 5, 11, 14\}, \{5, 7, 11, 14\}, \{5, 9, 11, 14\}\\
\{5, 12\} $\rightarrow$ \{2, 12\}, \{12\}, \{5, 10\}, \{5, 14\}, \{5, 15\}, \{3, 5, 12\}, \{4, 5, 12\}, \{5, 7, 12\}, \{5, 9, 12\}, \{5, 12, 13\}\\
\{5, 12, 13\} $\rightarrow$ \{2, 12, 13\}, \{12, 13\}, \{5, 13, 14\}, \{5, 13, 15\}, \{5, 12\}, \{3, 5, 12, 13\}, \{4, 5, 12, 13\}, \{5, 7, 12, 13\}, \{5, 9, 12, 13\}\\
\{5, 13\} $\rightarrow$ \{1, 13\}, \{2, 13\}, \{13\}, \{5, 11\}, \{5\}, \{3, 5, 13\}, \{4, 5, 13\}, \{5, 7, 13\}, \{5, 9, 13\}, \{5, 13, 14\}, \{5, 13, 15\}\\
\{5, 13, 14\} $\rightarrow$ \{1, 13, 14\}, \{2, 13, 14\}, \{13, 14\}, \{5, 11, 14\}, \{5, 14\}, \{5, 12, 13\}, \{5, 13\}, \{3, 5, 13, 14\}, \{4, 5, 13, 14\}, \{5, 7, 13, 14\}, \{5, 9, 13, 14\}, \{5, 13, 14, 15\}\\
\{5, 13, 14, 15\} $\rightarrow$ \{1, 13, 14, 15\}, \{2, 13, 14, 15\}, \{13, 14, 15\}, \{5, 14, 15\}, \{5, 13, 15\}, \{5, 13, 14\}, \{3, 5, 13, 14, 15\}, \{4, 5, 13, 14, 15\}, \{5, 7, 13, 14, 15\}, \{5, 9, 13, 14, 15\}\\
\{5, 13, 15\} $\rightarrow$ \{1, 13, 15\}, \{2, 13, 15\}, \{13, 15\}, \{5, 15\}, \{5, 12, 13\}, \{5, 13\}, \{3, 5, 13, 15\}, \{4, 5, 13, 15\}, \{5, 7, 13, 15\}, \{5, 9, 13, 15\}, \{5, 13, 14, 15\}\\
\{5, 14\} $\rightarrow$ \{1, 14\}, \{2, 14\}, \{14\}, \{5, 12\}, \{5\}, \{3, 5, 14\}, \{4, 5, 14\}, \{5, 7, 14\}, \{5, 9, 14\}, \{5, 13, 14\}, \{5, 14, 15\}\\
\{5, 14, 15\} $\rightarrow$ \{1, 14, 15\}, \{2, 14, 15\}, \{14, 15\}, \{5, 15\}, \{5, 11, 14\}, \{5, 14\}, \{3, 5, 14, 15\}, \{4, 5, 14, 15\}, \{5, 7, 14, 15\}, \{5, 9, 14, 15\}, \{5, 13, 14, 15\}\\
\{5, 15\} $\rightarrow$ \{1, 15\}, \{2, 15\}, \{15\}, \{5, 11\}, \{5, 12\}, \{5\}, \{3, 5, 15\}, \{4, 5, 15\}, \{5, 7, 15\}, \{5, 9, 15\}, \{5, 13, 15\}, \{5, 14, 15\}\\
\{6, 9\} $\rightarrow$ \{7, 9\}, \{8, 9\}, \{6\}, \{6, 9, 13\}, \{6, 9, 14\}, \{6, 9, 15\}, \{0\}\\
\{6, 9, 12\} $\rightarrow$ \{7, 9, 12\}, \{8, 9, 12\}, \{6, 12\}, \{6, 9, 14\}, \{6, 9, 15\}, \{6, 9, 12, 13\}\\
\{6, 9, 12, 13\} $\rightarrow$ \{7, 9, 12, 13\}, \{8, 9, 12, 13\}, \{6, 12, 13\}, \{6, 9, 13, 14\}, \{6, 9, 13, 15\}, \{6, 9, 12\}\\
\{6, 9, 13\} $\rightarrow$ \{7, 9, 13\}, \{8, 9, 13\}, \{6, 13\}, \{6, 9\}, \{6, 9, 13, 14\}, \{6, 9, 13, 15\}, \{0, 13\}\\
\{6, 9, 13, 14\} $\rightarrow$ \{7, 9, 13, 14\}, \{8, 9, 13, 14\}, \{6, 13, 14\}, \{6, 9, 14\}, \{6, 9, 12, 13\}, \{6, 9, 13\}, \{6, 9, 13, 14, 15\}, \{0, 13, 14\}\\
\{6, 9, 13, 14, 15\} $\rightarrow$ \{7, 9, 13, 14, 15\}, \{8, 9, 13, 14, 15\}, \{6, 13, 14, 15\}, \{6, 9, 14, 15\}, \{6, 9, 13, 15\}, \{6, 9, 13, 14\}, \{0, 13, 14, 15\}\\
\{6, 9, 13, 15\} $\rightarrow$ \{7, 9, 13, 15\}, \{8, 9, 13, 15\}, \{6, 13, 15\}, \{6, 9, 15\}, \{6, 9, 12, 13\}, \{6, 9, 13\}, \{6, 9, 13, 14, 15\}, \{0, 13, 15\}\\
\{6, 9, 14\} $\rightarrow$ \{7, 9, 14\}, \{8, 9, 14\}, \{6, 14\}, \{6, 9, 12\}, \{6, 9\}, \{6, 9, 13, 14\}, \{6, 9, 14, 15\}, \{0, 14\}\\
\{6, 9, 14, 15\} $\rightarrow$ \{7, 9, 14, 15\}, \{8, 9, 14, 15\}, \{6, 14, 15\}, \{6, 9, 15\}, \{6, 9, 14\}, \{6, 9, 13, 14, 15\}, \{0, 14, 15\}\\
\{6, 9, 15\} $\rightarrow$ \{7, 9, 15\}, \{8, 9, 15\}, \{6, 15\}, \{6, 9, 12\}, \{6, 9\}, \{6, 9, 13, 15\}, \{6, 9, 14, 15\}, \{0, 15\}\\
\{6, 12\} $\rightarrow$ \{7, 12\}, \{8, 12\}, \{6, 14\}, \{6, 15\}, \{3, 6, 12\}, \{6, 9, 12\}, \{6, 12, 13\}\\
\{6, 12, 13\} $\rightarrow$ \{7, 12, 13\}, \{8, 12, 13\}, \{6, 13, 14\}, \{6, 13, 15\}, \{6, 12\}, \{3, 6, 12, 13\}, \{6, 9, 12, 13\}\\
\{6, 13\} $\rightarrow$ \{7, 13\}, \{8, 13\}, \{6\}, \{3, 6, 13\}, \{6, 9, 13\}, \{6, 13, 14\}, \{6, 13, 15\}\\
\{6, 13, 14\} $\rightarrow$ \{7, 13, 14\}, \{8, 13, 14\}, \{6, 14\}, \{6, 12, 13\}, \{6, 13\}, \{3, 6, 13, 14\}, \{6, 9, 13, 14\}, \{6, 13, 14, 15\}\\
\{6, 13, 14, 15\} $\rightarrow$ \{7, 13, 14, 15\}, \{8, 13, 14, 15\}, \{6, 14, 15\}, \{6, 13, 15\}, \{6, 13, 14\}, \{3, 6, 13, 14, 15\}, \{6, 9, 13, 14, 15\}\\
\{6, 13, 15\} $\rightarrow$ \{7, 13, 15\}, \{8, 13, 15\}, \{6, 15\}, \{6, 12, 13\}, \{6, 13\}, \{3, 6, 13, 15\}, \{6, 9, 13, 15\}, \{6, 13, 14, 15\}\\
\{6, 14\} $\rightarrow$ \{7, 14\}, \{8, 14\}, \{6, 12\}, \{6\}, \{3, 6, 14\}, \{6, 9, 14\}, \{6, 13, 14\}, \{6, 14, 15\}\\
\{6, 14, 15\} $\rightarrow$ \{7, 14, 15\}, \{8, 14, 15\}, \{6, 15\}, \{6, 14\}, \{3, 6, 14, 15\}, \{6, 9, 14, 15\}, \{6, 13, 14, 15\}\\
\{6, 15\} $\rightarrow$ \{7, 15\}, \{8, 15\}, \{6, 12\}, \{6\}, \{3, 6, 15\}, \{6, 9, 15\}, \{6, 13, 15\}, \{6, 14, 15\}\\
\{7, 8\} $\rightarrow$ \{8\}, \{7\}, \{3, 7, 8\}, \{7, 8, 9\}, \{7, 8, 13\}, \{7, 8, 14\}, \{7, 8, 15\}\\
\{7, 8, 9\} $\rightarrow$ \{8, 9\}, \{7, 9\}, \{7, 8\}, \{7, 8, 9, 13\}, \{7, 8, 9, 14\}, \{7, 8, 9, 15\}\\
\{7, 8, 9, 10\} $\rightarrow$ \{8, 9, 10\}, \{7, 9, 10\}, \{7, 8, 10\}, \{7, 8, 9, 11\}, \{7, 8, 9, 12\}\\
\{7, 8, 9, 11\} $\rightarrow$ \{8, 9, 11\}, \{7, 9, 11\}, \{7, 8, 11\}, \{7, 8, 9, 10\}, \{7, 8, 9, 13\}, \{7, 8, 9, 15\}, \{7, 8, 9, 11, 14\}\\
\{7, 8, 9, 11, 14\} $\rightarrow$ \{8, 9, 11, 14\}, \{7, 9, 11, 14\}, \{7, 8, 11, 14\}, \{7, 8, 9, 13, 14\}, \{7, 8, 9, 14, 15\}, \{7, 8, 9, 11\}\\
\{7, 8, 9, 12\} $\rightarrow$ \{8, 9, 12\}, \{7, 9, 12\}, \{7, 8, 12\}, \{7, 8, 9, 10\}, \{7, 8, 9, 14\}, \{7, 8, 9, 15\}, \{7, 8, 9, 12, 13\}\\
\{7, 8, 9, 12, 13\} $\rightarrow$ \{8, 9, 12, 13\}, \{7, 9, 12, 13\}, \{7, 8, 12, 13\}, \{7, 8, 9, 13, 14\}, \{7, 8, 9, 13, 15\}, \{7, 8, 9, 12\}\\
\{7, 8, 9, 13\} $\rightarrow$ \{8, 9, 13\}, \{7, 9, 13\}, \{7, 8, 13\}, \{7, 8, 9, 11\}, \{7, 8, 9\}, \{7, 8, 9, 13, 14\}, \{7, 8, 9, 13, 15\}\\
\{7, 8, 9, 13, 14\} $\rightarrow$ \{8, 9, 13, 14\}, \{7, 9, 13, 14\}, \{7, 8, 13, 14\}, \{7, 8, 9, 11, 14\}, \{7, 8, 9, 14\}, \{7, 8, 9, 12, 13\}, \{7, 8, 9, 13\}, \{7, 8, 9, 13, 14, 15\}\\
\{7, 8, 9, 13, 14, 15\} $\rightarrow$ \{8, 9, 13, 14, 15\}, \{7, 9, 13, 14, 15\}, \{7, 8, 13, 14, 15\}, \{7, 8, 9, 14, 15\}, \{7, 8, 9, 13, 15\}, \{7, 8, 9, 13, 14\}\\
\{7, 8, 9, 13, 15\} $\rightarrow$ \{8, 9, 13, 15\}, \{7, 9, 13, 15\}, \{7, 8, 13, 15\}, \{7, 8, 9, 15\}, \{7, 8, 9, 12, 13\}, \{7, 8, 9, 13\}, \{7, 8, 9, 13, 14, 15\}\\
\{7, 8, 9, 14\} $\rightarrow$ \{8, 9, 14\}, \{7, 9, 14\}, \{7, 8, 14\}, \{7, 8, 9, 12\}, \{7, 8, 9\}, \{7, 8, 9, 13, 14\}, \{7, 8, 9, 14, 15\}\\
\{7, 8, 9, 14, 15\} $\rightarrow$ \{8, 9, 14, 15\}, \{7, 9, 14, 15\}, \{7, 8, 14, 15\}, \{7, 8, 9, 15\}, \{7, 8, 9, 11, 14\}, \{7, 8, 9, 14\}, \{7, 8, 9, 13, 14, 15\}\\
\{7, 8, 9, 15\} $\rightarrow$ \{8, 9, 15\}, \{7, 9, 15\}, \{7, 8, 15\}, \{7, 8, 9, 11\}, \{7, 8, 9, 12\}, \{7, 8, 9\}, \{7, 8, 9, 13, 15\}, \{7, 8, 9, 14, 15\}\\
\{7, 8, 10\} $\rightarrow$ \{8, 10\}, \{7, 10\}, \{7, 8, 11\}, \{7, 8, 12\}, \{3, 7, 8, 10\}, \{7, 8, 9, 10\}\\
\{7, 8, 11\} $\rightarrow$ \{8, 11\}, \{7, 11\}, \{7, 8, 10\}, \{7, 8, 13\}, \{7, 8, 15\}, \{3, 7, 8, 11\}, \{7, 8, 9, 11\}, \{7, 8, 11, 14\}\\
\{7, 8, 11, 14\} $\rightarrow$ \{8, 11, 14\}, \{7, 11, 14\}, \{7, 8, 13, 14\}, \{7, 8, 14, 15\}, \{7, 8, 11\}, \{3, 7, 8, 11, 14\}, \{7, 8, 9, 11, 14\}\\
\{7, 8, 12\} $\rightarrow$ \{8, 12\}, \{7, 12\}, \{7, 8, 10\}, \{7, 8, 14\}, \{7, 8, 15\}, \{3, 7, 8, 12\}, \{7, 8, 9, 12\}, \{7, 8, 12, 13\}\\
\{7, 8, 12, 13\} $\rightarrow$ \{8, 12, 13\}, \{7, 12, 13\}, \{7, 8, 13, 14\}, \{7, 8, 13, 15\}, \{7, 8, 12\}, \{3, 7, 8, 12, 13\}, \{7, 8, 9, 12, 13\}\\
\{7, 8, 13\} $\rightarrow$ \{8, 13\}, \{7, 13\}, \{7, 8, 11\}, \{7, 8\}, \{3, 7, 8, 13\}, \{7, 8, 9, 13\}, \{7, 8, 13, 14\}, \{7, 8, 13, 15\}\\
\{7, 8, 13, 14\} $\rightarrow$ \{8, 13, 14\}, \{7, 13, 14\}, \{7, 8, 11, 14\}, \{7, 8, 14\}, \{7, 8, 12, 13\}, \{7, 8, 13\}, \{3, 7, 8, 13, 14\}, \{7, 8, 9, 13, 14\}, \{7, 8, 13, 14, 15\}\\
\{7, 8, 13, 14, 15\} $\rightarrow$ \{8, 13, 14, 15\}, \{7, 13, 14, 15\}, \{7, 8, 14, 15\}, \{7, 8, 13, 15\}, \{7, 8, 13, 14\}, \{3, 7, 8, 13, 14, 15\}, \{7, 8, 9, 13, 14, 15\}\\
\{7, 8, 13, 15\} $\rightarrow$ \{8, 13, 15\}, \{7, 13, 15\}, \{7, 8, 15\}, \{7, 8, 12, 13\}, \{7, 8, 13\}, \{3, 7, 8, 13, 15\}, \{7, 8, 9, 13, 15\}, \{7, 8, 13, 14, 15\}\\
\{7, 8, 14\} $\rightarrow$ \{8, 14\}, \{7, 14\}, \{7, 8, 12\}, \{7, 8\}, \{3, 7, 8, 14\}, \{7, 8, 9, 14\}, \{7, 8, 13, 14\}, \{7, 8, 14, 15\}\\
\{7, 8, 14, 15\} $\rightarrow$ \{8, 14, 15\}, \{7, 14, 15\}, \{7, 8, 15\}, \{7, 8, 11, 14\}, \{7, 8, 14\}, \{3, 7, 8, 14, 15\}, \{7, 8, 9, 14, 15\}, \{7, 8, 13, 14, 15\}\\
\{7, 8, 15\} $\rightarrow$ \{8, 15\}, \{7, 15\}, \{7, 8, 11\}, \{7, 8, 12\}, \{7, 8\}, \{3, 7, 8, 15\}, \{7, 8, 9, 15\}, \{7, 8, 13, 15\}, \{7, 8, 14, 15\}\\
\{7, 9\} $\rightarrow$ \{6, 9\}, \{9\}, \{7\}, \{5, 7, 9\}, \{7, 8, 9\}, \{7, 9, 13\}, \{7, 9, 14\}, \{7, 9, 15\}\\
\{7, 9, 10\} $\rightarrow$ \{9, 10\}, \{7, 10\}, \{7, 9, 11\}, \{7, 9, 12\}, \{5, 7, 9, 10\}, \{7, 8, 9, 10\}\\
\{7, 9, 11\} $\rightarrow$ \{9, 11\}, \{7, 11\}, \{7, 9, 10\}, \{7, 9, 13\}, \{7, 9, 15\}, \{5, 7, 9, 11\}, \{7, 8, 9, 11\}, \{7, 9, 11, 14\}\\
\{7, 9, 11, 14\} $\rightarrow$ \{9, 11, 14\}, \{7, 11, 14\}, \{7, 9, 13, 14\}, \{7, 9, 14, 15\}, \{7, 9, 11\}, \{5, 7, 9, 11, 14\}, \{7, 8, 9, 11, 14\}\\
\{7, 9, 12\} $\rightarrow$ \{6, 9, 12\}, \{9, 12\}, \{7, 12\}, \{7, 9, 10\}, \{7, 9, 14\}, \{7, 9, 15\}, \{5, 7, 9, 12\}, \{7, 8, 9, 12\}, \{7, 9, 12, 13\}\\
\{7, 9, 12, 13\} $\rightarrow$ \{6, 9, 12, 13\}, \{9, 12, 13\}, \{7, 12, 13\}, \{7, 9, 13, 14\}, \{7, 9, 13, 15\}, \{7, 9, 12\}, \{5, 7, 9, 12, 13\}, \{7, 8, 9, 12, 13\}\\
\{7, 9, 13\} $\rightarrow$ \{6, 9, 13\}, \{9, 13\}, \{7, 13\}, \{7, 9, 11\}, \{7, 9\}, \{5, 7, 9, 13\}, \{7, 8, 9, 13\}, \{7, 9, 13, 14\}, \{7, 9, 13, 15\}\\
\{7, 9, 13, 14\} $\rightarrow$ \{6, 9, 13, 14\}, \{9, 13, 14\}, \{7, 13, 14\}, \{7, 9, 11, 14\}, \{7, 9, 14\}, \{7, 9, 12, 13\}, \{7, 9, 13\}, \{5, 7, 9, 13, 14\}, \{7, 8, 9, 13, 14\}, \{7, 9, 13, 14, 15\}\\
\{7, 9, 13, 14, 15\} $\rightarrow$ \{6, 9, 13, 14, 15\}, \{9, 13, 14, 15\}, \{7, 13, 14, 15\}, \{7, 9, 14, 15\}, \{7, 9, 13, 15\}, \{7, 9, 13, 14\}, \{5, 7, 9, 13, 14, 15\}, \{7, 8, 9, 13, 14, 15\}\\
\{7, 9, 13, 15\} $\rightarrow$ \{6, 9, 13, 15\}, \{9, 13, 15\}, \{7, 13, 15\}, \{7, 9, 15\}, \{7, 9, 12, 13\}, \{7, 9, 13\}, \{5, 7, 9, 13, 15\}, \{7, 8, 9, 13, 15\}, \{7, 9, 13, 14, 15\}\\
\{7, 9, 14\} $\rightarrow$ \{6, 9, 14\}, \{9, 14\}, \{7, 14\}, \{7, 9, 12\}, \{7, 9\}, \{5, 7, 9, 14\}, \{7, 8, 9, 14\}, \{7, 9, 13, 14\}, \{7, 9, 14, 15\}\\
\{7, 9, 14, 15\} $\rightarrow$ \{6, 9, 14, 15\}, \{9, 14, 15\}, \{7, 14, 15\}, \{7, 9, 15\}, \{7, 9, 11, 14\}, \{7, 9, 14\}, \{5, 7, 9, 14, 15\}, \{7, 8, 9, 14, 15\}, \{7, 9, 13, 14, 15\}\\
\{7, 9, 15\} $\rightarrow$ \{6, 9, 15\}, \{9, 15\}, \{7, 15\}, \{7, 9, 11\}, \{7, 9, 12\}, \{7, 9\}, \{5, 7, 9, 15\}, \{7, 8, 9, 15\}, \{7, 9, 13, 15\}, \{7, 9, 14, 15\}\\
\{7, 10\} $\rightarrow$ \{10\}, \{7, 11\}, \{7, 12\}, \{3, 7, 10\}, \{5, 7, 10\}, \{7, 8, 10\}, \{7, 9, 10\}\\
\{7, 11\} $\rightarrow$ \{11\}, \{7, 10\}, \{7, 13\}, \{7, 15\}, \{3, 7, 11\}, \{5, 7, 11\}, \{7, 8, 11\}, \{7, 9, 11\}, \{7, 11, 14\}\\
\{7, 11, 14\} $\rightarrow$ \{11, 14\}, \{7, 13, 14\}, \{7, 14, 15\}, \{7, 11\}, \{3, 7, 11, 14\}, \{5, 7, 11, 14\}, \{7, 8, 11, 14\}, \{7, 9, 11, 14\}\\
\{7, 12\} $\rightarrow$ \{6, 12\}, \{12\}, \{7, 10\}, \{7, 14\}, \{7, 15\}, \{3, 7, 12\}, \{5, 7, 12\}, \{7, 8, 12\}, \{7, 9, 12\}, \{7, 12, 13\}\\
\{7, 12, 13\} $\rightarrow$ \{6, 12, 13\}, \{12, 13\}, \{7, 13, 14\}, \{7, 13, 15\}, \{7, 12\}, \{3, 7, 12, 13\}, \{5, 7, 12, 13\}, \{7, 8, 12, 13\}, \{7, 9, 12, 13\}\\
\{7, 13\} $\rightarrow$ \{6, 13\}, \{13\}, \{7, 11\}, \{7\}, \{3, 7, 13\}, \{5, 7, 13\}, \{7, 8, 13\}, \{7, 9, 13\}, \{7, 13, 14\}, \{7, 13, 15\}\\
\{7, 13, 14\} $\rightarrow$ \{6, 13, 14\}, \{13, 14\}, \{7, 11, 14\}, \{7, 14\}, \{7, 12, 13\}, \{7, 13\}, \{3, 7, 13, 14\}, \{5, 7, 13, 14\}, \{7, 8, 13, 14\}, \{7, 9, 13, 14\}, \{7, 13, 14, 15\}\\
\{7, 13, 14, 15\} $\rightarrow$ \{6, 13, 14, 15\}, \{13, 14, 15\}, \{7, 14, 15\}, \{7, 13, 15\}, \{7, 13, 14\}, \{3, 7, 13, 14, 15\}, \{5, 7, 13, 14, 15\}, \{7, 8, 13, 14, 15\}, \{7, 9, 13, 14, 15\}\\
\{7, 13, 15\} $\rightarrow$ \{6, 13, 15\}, \{13, 15\}, \{7, 15\}, \{7, 12, 13\}, \{7, 13\}, \{3, 7, 13, 15\}, \{5, 7, 13, 15\}, \{7, 8, 13, 15\}, \{7, 9, 13, 15\}, \{7, 13, 14, 15\}\\
\{7, 14\} $\rightarrow$ \{6, 14\}, \{14\}, \{7, 12\}, \{7\}, \{3, 7, 14\}, \{5, 7, 14\}, \{7, 8, 14\}, \{7, 9, 14\}, \{7, 13, 14\}, \{7, 14, 15\}\\
\{7, 14, 15\} $\rightarrow$ \{6, 14, 15\}, \{14, 15\}, \{7, 15\}, \{7, 11, 14\}, \{7, 14\}, \{3, 7, 14, 15\}, \{5, 7, 14, 15\}, \{7, 8, 14, 15\}, \{7, 9, 14, 15\}, \{7, 13, 14, 15\}\\
\{7, 15\} $\rightarrow$ \{6, 15\}, \{15\}, \{7, 11\}, \{7, 12\}, \{7\}, \{3, 7, 15\}, \{5, 7, 15\}, \{7, 8, 15\}, \{7, 9, 15\}, \{7, 13, 15\}, \{7, 14, 15\}\\
\{8, 9\} $\rightarrow$ \{6, 9\}, \{9\}, \{8\}, \{4, 8, 9\}, \{7, 8, 9\}, \{8, 9, 13\}, \{8, 9, 14\}, \{8, 9, 15\}\\
\{8, 9, 10\} $\rightarrow$ \{9, 10\}, \{8, 10\}, \{8, 9, 11\}, \{8, 9, 12\}, \{4, 8, 9, 10\}, \{7, 8, 9, 10\}\\
\{8, 9, 11\} $\rightarrow$ \{9, 11\}, \{8, 11\}, \{8, 9, 10\}, \{8, 9, 13\}, \{8, 9, 15\}, \{4, 8, 9, 11\}, \{7, 8, 9, 11\}, \{8, 9, 11, 14\}\\
\{8, 9, 11, 14\} $\rightarrow$ \{9, 11, 14\}, \{8, 11, 14\}, \{8, 9, 13, 14\}, \{8, 9, 14, 15\}, \{8, 9, 11\}, \{4, 8, 9, 11, 14\}, \{7, 8, 9, 11, 14\}\\
\{8, 9, 12\} $\rightarrow$ \{6, 9, 12\}, \{9, 12\}, \{8, 12\}, \{8, 9, 10\}, \{8, 9, 14\}, \{8, 9, 15\}, \{4, 8, 9, 12\}, \{7, 8, 9, 12\}, \{8, 9, 12, 13\}\\
\{8, 9, 12, 13\} $\rightarrow$ \{6, 9, 12, 13\}, \{9, 12, 13\}, \{8, 12, 13\}, \{8, 9, 13, 14\}, \{8, 9, 13, 15\}, \{8, 9, 12\}, \{4, 8, 9, 12, 13\}, \{7, 8, 9, 12, 13\}\\
\{8, 9, 13\} $\rightarrow$ \{6, 9, 13\}, \{9, 13\}, \{8, 13\}, \{8, 9, 11\}, \{8, 9\}, \{4, 8, 9, 13\}, \{7, 8, 9, 13\}, \{8, 9, 13, 14\}, \{8, 9, 13, 15\}\\
\{8, 9, 13, 14\} $\rightarrow$ \{6, 9, 13, 14\}, \{9, 13, 14\}, \{8, 13, 14\}, \{8, 9, 11, 14\}, \{8, 9, 14\}, \{8, 9, 12, 13\}, \{8, 9, 13\}, \{4, 8, 9, 13, 14\}, \{7, 8, 9, 13, 14\}, \{8, 9, 13, 14, 15\}\\
\{8, 9, 13, 14, 15\} $\rightarrow$ \{6, 9, 13, 14, 15\}, \{9, 13, 14, 15\}, \{8, 13, 14, 15\}, \{8, 9, 14, 15\}, \{8, 9, 13, 15\}, \{8, 9, 13, 14\}, \{4, 8, 9, 13, 14, 15\}, \{7, 8, 9, 13, 14, 15\}\\
\{8, 9, 13, 15\} $\rightarrow$ \{6, 9, 13, 15\}, \{9, 13, 15\}, \{8, 13, 15\}, \{8, 9, 15\}, \{8, 9, 12, 13\}, \{8, 9, 13\}, \{4, 8, 9, 13, 15\}, \{7, 8, 9, 13, 15\}, \{8, 9, 13, 14, 15\}\\
\{8, 9, 14\} $\rightarrow$ \{6, 9, 14\}, \{9, 14\}, \{8, 14\}, \{8, 9, 12\}, \{8, 9\}, \{4, 8, 9, 14\}, \{7, 8, 9, 14\}, \{8, 9, 13, 14\}, \{8, 9, 14, 15\}\\
\{8, 9, 14, 15\} $\rightarrow$ \{6, 9, 14, 15\}, \{9, 14, 15\}, \{8, 14, 15\}, \{8, 9, 15\}, \{8, 9, 11, 14\}, \{8, 9, 14\}, \{4, 8, 9, 14, 15\}, \{7, 8, 9, 14, 15\}, \{8, 9, 13, 14, 15\}\\
\{8, 9, 15\} $\rightarrow$ \{6, 9, 15\}, \{9, 15\}, \{8, 15\}, \{8, 9, 11\}, \{8, 9, 12\}, \{8, 9\}, \{4, 8, 9, 15\}, \{7, 8, 9, 15\}, \{8, 9, 13, 15\}, \{8, 9, 14, 15\}\\
\{8, 10\} $\rightarrow$ \{10\}, \{8, 11\}, \{8, 12\}, \{3, 8, 10\}, \{4, 8, 10\}, \{7, 8, 10\}, \{8, 9, 10\}\\
\{8, 11\} $\rightarrow$ \{11\}, \{8, 10\}, \{8, 13\}, \{8, 15\}, \{3, 8, 11\}, \{4, 8, 11\}, \{7, 8, 11\}, \{8, 9, 11\}, \{8, 11, 14\}\\
\{8, 11, 14\} $\rightarrow$ \{11, 14\}, \{8, 13, 14\}, \{8, 14, 15\}, \{8, 11\}, \{3, 8, 11, 14\}, \{4, 8, 11, 14\}, \{7, 8, 11, 14\}, \{8, 9, 11, 14\}\\
\{8, 12\} $\rightarrow$ \{6, 12\}, \{12\}, \{8, 10\}, \{8, 14\}, \{8, 15\}, \{3, 8, 12\}, \{4, 8, 12\}, \{7, 8, 12\}, \{8, 9, 12\}, \{8, 12, 13\}\\
\{8, 12, 13\} $\rightarrow$ \{6, 12, 13\}, \{12, 13\}, \{8, 13, 14\}, \{8, 13, 15\}, \{8, 12\}, \{3, 8, 12, 13\}, \{4, 8, 12, 13\}, \{7, 8, 12, 13\}, \{8, 9, 12, 13\}\\
\{8, 13\} $\rightarrow$ \{6, 13\}, \{13\}, \{8, 11\}, \{8\}, \{3, 8, 13\}, \{4, 8, 13\}, \{7, 8, 13\}, \{8, 9, 13\}, \{8, 13, 14\}, \{8, 13, 15\}\\
\{8, 13, 14\} $\rightarrow$ \{6, 13, 14\}, \{13, 14\}, \{8, 11, 14\}, \{8, 14\}, \{8, 12, 13\}, \{8, 13\}, \{3, 8, 13, 14\}, \{4, 8, 13, 14\}, \{7, 8, 13, 14\}, \{8, 9, 13, 14\}, \{8, 13, 14, 15\}\\
\{8, 13, 14, 15\} $\rightarrow$ \{6, 13, 14, 15\}, \{13, 14, 15\}, \{8, 14, 15\}, \{8, 13, 15\}, \{8, 13, 14\}, \{3, 8, 13, 14, 15\}, \{4, 8, 13, 14, 15\}, \{7, 8, 13, 14, 15\}, \{8, 9, 13, 14, 15\}\\
\{8, 13, 15\} $\rightarrow$ \{6, 13, 15\}, \{13, 15\}, \{8, 15\}, \{8, 12, 13\}, \{8, 13\}, \{3, 8, 13, 15\}, \{4, 8, 13, 15\}, \{7, 8, 13, 15\}, \{8, 9, 13, 15\}, \{8, 13, 14, 15\}\\
\{8, 14\} $\rightarrow$ \{6, 14\}, \{14\}, \{8, 12\}, \{8\}, \{3, 8, 14\}, \{4, 8, 14\}, \{7, 8, 14\}, \{8, 9, 14\}, \{8, 13, 14\}, \{8, 14, 15\}\\
\{8, 14, 15\} $\rightarrow$ \{6, 14, 15\}, \{14, 15\}, \{8, 15\}, \{8, 11, 14\}, \{8, 14\}, \{3, 8, 14, 15\}, \{4, 8, 14, 15\}, \{7, 8, 14, 15\}, \{8, 9, 14, 15\}, \{8, 13, 14, 15\}\\
\{8, 15\} $\rightarrow$ \{6, 15\}, \{15\}, \{8, 11\}, \{8, 12\}, \{8\}, \{3, 8, 15\}, \{4, 8, 15\}, \{7, 8, 15\}, \{8, 9, 15\}, \{8, 13, 15\}, \{8, 14, 15\}\\
\{9, 10\} $\rightarrow$ \{10\}, \{9, 11\}, \{9, 12\}, \{4, 9, 10\}, \{5, 9, 10\}, \{7, 9, 10\}, \{8, 9, 10\}\\
\{9, 11\} $\rightarrow$ \{11\}, \{9, 10\}, \{9, 13\}, \{9, 15\}, \{4, 9, 11\}, \{5, 9, 11\}, \{7, 9, 11\}, \{8, 9, 11\}, \{9, 11, 14\}\\
\{9, 11, 14\} $\rightarrow$ \{11, 14\}, \{9, 13, 14\}, \{9, 14, 15\}, \{9, 11\}, \{4, 9, 11, 14\}, \{5, 9, 11, 14\}, \{7, 9, 11, 14\}, \{8, 9, 11, 14\}\\
\{9, 12\} $\rightarrow$ \{12\}, \{9, 10\}, \{9, 14\}, \{9, 15\}, \{4, 9, 12\}, \{5, 9, 12\}, \{7, 9, 12\}, \{8, 9, 12\}, \{9, 12, 13\}\\
\{9, 12, 13\} $\rightarrow$ \{12, 13\}, \{9, 13, 14\}, \{9, 13, 15\}, \{9, 12\}, \{4, 9, 12, 13\}, \{5, 9, 12, 13\}, \{7, 9, 12, 13\}, \{8, 9, 12, 13\}\\
\{9, 13\} $\rightarrow$ \{13\}, \{9, 11\}, \{9\}, \{4, 9, 13\}, \{5, 9, 13\}, \{7, 9, 13\}, \{8, 9, 13\}, \{9, 13, 14\}, \{9, 13, 15\}\\
\{9, 13, 14\} $\rightarrow$ \{13, 14\}, \{9, 11, 14\}, \{9, 14\}, \{9, 12, 13\}, \{9, 13\}, \{4, 9, 13, 14\}, \{5, 9, 13, 14\}, \{7, 9, 13, 14\}, \{8, 9, 13, 14\}, \{9, 13, 14, 15\}\\
\{9, 13, 14, 15\} $\rightarrow$ \{13, 14, 15\}, \{9, 14, 15\}, \{9, 13, 15\}, \{9, 13, 14\}, \{4, 9, 13, 14, 15\}, \{5, 9, 13, 14, 15\}, \{7, 9, 13, 14, 15\}, \{8, 9, 13, 14, 15\}\\
\{9, 13, 15\} $\rightarrow$ \{13, 15\}, \{9, 15\}, \{9, 12, 13\}, \{9, 13\}, \{4, 9, 13, 15\}, \{5, 9, 13, 15\}, \{7, 9, 13, 15\}, \{8, 9, 13, 15\}, \{9, 13, 14, 15\}\\
\{9, 14\} $\rightarrow$ \{14\}, \{9, 12\}, \{9\}, \{4, 9, 14\}, \{5, 9, 14\}, \{7, 9, 14\}, \{8, 9, 14\}, \{9, 13, 14\}, \{9, 14, 15\}\\
\{9, 14, 15\} $\rightarrow$ \{14, 15\}, \{9, 15\}, \{9, 11, 14\}, \{9, 14\}, \{4, 9, 14, 15\}, \{5, 9, 14, 15\}, \{7, 9, 14, 15\}, \{8, 9, 14, 15\}, \{9, 13, 14, 15\}\\
\{9, 15\} $\rightarrow$ \{15\}, \{9, 11\}, \{9, 12\}, \{9\}, \{4, 9, 15\}, \{5, 9, 15\}, \{7, 9, 15\}, \{8, 9, 15\}, \{9, 13, 15\}, \{9, 14, 15\}\\
\{11, 14\} $\rightarrow$ \{13, 14\}, \{14, 15\}, \{11\}, \{3, 11, 14\}, \{4, 11, 14\}, \{5, 11, 14\}, \{7, 11, 14\}, \{8, 11, 14\}, \{9, 11, 14\}\\
\{12, 13\} $\rightarrow$ \{13, 14\}, \{13, 15\}, \{12\}, \{3, 12, 13\}, \{4, 12, 13\}, \{5, 12, 13\}, \{7, 12, 13\}, \{8, 12, 13\}, \{9, 12, 13\}\\
\{13, 14\} $\rightarrow$ \{11, 14\}, \{14\}, \{12, 13\}, \{13\}, \{3, 13, 14\}, \{4, 13, 14\}, \{5, 13, 14\}, \{7, 13, 14\}, \{8, 13, 14\}, \{9, 13, 14\}, \{13, 14, 15\}\\
\{13, 14, 15\} $\rightarrow$ \{14, 15\}, \{13, 15\}, \{13, 14\}, \{3, 13, 14, 15\}, \{4, 13, 14, 15\}, \{5, 13, 14, 15\}, \{7, 13, 14, 15\}, \{8, 13, 14, 15\}, \{9, 13, 14, 15\}\\
\{13, 15\} $\rightarrow$ \{15\}, \{12, 13\}, \{13\}, \{3, 13, 15\}, \{4, 13, 15\}, \{5, 13, 15\}, \{7, 13, 15\}, \{8, 13, 15\}, \{9, 13, 15\}, \{13, 14, 15\}\\
\{14, 15\} $\rightarrow$ \{15\}, \{11, 14\}, \{14\}, \{3, 14, 15\}, \{4, 14, 15\}, \{5, 14, 15\}, \{7, 14, 15\}, \{8, 14, 15\}, \{9, 14, 15\}, \{13, 14, 15\}

\normalsize

To be completed...

\section*{Supplement references}
 
\printbibliography[heading = none]

\end{refsection}

\end{document}